\documentclass[twocolumn, superscriptaddress, prb, preprintnumbers, showpacs, floatfix]{revtex4-1}

\usepackage{ifthen} 
\usepackage{graphicx}
\usepackage{amsmath}
\usepackage{braket}
\usepackage{cancel}
\usepackage{color}
\def\rd#1{#1}

\begin{document}

\title{Optimization of constrained density functional theory}

\author{David D. O'Regan}
\email{daithi.o.riogain@tcd.ie}
\affiliation{School of Physics, CRANN and AMBER, 
Trinity College Dublin, Dublin 2, Ireland}

\author{Gilberto Teobaldi}
\email{g.teobaldi@liverpool.ac.uk}
\affiliation{Stephenson Institute for Renewable Energy and  Department of Chemistry, The University of Liverpool, L69 3BX Liverpool, United Kingdom}
\affiliation{Beijing Computational Science Research Center, Beijing 100094, China}

\date{\today{}}

\begin{abstract}

Constrained density functional theory (cDFT) is a versatile electronic structure method \rd{that enables} ground-state calculations to be performed subject to physical constraints. \rd{It thereby broadens} their applicability and utility.
\rd{Automated Lagrange multiplier optimisation is necessary
for multiple constraints to be applied efficiently in cDFT, for 
it to be used in tandem with geometry optimization, or  with 
molecular dynamics.
In order to facilitate this, 
we comprehensively develop the connection between
cDFT energy derivatives and  response 
functions, providing a rigorous assessment 
of the uniqueness and character of cDFT stationary points
while accounting for electronic interactions and  screening.
In particular, we provide a  
new, non-perturbative proof that stable stationary points of
linear density constraints 
occur only at energy maxima with respect to 
their Lagrange multipliers. 
We show that multiple  solutions, hysteresis, and energy
discontinuities may occur in cDFT. 
Expressions are derived, in terms of convenient 
by-products of cDFT optimization,
for  quantities such as the dielectric function 
and a condition number quantifying
ill-definition in multi-constraint cDFT.}

\end{abstract}


\maketitle

\section{Background}
\label{Sec:Introduction}

Constrained density functional theory 
(cDFT)~\cite{doi10.1021/cr200148b} is a generalization of 
density functional theory (DFT)~\cite{PhysRev.136.B864, 
PhysRev.140.A1133} in which external constraints are applied
in order to simulate excitation processes, 
to calculate response properties, or to impose a physical condition 
that is not met by the unconstrained approximate 
exchange-correlation functional.
Such constraints may be applied to 
expectation values of
the charge or spin density, their
sums, differences, and moments within 
pre-defined spatial regions~\cite{doi:10.1021/ct0503163,
PhysRevA.72.024502,  
doi:10.1038/nmat1885, doi:10.1021/jp103153a,
PhysRevLett.53.2512,
doi:10.1021/jp061848y, :/content/aip/journal/jcp/131/6/10.1063/1.3190169, PhysRevLett.97.028303,
doi:10.1021/jp204962k,  PhysRevB.77.115421,
:/content/aip/journal/jcp/124/2/10.1063/1.2145878,
doi:10.1021/ic700871f, 
:/content/aip/journal/jcp/129/11/10.1063/1.2978168}.
\rd{They are necessarily} chosen on the basis of physical intuition and 
experience.
Constraining potentials 
that are non-local or orbital-dependent 
may also be introduced, moving beyond formal 
DFT~\cite{doi:10.1021/ct0503163}.
cDFT enables individual excited states to be studied within the well-established framework of ground-state DFT~\cite{PhysRevA.72.024502,
doi:10.1021/ct0503163,doi:10.1038/nmat1885,
doi:10.1021/ja076125m,
doi:10.1021/jp103153a,doi:10.1021/ct100601n, PhysRevB.93.165102}, 
\rd{particularly} those excited states which may be 
represented as the ground-state for some 
potential.
While these excitations are not guaranteed
 to match the neutral excitations of the system, yielded 
by time-dependent DFT\rd{~\cite{PhysRevLett.52.997}, for example,}
their description may 
nonetheless benefit from
physical conditions, such as charge
transfer, which \rd{may be absent from the 
approximate functional but reintroduced using cDFT}.
As such, cDFT offers important insights that are 
challenging to obtain otherwise~\cite{PhysRevA.72.024502,
doi:10.1021/ja076125m,
doi:10.1021/ct100508y,PhysRevB.77.195126, PhysRevLett.97.028303}.

In practice, cDFT has proven to be a very efficient approach for simulating neutral excitations in molecular systems, particularly in cases where a clear spatial delineation may be made between charge (or spin) donor and acceptor 
regions~\cite{PhysRevA.72.024502,doi:10.1021/ct0503163,doi:10.1021/ja076125m, 
doi:10.1038/nmat1885, doi:10.1021/jp103153a, doi:10.1021/ct100601n, 
PhysRevB.93.045130, PhysRevB.88.165112, PhysRevB.93.165102, doi:10.1021/ct100508y, 
PhysRevB.77.195126}.
cDFT is a significant asset, therefore, \rd{to} the simulation 
of exciton formation,
where the incorrect long-ranged behaviour of 
conventional local or semi-local exchange-correlation 
functionals may be partially corrected by 
using appropriately constructed constraints~\cite{PhysRevLett.101.026804,
PhysRevB.88.165112, PhysRevB.93.045130, PhysRevB.91.195438, PhysRevB.93.165102}. 
It has also been used to calculate
electron transfer~\cite{PhysRevLett.97.028303,:/content/aip/journal/jcp/125/16/10.1063/1.2360263,
doi:10.1021/jp912049p,Oberhofer09,Oberhofer10jcp,
Kubas14jcp,Kubas15pccp,
Oberhofer10acie, Oberhofer12, McKenna12prb, Blumberger13},
excitation energy transfer~\cite{doi:10.1021/jp106989t,doi:10.1038/nchem.1945},
and exchange coupling
parameters~\cite{:/content/aip/journal/jcp/124/2/10.1063/1.2145878,doi:10.1021/ic700871f,
PhysRevB.78.165108}
for use in model Hamiltonians,
as well as Coulomb interaction parameters for methods such as 
DFT+$U$~\cite{PhysRevB.44.943,0953-8984-10-31-012,0022-3719-19-31-004,
Meider1999339,PhysRevB.39.9028,PhysRevB.67.153106}.
%
\rd{Moreover, cDFT has} been shown to provide 
an effective \rd{correction} for
\rd{the} self-interaction error \rd{exhibited by} 
approximate functionals when calculating
diabatic free-energy surfaces for electron-transfer
reactions~\cite{PhysRevLett.97.028303}.
%
%
As a promising antidote to static correlation error in approximate functionals, cDFT has been used to generate small, efficient basis sets for configuration interaction calculations, \rd{by 
enabling the most relevant charge and spin 
states to be straightforwardly sampled~\cite{:/content/aip/journal/jcp/127/16/10.1063/1.2800022,
:/content/aip/journal/jcp/130/3/10.1063/1.3059784, PhysRevB.77.195126,
:/content/aip/journal/jcp/133/6/10.1063/1.3470106,
:/content/aip/journal/jcp/140/18/10.1063/1.4862497}.}
For a recent comprehensive review \rd{of cDFT, we refer
the reader to} 
Ref.~\onlinecite{doi10.1021/cr200148b}.

\rd{cDFT may yet play unforeseen roles in future
first-principles atomistic simulation.
As the field moves increasingly towards the automated
construction and interrogation of materials databases generated
using high-throughput DFT
approaches~\cite{Jain20112295,curtarolo},
 for example, it could be used
 in the large-scale screening of candidate charge-transfer
and energy-transfer materials or, as we describe below, 
to screen for the average local microscopic dielectric functions 
of complex materials and interfaces.}
In order for the great utility and potential of 
the cDFT approach to be fully \rd{and routinely}
realized in the simulation of charge-transfer excitations, 
and in degenerate or strongly-interacting systems, it must be 
 efficiently automated, reliable, and convenient for users.
For this, robust optimisation algorithms for the Lagrange-multipliers 
enforcing the constraint functionals of cDFT are desirable, 
and indeed necessary in cases of multiple 
simultaneous constraints {\rd{being applied,} such 
as on charge and spin~\cite{doi10.1021/cr200148b,
doi:10.1021/jp061848y,doi:10.1021/ct0503163,
:/content/aip/journal/jcp/131/6/10.1063/1.3190169,
doi:10.1021/ct100601n}.
Additionally, automated Lagrange multiplier \rd{updates 
at each ionic configuration are indispensable} 
when performing  geometry 
optimisation~\cite{PhysRevA.72.024502,
doi:10.1021/ct0503163,doi:10.1038/nmat1885,
doi:10.1021/ja076125m,
doi:10.1021/jp103153a,doi:10.1021/ct100601n}
or molecular dynamics~\cite{PhysRevLett.97.028303, 
Oberhofer10jcp, 
:/content/aip/journal/jcp/131/6/10.1063/1.3190169,
Oberhofer10acie, 
doi:10.1021/ct200570u} 
in tandem with cDFT.

Critical to both the theoretical underpinning and \rd{viability} of 
cDFT \rd{optimization} is the nature of the energy landscape with respect to  
the Lagrange multipliers that determine the 
strength of \rd{its} constraining potentials.
In particular, certainty about the nature and uniqueness of 
any stationary points at which the constraints are satisfied
is \rd{a} prerequisite to efficiently locating them numerically.
Wu and Van Voorhis (W\&VV)~\cite{PhysRevA.72.024502}
carried out the pioneering and enabling 
work in this area, analysing the
relevant derivatives, and their principal
results have been \rd{subsequently} synopsized in numerous 
works~\cite{doi10.1021/cr200148b, 
doi:10.1021/ct0503163, 
PhysRevB.88.165112, doi:10.1021/ct100601n, 
:/content/aip/journal/jcp/142/23/10.1063/1.4922378}.
It was concluded by W\&VV~\cite{PhysRevA.72.024502}
\rd{on the basis of non-degenerate perturbation theory}
that a non-trivial stationary point, 
\rd{for an arbitrary constraint on the electron density}, arises 
only at a maximum of the total-energy with 
respect to a cDFT Lagrange multiplier, 
and that this solution is unique. 
This is a central result in cDFT, 
\rd{suggesting the feasibility of its
routine automated optimization,} 
which has been  extended to multivariate cases in 
Refs.~\onlinecite{doi:10.1021/ct0503163} 
and~\onlinecite{doi10.1021/cr200148b}.

\section{Introduction and Motivation}
\label{Sec:Motivation}

\rd{In this work, we rigorously generalize the latter result, 
building upon the foundation provided by
W\&VV's cDFT
stationary point classification,
first showing that 
the  analysis 
becomes inconclusive
when electronic screening effects are considered.
Specifically, we find that while the cDFT energy 
curvature~\footnote{The ``curvature'' is used 
here as a convenient shorthand for the second derivative. 
We do not imply the geometric 
curvature, which equals the second derivative only at 
stationary points.} 
formula derived by W\&VV is 
appropriate for updating cDFT Lagrange multipliers during the
density update step, or inner loop, of 
self-consistent field DFT 
algorithms~\cite{wuprivate,wu:2498,doi:10.1021/ct0503163}, 
it is not applicable
to the self-consistently relaxed total-energy relevant to 
the global classification of cDFT solutions.
Fig.~\ref{fig2} illustrates the large discrepancy between
the cDFT energy curvatures calculated using W\&VV's
formula (solid circles) and those evaluated using 
finite-differences (open squares), hence the necessity 
to revisit the topic here.
}

\begin{figure}
\begin{center}
\includegraphics[width=0.985\columnwidth]{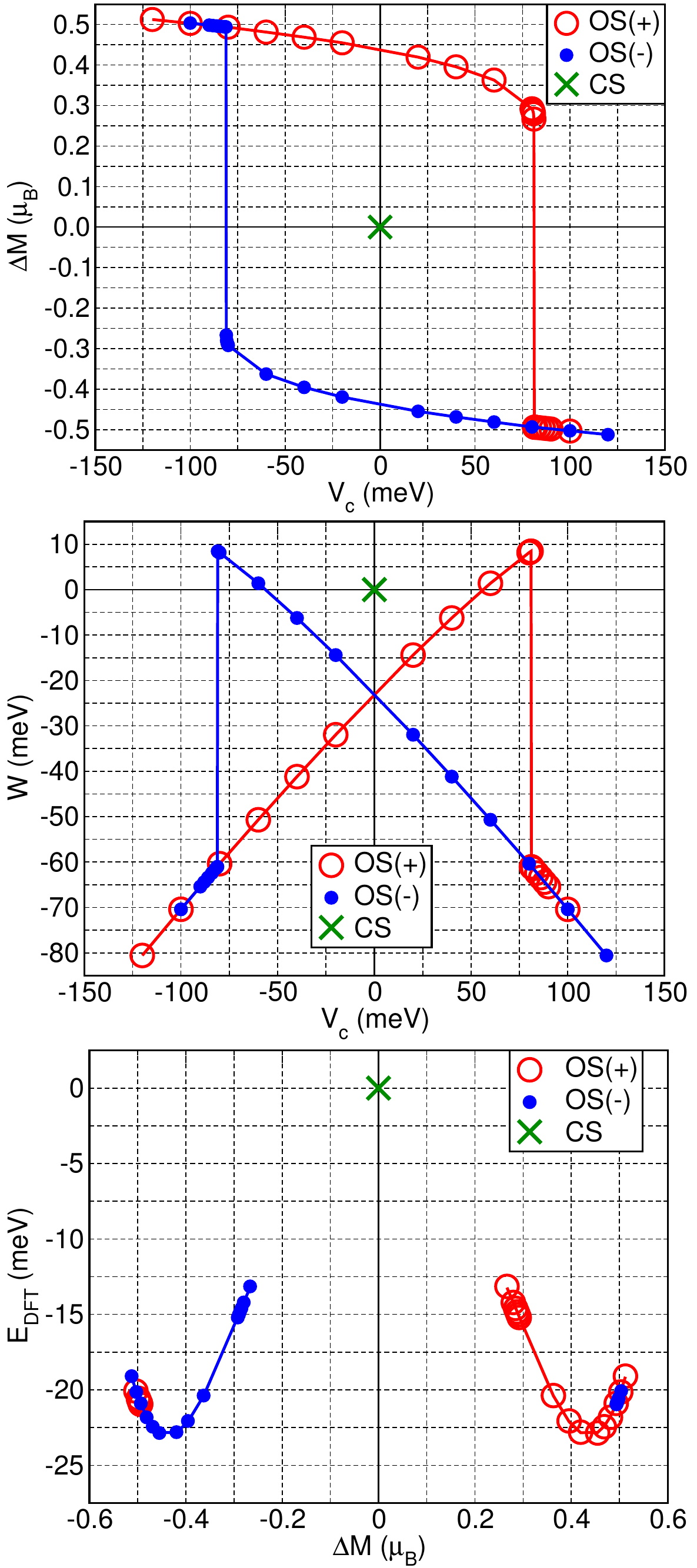}
\caption{(Color online) Multiple solutions and 
hysteresis in spin-constrained, 
open-shell, stretched molecular hydrogen. 
Shown are (top) the magnetic moment difference $\Delta M$  
between constrained regions versus the Lagrange multiplier $V_\textit{c}$, 
(middle) the cDFT total-energy $W$ versus $V_\textit{c}$
corresponding to a target moment difference 
$\Delta M_\textit{c} = 0~\mu_B$, 
and (bottom) the DFT energy component $E_\textrm{DFT}$ versus
$\Delta M$. ``OS(-)'' indicates data obtained 
using the open-shell ground-state with negative
 $\Delta M$ at $ \left( V_\textit{c} = 0 \right) $,
``OS(+)'' the symmetry-related open-shell solution with positive
$\Delta M \left( V_\textit{c} = 0 \right) $, 
with data adapted from the latter,
and ``CS"  the meta-stable closed-shell solution.
}
\label{fig0}
\end{center}
\end{figure}

\rd{In addition, Fig.~\ref{fig0} 
shows cDFT data 
that exhibits effects not hitherto discussed
in the relevant literature, to our knowledge, namely
multiple solutions, hysteresis, and energy discontinuities, 
and thus further motivates the present study.
Here, the ONETEP linear-scaling simulation 
code~\cite{:/content/aip/journal/jcp/122/8/10.1063/1.1839852},
as discussed in Section~\ref{Sec:Numerical}, was used to carry out
cDFT calculations on the hydrogen molecule stretched to 
an internuclear distance of  $3.2~\mathrm{a}_0$, 
which, using the PBE functional~\cite{PhysRevLett.77.3865} 
with no spin-orbit coupling,
lies just beyond the Coulson-Fischer point 
at which an open-shell singlet ground-state becomes 
favoured~\cite{cfpoint, doi:10.1021/jp0534479}.
A constraint was placed on the difference 
of spin magnetic moments, $\Delta M$, between the two hydrogen 
atoms, defined on the basis of their isolated $1s$ valence pseudo-orbitals. 
The constraint target was set to $\Delta M_\textit{c} = 0~\mu_B$,
and the Lagrange multiplier $V_\textit{c}$ was defined such that
increasing its value increased the spin-dependent potential
acting to decrease $\Delta M$. 
It was found that the unpolarized, 
closed shell  ``CS'' state is meta-stable,
starting from which any non-zero value 
of $V_\textit{c}$ initiates a collapse to 
one of two symmetry-related, degenerate open-shell ground-states,
``OS(-)'' and ``OS(+)'' defined in the caption of Fig.~\ref{fig0}.
}

\rd{
The constraint proved capable of traversing between the 
two energy basins associated with these states
(the ``CS'' state lies at the saddle point connecting them),
insofar as that it caused abrupt switching between states at a critical 
constraining potential  of
$V_\textit{c}  \approx \pm 81.1$~meV. The top
panel of Fig.~\ref{fig0} shows that multiple energy minima 
are possible for particular values of the cDFT Lagrange multiplier
$V_\textit{c}$, that certain ranges of the constraint target 
(in this case $\Delta M_\textit{c}$) may be inaccessible to cDFT, 
and that its response diverges simultaneously
with an energy discontinuity
at the transition between states. The middle panel shows
that the total-energy $W$ may exhibit non-differentiable cusps 
at such points, close to which cDFT optimisation is impracticable. 
The bottom panel shows the DFT contribution 
to the total-energy, $E_\textrm{DFT}$, as a function of the constrained
quantity, and the two symmetry-related  ``OS'' 
energy basins in question.
Geometrically, it is inevitable 
that at least some 
solutions within the inaccessible interval connecting 
these two curves would, in principle, exhibit negative curvatures.
We will show that such solutions, unless meta-stable 
like the ``CS'' state, are unstable due to  
anomalous response function sign, and thus cannot be
realized using cDFT.}

\rd{Prompted by these results, in the following we
provide generalized energy curvature
expressions which ensure that the stable stationary points
of non-trivial linear constraints 
in the density may occur only at
maxima of the total-energy with 
respect to their Lagrange multipliers, 
thereby cementing the theoretical 
basis of automated cDFT optimization. 
Our approach is general up to arbitrary
orders in response and
it also lifts the assumption of orbital 
non-degeneracy made in W\&VV's treatment.
It also allows for the maximizability of the cDFT energy to
imply the uniqueness of such maxima only
when the unconstrained system is devoid of multiple electronic minima,
a specific counterexample of which is demonstrated by
Fig~\ref{fig0}.
The  consistency  of our analytical approach is 
 validated throughout this work by means of numerically 
verified equalities, in some cases newly established, between
integrated linear-response functions and 
components of the cDFT total-energy curvature. 
As such, the present work
represents a timely, comprehensive treatment 
of density response  from the energy landscape perspective.
We provide energy-curvature relationships for each of the integrated 
response and inverse-response functions, both interacting and
non-interacting, which we generalize  to multiple constraints.}

\rd{These response functions are by-products of
cDFT Lagrange multiplier optimization, and they 
can be used, for example, to calculate the average
dielectric constant in a particular 
region, which is of critical importance for supercell convergence 
acceleration~\cite{PhysRevB.93.165102} and 
implicit solvation~\cite{0295-5075-95-4-43001,
:/content/aip/journal/jcp/136/6/10.1063/1.3676407} schemes.
A formula for the cDFT optimization 
condition number is provided, 
intended for  identifying systems 
in which extremization of the cDFT total-energy is to be
expected to present challenges, 
and in which Newton's method may 
be apposite, as has 
previously been suggested within the cDFT context in  
Refs.~\onlinecite{doi:10.1021/ct0503163, 
:/content/aip/journal/jcp/131/6/10.1063/1.3190169, 
doi10.1021/cr200148b}. }

\rd{Given the  broad span of the physical and 
mathematical issues necessarily considered 
and revised, and to favor the 
readability of the article, it is organized into Sections, as follows.
In Section~\ref{Sec:Gradients}, we present 
and discuss the previously 
overlooked role of dielectric screening in 
the energy derivatives in self-consistent cDFT.
In Section~\ref{Sec:Numerical}, we introduce 
the formal connections between the Lagrange multiplier 
curvature of  the different contributions to the cDFT total-energy  
and the integrated electronic response function, 
both in the interacting and non-interacting cases. 
These newly derived formulae are then applied to
globally characterize the stationary points of cDFT, 
 both in the single constraint (Section~\ref{Sec:curvaturesign}) 
and multiple constraint (Section~\ref{Sec:multiple}) cases. 
In Section~\ref{Sec:Conclusion}, we provide a synopsis of 
our main findings and conclusions.}

\section{Dielectric screening of 
Constrained DFT
energy derivatives}
\label{Sec:Gradients}

We begin \rd{our analysis with} the 
cDFT~\cite{doi10.1021/cr200148b} constraint functional 
\rd{as per} the definition 
and notation introduced in the \rd{founding} 
article on cDFT automation by W\&VV~\cite{PhysRevA.72.024502}.
We consider an electronic system 
treated using Kohn-Sham 
density-functional theory (DFT)~\cite{PhysRev.136.B864, PhysRev.140.A1133}, subject to an arbitrary 
constraint on its electron density (see W\&VV's Eq.~1, Ref.~\onlinecite{PhysRevLett.53.2512} and 
footnote~\footnote{A single, strictly local occupancy constraint 
is considered in Ref.~\onlinecite{PhysRevA.72.024502}.
We retain these restrictions so as not
to obscure the fundamental aspects under consideration.
These conditions are typically lifted in practical 
cDFT calculations,
bringing us  into  multivariate optimisation of constrained
Kohn-Sham spin-density functional theory, 
which may also be non-local or orbital dependent.})
of the \rd{linear}  form 
\begin{align}
\label{Eq:constraint}
C \left[ \rho \right] = N \left[ \rho \right] - N_\textit{c}; \;\;
N \left[ \rho \right] = 
\sum_\sigma \int   w_\textit{c}^\sigma \left( \mathbf{r} \right)
\rho^\sigma  \left( \mathbf{r} \right) \; d\mathbf{r} .
\end{align}
Here, $\rho^\sigma \left( \mathbf{r} \right) $ is the electronic density
of spin $\sigma$, $w_\textit{c}^\sigma \left( \mathbf{r} \right)$ is an arbitrary
local weight function  describing a spatial 
region of particular interest in the system, and
 $N_\textit{c}$ is the target electron number to be enforced on that region.
In order to apply this constraint, a term is added to the conventional DFT
total-energy, $E_\textrm{DFT} \left[ \rho \right]$, to build the functional 
given by (c.f. W\&VV's Eq.~4) 
\begin{align}
\label{Eq:W}
W \left[ \rho, V_\textit{c}\right] ={}& 
E_\textit{c}  \left[ \rho, V_\textit{c}\right]
+  E_\textrm{DFT}\left[ \rho \right], \;\; \mbox{with} \\ 
E_\textit{c}  \left[ \rho, V_\textit{c}\right] ={}&  
V_\textit{c} C\left[ \rho \right],
\end{align}
 where $V_\textit{c}$ is the Lagrange multiplier. 
Minimizing $W$ with respect to the density via the Kohn-Sham orbitals
$\phi_{i \sigma} $, 
for a given $V_\textit{c}$, 
under the condition that these orbitals 
are orthonormalized for each spin, 
gives rise to the Kohn-Sham equations~\cite{PhysRev.140.A1133} including a constraining potential $V_\textit{c}  w_\textit{c}^\sigma \left( \mathbf{r} \right)$. 
This minimization, \rd{which is equivalent to} solving the constrained Kohn-Sham equations, 
does not correspond to the free extremization   
 $\delta W / \delta \phi^\ast_{i \sigma} = 0$
\rd{ invoked} by W\&VV, since the functional derivative 
 cannot encode the orbital orthonormality constraint.
Rather, it instead 
corresponds to extremizing the 
Lagrangian  
\begin{align}
\label{Eq:Lagrangian}
\Omega \left[ \rho, V_\textit{c} \right] ={}&
W \left[ \rho, V_\textit{c}  \right]  \\ \nonumber
&-  \sum_\sigma \sum_{ i j}^{N_\sigma} \varepsilon_{i j \sigma}
\left( \int \phi_{i \sigma}^{ \ast} \left( \mathbf{r} \right) 
\phi_{j \sigma} \left( \mathbf{r} \right)  \; d\mathbf{r} - \delta_{i j} \right).
\end{align}
Here, we have assumed that the system is 
Kohn-Sham insulating, for simplicity, 
with $N^{\sigma}$ electrons per spin $\sigma$. 
Following  application of the condition 
$\delta \Omega / \delta \phi^\ast_{i \sigma} = 0$, 
we may perform a unitary transformation
 among the resulting equations.
This  also transforms the orbitals, 
yet it presents no \rd{difficulties} since 
$\Omega \left[ \rho, V_\textit{c} \right] $ is 
invariant under such transformations for  density functionals. 
Diagonalising the matrix 
Lagrange multiplier $ \varepsilon_{i j \sigma} $, thereby, 
returns the Kohn-Sham cDFT
equations of W\&VV's Eq.~5, with
eigenvalues $  \varepsilon_{i \sigma}$.
 Thus, we may succinctly write, at the physically relevant minimum
 of W, the expression $\delta W / \delta \phi^\ast_{i \sigma} = 
\hat{H}_\sigma \phi_{i \sigma} $
 or indeed its complex conjugate $\delta W / \delta \phi_{i \sigma} = 
 \phi^\ast_{i \sigma} \hat{H}_\sigma  $, where
$\hat{H}_\sigma$ is the Kohn-Sham cDFT
Hamiltonian for spin $\sigma$.
\rd {Following} W\&VV, we \rd{may} next define the function
$W\left( V_\textit{c}\right)$
as the evaluation of $W \left[ \rho, V_\textit{c}  \right]$ 
using the density generated by 
the \rd{orthonormal} orbitals 
which solve the Kohn-Sham equations
including the constraining potential 
$V_\textit{c}  w_\textit{c}^\sigma \left( \mathbf{r} \right)$ or, 
for the avoidance of doubt concerning Kohn-Sham excited states,
\rd{as} the physical minimum of the total-energy for a given $ V_\textit{c}$.


\subsection{Total-energy first derivative}

\rd{We now begin to analyze the derivatives of
$W\left( V_\textit{c}\right)$
required for the location of cDFT solutions.
Partial derivatives couple only 
explicit dependencies, and are sufficient for optimising
the Lagrange multiplier during the
density update step, or inner loop, of 
self-consistent field DFT 
algorithms~\cite{wu:2498,doi:10.1021/ct0503163}. 
In order to determine the 
character of cDFT stationary points globally, 
on the other hand, we must consider  total derivatives, which 
include orbital and density relaxation effects.} In 
 simulations where a number of constraints are 
simultaneously applied, 
their \rd{Lagrange multipliers} are independent 
variables, \rd{so that} the Hessian
of interest is the matrix of mixed second total 
derivatives. These simulations are 
discussed in Section~\ref{Sec:multiple}.

The first total derivative of the cDFT total-energy with
respect to the Lagrange multiplier, $V_\textit{c}$, is given by
\begin{align}
\label{Eq:first}
\frac{d W}{d V_\textit{c}} ={}& \sum_\sigma \sum_i^{N_\sigma}
\textrm{Tr} \left[ \frac{\delta W}{\delta \phi_{i \sigma}^{\ast}} 
\frac{d \phi_{i \sigma}^{\ast}}{d V_\textit{c}} 
+ \textit{c.c.}\right]
+\frac{\partial W}{\partial V_\textit{c}} \nonumber \\
={}&
\sum_\sigma \sum_i^{N_\sigma}
\textrm{Tr}  \left[ \left( \hat{H}_\sigma \phi_{i \sigma}  \right)
\frac{d \phi_{i \sigma}^{\ast}}{d V_\textit{c}} 
+ \textit{c.c.}\right] \nonumber \\ 
&+  \left( \sum_\sigma \int  w_\textit{c}^\sigma \left( \mathbf{r} \right)
\rho^\sigma \left( \mathbf{r} \right) \; d\mathbf{r} - N_\textit{c}\right),
\end{align}
where the trace symbol $\textrm{Tr}$  denotes an
integral over space since the operators are all local, 
and  $\textit{c.c.}$ represents the complex conjugate
of the preceding term.
For any value of $V_\textit{c}$, the orbitals
that generate the density minimizing  
\rd{$\Omega \left[ \rho, V_\textit{c}  \right]$} 
are unique up to 
unitary transformations, and we are free to 
choose the set that diagonalize the 
Hamiltonian. This allows us to simplify the 
latter expression, \rd{since it guarantees that the 
orbital-coupling term}
\begin{align}
\label{Eq:perturb}
\textrm{Tr} \left[  \left( \hat{H}_\sigma \phi_{i \sigma}  \right)
\frac{d \phi_{i \sigma}^{\ast}}{d V_\textit{c}} \right]
={}&
\varepsilon_{i \sigma}
\textrm{Tr} \left[  \phi_{i \sigma} 
\frac{d \phi_{i \sigma}^{\ast}}{d V_\textit{c}} \right]
  \\
={}& \varepsilon_{i \sigma} \int \phi_{i \sigma} \left( \mathbf{r} \right)
\sum_{a \ne i} \phi_{a \sigma}^\ast  \left( \mathbf{r} \right)
\; d\mathbf{r} \nonumber \\
&\times
 \int
\frac{\phi_{a \sigma}^\ast  \left( \mathbf{r'} \right)
\frac{d v_\sigma^\textrm{KS} \left( \mathbf{r'} \right) }{d V_\textit{c}}
\phi_{i \sigma}  \left( \mathbf{r'} \right)
}{\varepsilon_{i \sigma}-\varepsilon_{a \sigma}} 
 \; d\mathbf{r'}  \nonumber
\end{align}
evaluates to zero by virtue of the 
orthonormality of $\phi_{i \sigma} $ 
and $\phi_{a \sigma}$ for $a\ne i$. 
Here, $v_\sigma^\textrm{KS} $ is the Kohn-Sham potential,
i.e., the total effective potential which enters
\rd{density-functional} perturbation 
theory~\cite{PhysRevLett.58.1861, 
RevModPhys.73.515}.

\rd{If focusing on the self-consistent 
field DFT inner loop as per W\&VV, we 
may neglect screening effects and, thereby, assert that}
the change in total potential equals the  external perturbation 
$\delta \hat{v}^\textrm{external}_\sigma $, and then
$\delta \hat{v}_\sigma^\textrm{KS} = 
\delta \hat{v}_\sigma^\textrm{external}
= \hat{w}^\sigma_\textit{c}\delta V_\textit{c} $, whence
$d v_\sigma^\textrm{KS} \left( \mathbf{r} \right) / d V_\textit{c}
=  \hat{w}^\sigma_\textit{c} \left( \mathbf{r} \right) $ 
in the above expression.
More generally, however, an account of electronic screening of the
perturbation is necessary, 
and such effects are encapsulated in
the inverse microscopic dielectric function
defined by
\rd{$\epsilon^{-1}_{\sigma \sigma'} \left( \mathbf{r},  \mathbf{r'} \right) = 
d v_\sigma^\textrm{KS} \left( \mathbf{r} \right) /
d v_{\sigma'}^\textrm{external}  \left( \mathbf{r'} \right) 
$}~\footnote{We use a lunate epsilon $\epsilon$ for the
microscopic dielectric function in order to distinguish from it
from the Kohn-Sham eigenvalues $\varepsilon$}. 
For \rd{pure (density-constrained
rather than non-local)} cDFT, we may write that
\begin{align}
\frac{ d v_\sigma^\textrm{KS} \left( \mathbf{r} \right) }{ d V_\textit{c} }
={}&
\sum_{\sigma'} \int 
\frac{ d v_\sigma^\textrm{KS} \left( \mathbf{r} \right) }{ 
d v_{\sigma'}^\textrm{external} \left( \mathbf{r'} \right) }
\frac{ d v_{\sigma'}^\textrm{external} \left( \mathbf{r'} \right) }{
d V_\textit{c} } \; d\mathbf{r'}
\nonumber \\  \nonumber 
={}&
\sum_{\sigma'} \int 
\epsilon^{-1}_{\sigma \sigma'} \left( \mathbf{r} , \mathbf{r'} \right)
w^{\sigma'}_\textit{c}  \left( \mathbf{r'} \right)
\; d\mathbf{r'}
\\  
\equiv{}&
\left( \epsilon^{-1} w_\textit{c} \right)^\sigma 
\left( \mathbf{r} \right). 
\label{Eq:dielectric}
\end{align}
For cDFT with non-local
potentials, \rd{the symmetric form 
$ \left( \epsilon^{-1/2} w_\textit{c} \epsilon^{-1/2}  \right)^\sigma \left( \mathbf{r},  \mathbf{r'} \right) $ 
may be used in place of the latter 
in order to ensure that the potential remains Hermitian}.


\begin{figure}
\begin{center}
\includegraphics[width=1\columnwidth]{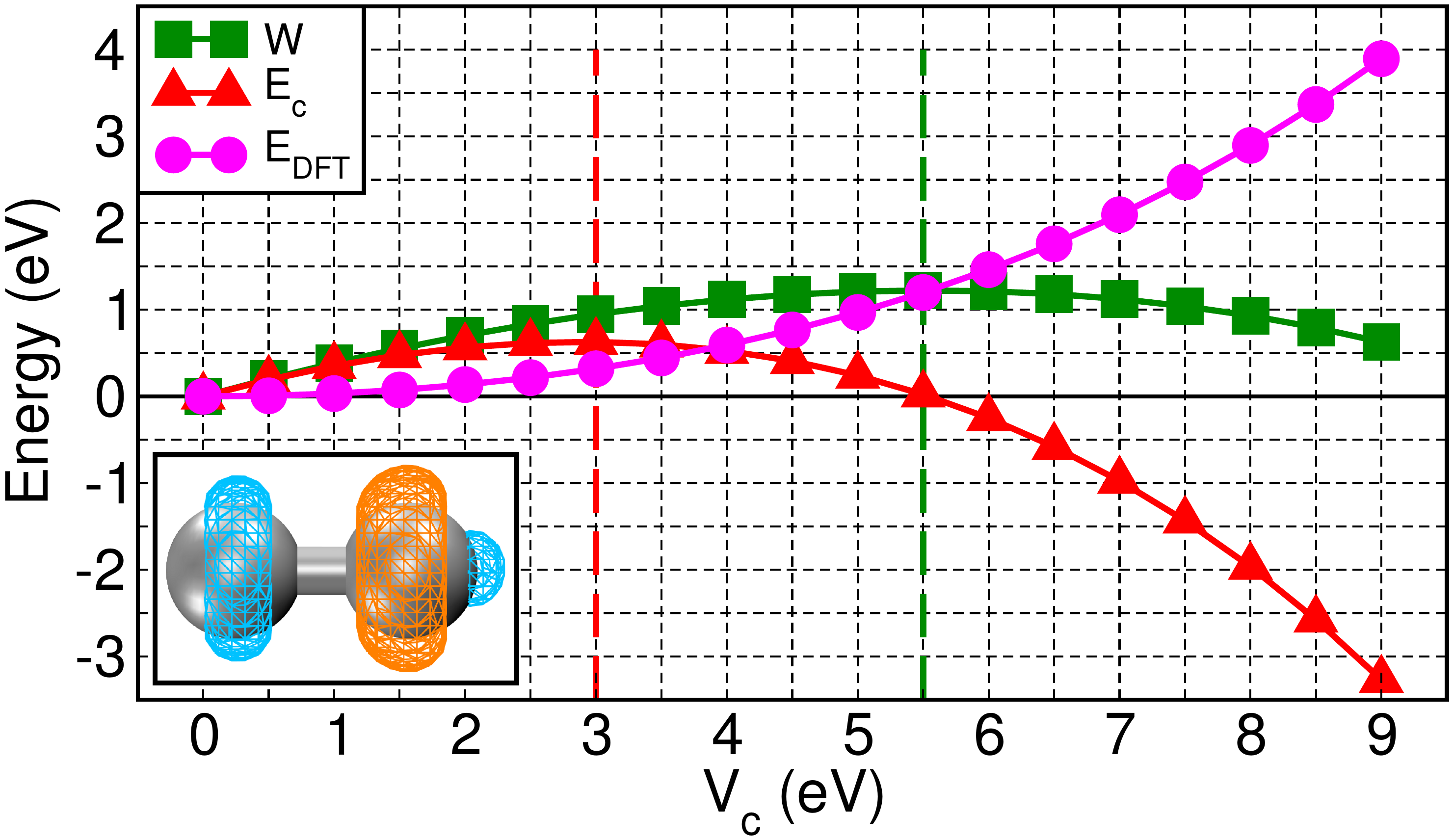}
\caption{(Color online) The cDFT total-energy $W$, and its
constraint, $E_\textit{c}$, and DFT,
$E_\textrm{DFT}$, components,
as a function of the Lagrange multiplier $V_\textit{c}$,
for a charge-constrained nitrogen molecule.
The $V_\textit{c}$ values \rd{for the data points} 
at which $E_\textit{c}$ and $W$ attain
maxima are shown with dashed vertical
red and green lines, respectively.
Inset: the left-hand atom is constrained 
to lose charge with respect to its ground-state population, 
using an on-atom population analysis 
combining  $s$ and $p$ orbitals. With respect to the
ground-state density, charge depleted regions 
are shown with a cyan charge-difference isosurface
and the charge augmented region is shown with an orange isosurface.
The unconstrained right-hand atom exhibits  strong polarization,
and depletion in the lone-pair region.}
\label{fig1}
\end{center}
\end{figure}

Screening effects notwithstanding, all cancels to zero in  
the total derivative of  $W \left(  V_\textit{c}  \right)$ \rd{except for
the explicit constraint contribution on the final line of Eq.~\ref{Eq:first},
namely $\partial W / \partial V_\textit{c} = C$, which then 
evaluates to zero when the constraint is satisfied. 
Thus, the task of enforcing the constraint condition is transformed
into that
of locating the stationary points of $W$ with respect to 
$V_\textit{c}$. As an aside, this result mirrors W\&VV's Eq.~6, 
which was derived in a slightly different way 
by invoking  $\delta W / \delta \phi^\ast_{i \sigma} = 0$
rather than  $\delta W / \delta \phi^\ast_{i \sigma} = 
\hat{H}_\sigma \phi_{i \sigma}$, 
the numerical distinction between which is vanishing 
due to Kohn-Sham orbital orthonormality.
}
\rd{Finally, we note that the vanishing trace in Eq.~\ref{Eq:first}
may be partitioned into two 
contributions which cancel, for any value of $V_\textit{c}$, 
by means of the expression}
\begin{align} 
\frac{\delta W}{\delta \phi_{i \sigma}^{\ast}} 
={}& \frac{\delta  E_\textit{c} }{\delta \phi_{i \sigma}^{\ast}} + 
 \frac{\delta  E_\textrm{DFT} }{\delta \phi_{i \sigma}^{\ast}} = 
V_\textit{c}  \frac{\delta  C}{\delta \phi_{i \sigma}^{\ast}} 
+
\frac{\delta E_\textrm{DFT}}{\delta \phi_{i \sigma}^{\ast}} \nonumber \\
\Rightarrow \quad 0 ={}&
V_\textit{c}   \textrm{Tr} \left[ \frac{\delta C}{\delta \phi_{i \sigma}^{\ast}} 
\frac{d \phi_{i \sigma}^{\ast}}{d V_\textit{c}} \right]
+
\textrm{Tr} \left[ \frac{\delta E_\textrm{DFT}}{\delta \phi_{i \sigma}^{\ast}} 
\frac{d \phi_{i \sigma}^{\ast}}{d V_\textit{c}} \right].
\end{align}
 Thus, both the DFT energy $E_\textrm{DFT}$ 
and the constraint  energy $E_\textit{c}$
may individually contribute substantially
to the derivatives of $W\left( V_\textit{c}\right)$, as shown in Fig.~\ref{fig1}.
As a result, \rd{a stationary total-energy $W$  
with respect to the cDFT
Lagrange multiplier does not 
imply a stationary
constraint contribution \rd{$E_\textit{c}$} alone, and vice versa.}




\subsection{Total-energy second derivative
by means of non-degenerate perturbation theory}

The second  derivative or ``curvature'' 
of $W\left( V_\textit{c}\right)$, 
is required to classify any  stationary point, 
or points, at which the constraint is satisfied. 
\rd{In cases where the second derivative vanishes,
higher derivatives may also be needed.
We consider here the self-consistent cDFT energy landscape, 
rather than the unscreened problem specific to the inner
loop of self-consistent field DFT 
codes~\cite{wuprivate,wu:2498,doi:10.1021/ct0503163}, 
and the resulting 
curvature differs from that of
W\&VV's treatment in magnitude and, potentially,  in sign}.
Following from Eq.~\ref{Eq:first} and applying the
product rule for differentiation where necessary,
we may write that
\begin{align}
\label{Eq:second}
\frac{d^2 W}{d V_\textit{c}^2} ={}& 
\frac{d }{d V_\textit{c}}  
\sum_\sigma \sum_i^{N_\sigma}
\textrm{Tr}  \left[ \left( \hat{H}_\sigma \phi_{i \sigma}  \right)
\frac{d \phi_{i \sigma}^{\ast}}{d V_\textit{c}} 
+ \textit{c.c.}\right] + \frac{d C}{d V_\textit{c}} 
  \nonumber \\ ={}& \nonumber
\sum_\sigma \sum_i^{N_\sigma}
 \textrm{Tr}  \left[ \left( \frac{d \hat{H}_{\sigma} }{d V_\textit{c}} 
 \phi_{i \sigma}  \right)
\frac{d \phi_{i \sigma}^{\ast}}{d V_\textit{c}} 
+ \textit{c.c.}\right] \\   \nonumber
&+ \sum_\sigma \sum_i^{N_\sigma}
\textrm{Tr} \left[ \left( \hat{H}_{\sigma} 
  \frac{d \phi_{i \sigma} }{d V_\textit{c}} \right)
\frac{d \phi_{i \sigma}^{\ast}}{d V_\textit{c}} 
+ \textit{c.c.}\right]  \\  \nonumber
&\quad\;\;+ \sum_\sigma \sum_i^{N_\sigma}
 \textrm{Tr}  \left[ \left( \hat{H}_\sigma \phi_{i \sigma} \right)
\frac{d^2 \phi_{i \sigma}^{\ast}}{d V_\textit{c}^2} 
+ \textit{c.c.}\right] \\ 
&\quad\;\;\quad\;\; +  \sum_\sigma \int  w_\textit{c}^\sigma \left( \mathbf{r} \right)
 \frac{d \rho^\sigma \left( \mathbf{r} \right)  }{d V_\textit{c}} \; d\mathbf{r}.
\end{align}
\rd{This makes}  the contributions arising at second order in
perturbation theory \rd{explicit}.
These \rd{rather cumbersome terms} may be circumvented 
by noting that the eigencondition
$\hat{H}_\sigma  \phi_{i \sigma} = \varepsilon_{i \sigma} 
 \phi_{i \sigma}$ holds continuously as we vary the parameter
$ V_\textit{c} $. \rd{As a result,} 
the quantity expressed in Eq.~\ref{Eq:perturb} vanishes
 for all $V_\textit{c}$, and so we may write that
 \begin{equation}
 \frac{d }{d V_\textit{c}} 
 \textrm{Tr} \left[  \left( \hat{H}_\sigma \phi_{i \sigma}  \right)
\frac{d \phi_{i \sigma}^{\ast}}{d V_\textit{c}} \right]
 = 0.
 \label{Eq:simply}
 \end{equation}
Thus, all terms in Eq.~\ref{Eq:second} numerically cancel
except for the final term, $d C / d V_\textit{c} $.
This may then be re-written using
non-degenerate (only where \rd{applicable}) 
first-order perturbation theory, after 
the present Eq.~\ref{Eq:perturb},  since
\begin{align}
\label{Eq:bare}
{}& \frac{d C}{ d V_\textit{c}} = \frac{d}{d V_\textit{c}} \sum_\sigma  \int
w_\textit{c}^\sigma \left( \mathbf{r} \right) \rho^\sigma 
 \left( \mathbf{r} \right) \; d\mathbf{r} \nonumber \\
={}& \sum_\sigma \sum_i^{N_\sigma} \nonumber
 \int  w_\textit{c}^\sigma 
\left( \mathbf{r} \right)
\phi_{i \sigma}^{\ast} \left( \mathbf{r} \right)
 \frac{d \phi_{i \sigma}  \left( \mathbf{r} \right)}{d V_\textit{c}} 
\; d\mathbf{r} 
+ \textit{c.c.} \\
={}& 
\sum_\sigma \sum_i^{N_\sigma}
\int  w_\textit{c}^\sigma \left( \mathbf{r'} \right)
\phi_{i \sigma}^\ast \left( \mathbf{r} \right)
\sum_{a \ne i} \phi_{a \sigma}  \left( \mathbf{r} \right)
\; d\mathbf{r} \nonumber 
\\ &\times \int
\frac{\phi_{a \sigma}^\ast  \left( \mathbf{r'} \right)
 \frac{d v^{\textrm{KS}}_{ \sigma} \left( \mathbf{r'} \right) }{d V_\textit{c}}
\phi_{i \sigma}  \left( \mathbf{r'} \right)
}{\varepsilon_{i \sigma}-\varepsilon_{a \sigma}} 
 \; d\mathbf{r'} + \textit{c.c.}  \nonumber \\
={}& 
 \sum_\sigma \sum_i^{N_\sigma} \sum_{a \ne i}
\frac{1}{\varepsilon_{i \sigma}-\varepsilon_{a \sigma}}
\nonumber \\ &\times \left(
\left( \int \phi_{a \sigma}^\ast  \left( \mathbf{r} \right)
 w_\textit{c}^\sigma \left( \mathbf{r} \right)
\phi_{i \sigma}  \left( \mathbf{r} \right) \; d\mathbf{r} \right) \right. 
 \\ &\quad \times \left.
\left( \int \phi_{i \sigma}^\ast  \left( \mathbf{r'} \right)
\left( \varepsilon^{-1} w_\textit{c} \right)^\sigma \left( \mathbf{r'} \right)
\phi_{a \sigma}  \left( \mathbf{r'} \right) \; d\mathbf{r'} \right) \right)
+ \textit{c.c.} \nonumber .
\end{align}
\rd{The latter} expression reduces to W\&VV's Eq.~7  
if screening effects are neglected, that is if we \rd{set} 
$\varepsilon^{-1}_{\sigma \sigma'}
\left( \mathbf{r}, \mathbf{r'} \right)
= \delta _{\sigma \sigma'} \delta\left( \mathbf{r} - \mathbf{r'} \right) $. 
Then, as noted by W\&VV, the anti-symmetry of the  summand
implies \rd{both} that contributions from 
$a \le {N_\sigma}$ cancel to zero and may be omitted, 
\rd{and that the total is strictly non-positive.
This condition holds  in \rd{non-degenerate,  
linearly-responding cases} of
cDFT Lagrange multiplier optimisation
carried out within the potential-update 
loop of self-consistent field DFT codes, where  
W\&VV's result ensures the sign of the energy 
curvature for any fixed Kohn-Sham potential.
}

\rd{Globally speaking, however, it appears
that the sign of the energy curvature 
cannot  be inferred directly from the symmetries of
Eq.~\ref{Eq:bare}.}  
The necessarily real-valued 
screened weight function $\left( \varepsilon^{-1} w_\textit{c} 
\right)^\sigma \left( \mathbf{r} \right)$ may locally vary, 
even in sign, with respect to  
$ w_\textit{c} \left( \mathbf{r} \right)$, in a 
complex, system-dependent manner, 
\rd{typically  causing an average net attenuation of the constraining
potential}. 
Even if we may assume
that $0 < \left( \varepsilon^{-1} w_\textit{c} 
\right)^\sigma \left( \mathbf{r} \right)
 < w_\textit{c} \left( \mathbf{r} \right)$ holds everywhere,
 for a particular system,
and that the orbitals are filled according
to the Aufbau principle in Eq.~\ref{Eq:bare}, 
the form of this sum offers no  guarantee 
regarding the  sign of $d^2 W / d V_\textit{c}^2$.

On the other hand,  experience and extensive 
literature (see  review Ref.~\onlinecite{doi10.1021/cr200148b}) 
\rd{yield observations} that $d^2 W / d V_\textit{c}^2$ 
is  negative for 
density constraints applied to a wide variety of systems.
As now we go on to numerically confirm, 
Eq.~\ref{Eq:simply} provides that 
this curvature  reduces, for ground-states, to the
interacting density response function of
the system, doubly integrated with $w^\sigma_\textit{c}$.
\rd{On this basis, we will show in 
Section~\ref{Sec:curvaturesign} that any solutions of
non-negative curvature are 
meta-stable 
(e.g., the ``CS'' state  of Fig.~\ref{fig0})
or unstable (e.g., the  inaccessible
region in the vicinity of ``CS'') 
with respect to small perturbations, 
so that such curvatures cannot be directly  
computed and
plotted.}

The presence of the inverse microscopic dielectric
function in Eq.~\ref{Eq:bare} renders it \rd{unsuitable} 
for use in accelerating the convergence
of automated cDFT
Lagrange multiplier optimization.}  Even by using a  
finite-difference method for Eq.~\ref{Eq:dielectric}, 
\rd{or indeed in the absence of screening effects,} 
a  converged sum over unoccupied states 
at each cDFT Lagrange multiplier optimization step 
is computationally demanding 
\rd{and conceptually undesirable in DFT.
To overcome these drawbacks, in the following 
Section we present
an alternative approach for characterizing cDFT energy curvatures, 
in the framework of response  theory.
}

\section{cDFT Energy curvatures from the response 
function perspective}
\label{Sec:Numerical}

\rd{As an alternative   to 
classifying the cDFT energy landscape by means of
summation over unoccupied states, in what follows we 
express the relevant energy curvatures 
in terms of more computationally
convenient integrated response functions, which depend only on the
density or occupied states.
This Section is intended to provide a 
comprehensive treatment of the relationship between
density response functions and energies in cDFT.
We extend our principal results  
to the multivariate regime of multiple simultaneous constraints
in Section~\ref{Sec:multiple}.
These response functions
are convenient by-products of automated cDFT optimization, 
and they are
experimental observables in simulations where
the constraining potentials are those of physical fields. 
For a cDFT potential representing a uniform electric field, for example, 
$w^\sigma_\textit{c} \left( \mathbf{r} \right)$
has a constant gradient, and the integrated 
interacting response function represents the high-frequency 
electric dipole-dipole polarizability $\mathbf{\alpha}_\infty$.
Another example is the microscopic dielectric 
constant averaged over a region, 
which is an important ingredient for supercell convergence 
acceleration~\cite{PhysRevB.93.165102} and 
implicit solvation~\cite{0295-5075-95-4-43001,
:/content/aip/journal/jcp/136/6/10.1063/1.3676407} schemes.
The results provided below  allow such 
response functions to be calculated, and even updated in self-consistent
schemes.}

\subsection{The integrated interacting response function}

\begin{figure}
\begin{center}
\includegraphics[width=1\columnwidth]{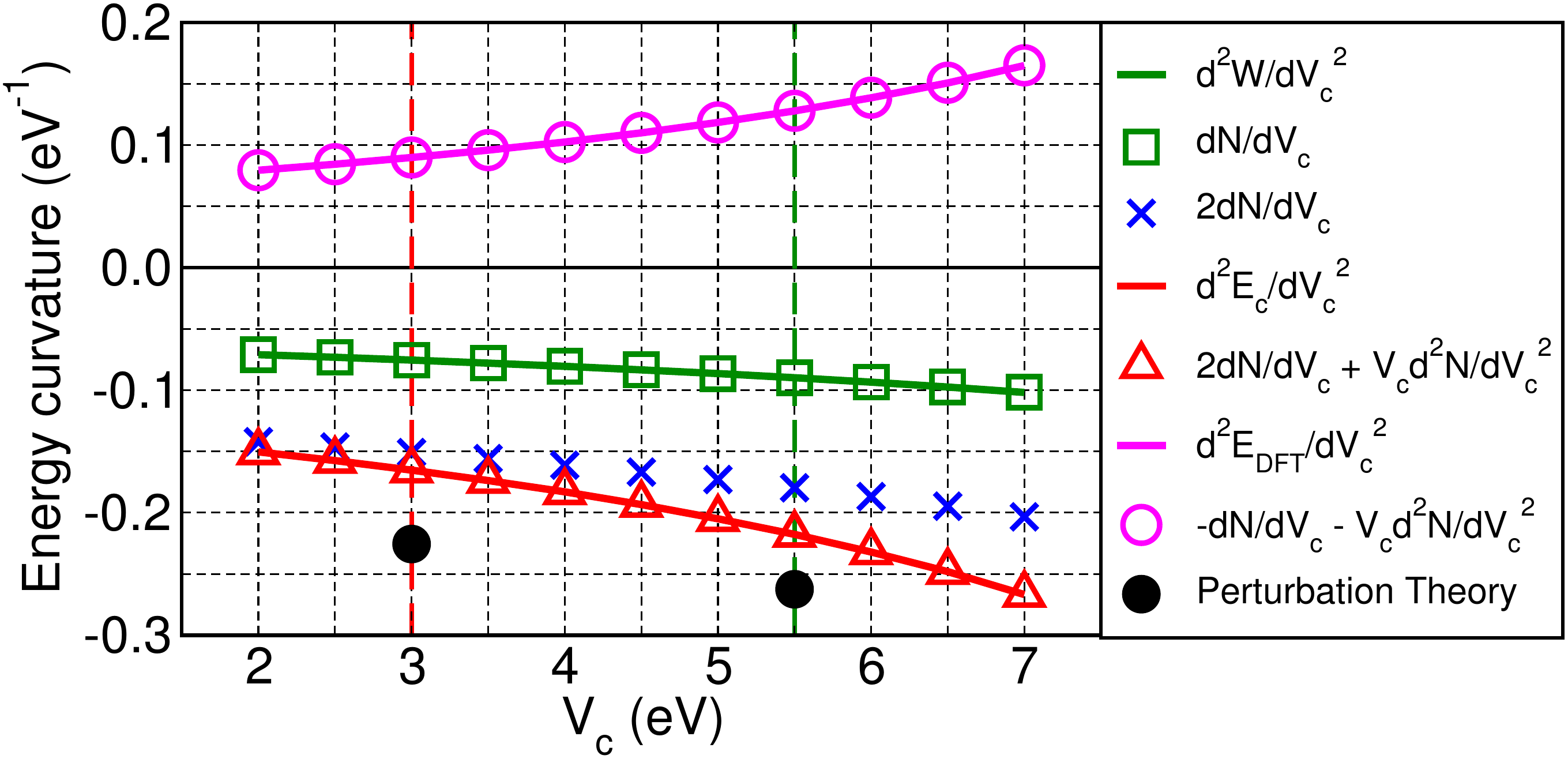}
\caption{(Color online) The numerically evaluated curvature 
(second derivatives) of the cDFT total-energy $W$, and its
constraint, $E_\textit{c}$, and DFT,
$E_\textrm{DFT}$, components, 
with respect to the Lagrange multiplier, $V_\textit{c}$, for the 
 system
shown in Fig.~\ref{fig1}.
Dashed vertical lines have the same significance as in Fig.~\ref{fig1}.
The Lagrange multiplier 
derivatives of the constrained occupancy $N$, which 
correspond to the latter curvatures, are shown,
as well as the linear-response approximation
to one of them (crosses). Evaluated values of the 
unscreened sum-over-states perturbation theory expression
for  $d^2 W / d V_\textit{c}^2$, following W\&VV,
 are also shown (solid circles).}
\label{fig2}
\end{center}
\end{figure}

\rd{The interacting linear response function 
$\chi^{\sigma \sigma'} 
\left( \mathbf{r} , \mathbf{r'}  \right)$
 measures the response in the density
due to a small change in the applied potential.
Since, by definition,
\begin{align}
d \rho^\sigma \left( \mathbf{r} \right)
&{}= \int 
\chi^{\sigma \sigma'} 
\left( \mathbf{r} , \mathbf{r'}  \right)
d v^\textrm{external}_{\sigma'} \left( \mathbf{r'} \right) \; d\mathbf{r'}
\nonumber \\
&{}= \sum_{\sigma'} \int 
\chi^{\sigma \sigma'} 
\left( \mathbf{r} , \mathbf{r'}  \right)
w_\textit{c}^{\sigma'} \left( \mathbf{r'} \right) d V_\textit{c}^{\sigma'} \; d\mathbf{r'}
,
\end{align}
one further integration over $w_\textit{c}^{\sigma} \left( \mathbf{r} \right)
 d\mathbf{r}$ 
allows us to define 
the integrated interacting density response function as
%
\begin{align}
\label{Eq:mydefintegratedchi}
\chi^{\sigma \sigma'} \equiv{}&
\iint   w_\textit{c}^\sigma \left( \mathbf{r} \right)
\chi^{\sigma \sigma'} \left( \mathbf{r} , \mathbf{r'} \right)
w_\textit{c}^{\sigma'} \left( \mathbf{r'} \right) 
d\mathbf{r} \; d\mathbf{r'}  \\ \nonumber
={}& 
\frac{dN^\sigma }{ dV^{\sigma'}_\textit{c} }
 =
\frac{
d \int   w_\textit{c}^{\sigma} \left( \mathbf{r} \right)
\rho^{\sigma} \left( \mathbf{r}  \right) 
\; d\mathbf{r} 
}{d 
\int   w_\textit{c}^{\sigma'} \left( \mathbf{r'} \right)
V^{\sigma'}_\textit{c}
\; d\mathbf{r'} }  \int w_\textit{c}^{\sigma'} \left( \mathbf{r''} \right)
\; d\mathbf{r''},
\end{align}
where we have made it explicit that $d V_\textit{c}$ is the average
change in external potential over the subspace.} 
For constraints defined using 
\rd{more general measures} of the density
or density matrix, \rd{such expressions } may be 
generalised straightforwardly
by replacing the local weighting functions 
$w_\textit{c}^{\sigma} \left( \mathbf{r} \right) $
by   non-local \rd{projection operators}.

\rd{The orthonormality preservation condition expressed in
Eq.~\ref{Eq:simply}
guarantees the simplification of Eq.~\ref{Eq:second}
to Eq.~\ref{Eq:bare} for all values of $V_\textit{c}$, and so provides 
that}
\begin{align}
\frac{d^2 W }{ dV_\textit{c}^2 } 
= \frac{ dC }{ dV_\textit{c} } = 
\frac{ d N }{ d V_\textit{c} } \rd{ \equiv \chi
= \frac{1}{2} \sum_{\sigma \sigma'} \chi^{\sigma \sigma'}  ,}
\label{Eq:d2WdVc2}
\end{align}
\rd{where the $1/2$ is specific to collinear spins. 
Taking the constraint contribution alone, on the other hand,  
we find that the second-order response function survives, since}
\begin{align}
\frac{d^2 E_\textit{c} }{d V_\textit{c}^2}
={}&  \frac{d^2 \left( V_\textit{c} C \right) }{d V_\textit{c}^2}
= \frac{d }{d V_\textit{c}} \left[ C + V_\textit{c}
\frac{d C}{d V_\textit{c}} \right] \nonumber \\
={}& 
 2 \frac{d N}{d V_\textit{c}}  + V_\textit{c} \frac{d^2 N}{d V_\textit{c}^2 }
\rd{ = 2 \chi + V_\textit{c} \frac{d \chi}{d V_\textit{c}}. }
\end{align} 
\rd{Combining these two results, we  deduce 
a general result for the DFT energy component  curvature, given by}
\begin{align}
\frac{d^2 E_\textrm{DFT} }{d V_\textit{c}^2}
={}& 
\frac{d^2 \left( W - E_\textit{c} \right) }{d V_\textit{c}^2}
=
\rd{- \chi  - V_\textit{c} \frac{d \chi}{d V_\textit{c} }} .
\end{align} 

\rd{In order to check the validity of our approach for 
analyzing functional interdependencies and derivatives,
and to numerically illustrate our analytical findings,}
we performed a cDFT study on the  nitrogen molecule
shown in Fig.~\ref{fig1}. 
This  serves to illustrate \rd{a case in which, if}
the constraint energy $E_\textit{c}$ vanishes for a  
finite Lagrange multiplier $V_\textit{c}$, the total-energy $W$ 
achieves a maximum with respect to \rd{$V_\textit{c}$}. 
The total and constraint energies
$W$ and $E_\textit{c}$ exhibit different
negative curvatures, and their 
the difference, the DFT component $E_\textrm{DFT}$,
necessarily exhibits a positive  curvature \rd{around the  ground-state
minimum at $V_\textit{c}=0$~eV.}
For each value of the charge-constraining
 Lagrange multiplier, well converged 
BLYP~\cite{PhysRevA.38.3098,PhysRevB.37.785}
ground-state energies and densities, 
with pseudized $1s$ states,
were calculated  using the 
ONETEP linear-scaling Kohn-Sham
DFT code~\cite{:/content/aip/journal/jcp/122/8/10.1063/1.1839852}.
This code solves for the ground-state by  
optimising a minimal set of nonorthogonal 
generalized Wannier functions~\cite{PhysRevB.66.035119} in situ.
Each of these functions is expanded in an underlying variational
plane-wave equivalent basis set and truncated within a
prescribed cut-off sphere, in this particular case to a  
radius of $10~\mathrm{a}_0$. This approach
has  been shown to offer finite-difference
linear response properties with 
an accuracy matching that of
conventional plane-wave DFT~\cite{PhysRevB.85.193101}.
The constrained population was defined
using the four $2s$ and $2p$ valence 
pseudo-orbitals of the isolated 
atom~\footnote{The resulting constraint  
acts on the Kohn-Sham density-matrix rather than 
on the density. Our analytical findings extends 
to that case with minor notational changes.}.
In the dimer, the resulting unconstrained ground state atomic 
occupancy was approximately $6.5$~e, due to overlap between
 pseudo-orbitals.
The charge of one of the nitrogen atoms was constrained,
with the target occupancy set to $N_\textit{c} = 6.0$~e.

\rd{In Fig.~\ref{fig2}, we note a very precise
numerical correspondence between
the total-energy curvature, its constraint, and DFT components, 
and their respectively predicted reformulations in terms
of first and second order  integrated interacting response functions, 
$\chi = d N / d V_\textit{c}$ and $d \chi / d V_\textit{c} $.
This serves to confirm  that the orthonormality 
preservation condition of Eq.~\ref{Eq:simply} holds for
all $V_\textit{c}$.
The failure of the linear-response approximation
$ 2 \chi = 2 d N /d V_\textit{c}$  (shown with blue crosses) to 
$d^2 E_\textit{c} / d V_\textit{c}$
is somewhat discouraging for the 
application of root-finding algorithms on 
$E_\textit{c}$ in order to optimize the cDFT potential, 
otherwise
a plausible alternative or compliment to extremizing $W$.}

Calculated \rd{values for} the unscreened
sum-over-states perturbation theory result for
$d^2 W / d V_\textit{c}^2$
following W\&VV,  corresponding to the \rd{omission} of 
screening in Eq.~\ref{Eq:bare}, 
\rd{are also shown in Fig.~\ref{fig2} (solid circles).}
For this, we generated optimized 
conduction band states using the method described in
Ref.~\onlinecite{PhysRevB.84.165131}, 
with the conduction band Wannier function 
cutoff radii set to 
$14~\mathrm{a}_0$.
This enabled us to numerically confirm that 
unscreened perturbation theory 
does not generally
\rd{match the self-consistent 
total energy curvature, nor that of its constraint or DFT 
energy contributions individually.
The anti-symmetry of the unscreened summand guarantees that it
 monotonically decreases
with an increasing number of conduction band states, 
so that the difference 
between the measured energy curvature
and the unscreened sum-over-states 
also monotonically increases with the number of states.}

\subsection{The integrated non-interacting response function
and dielectric function}

\begin{figure}
\begin{center}
\includegraphics[width=1\columnwidth]{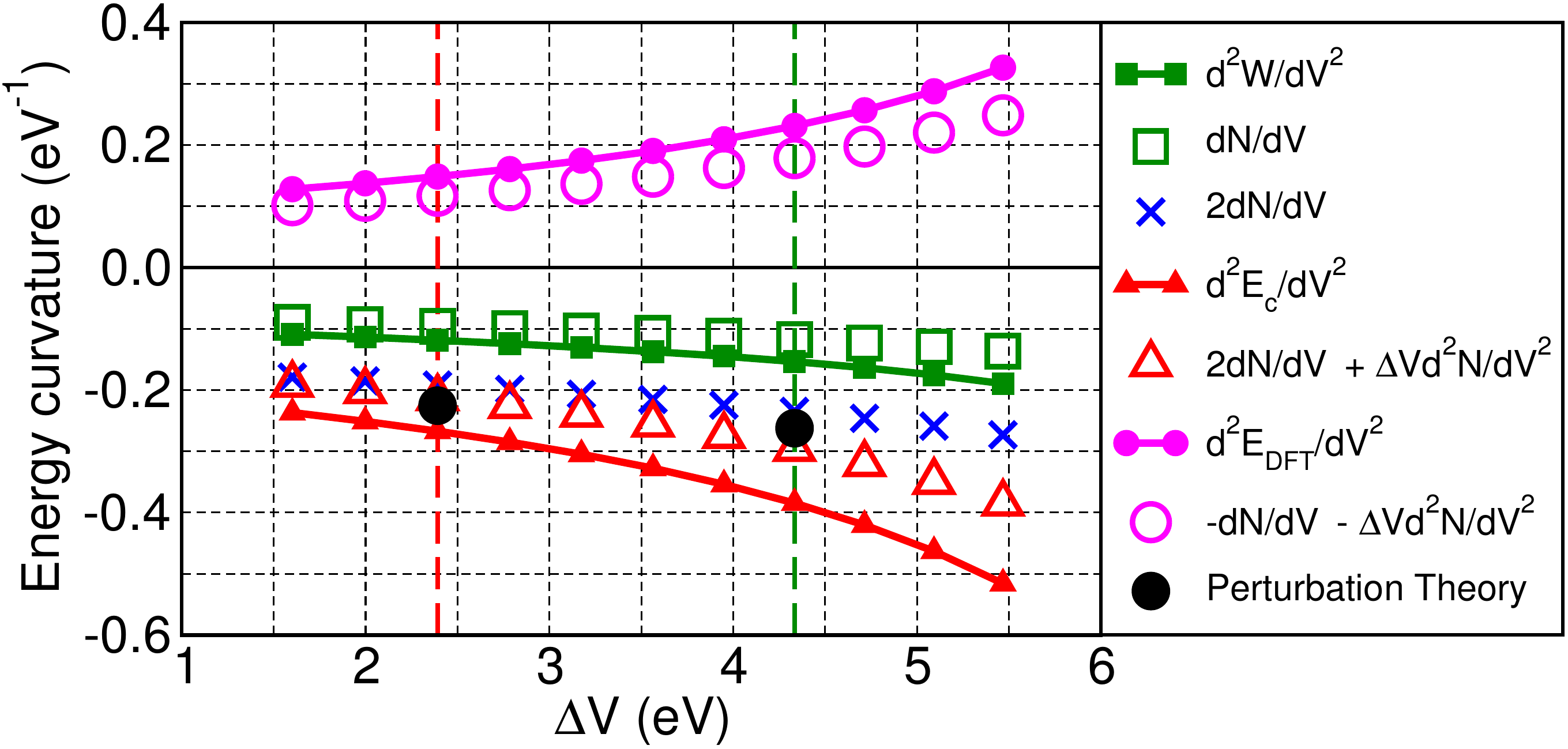}
\caption{(Color online) Results of the  calculations 
 shown in Fig.~\ref{fig2},  where  derivatives
are instead taken  with respect to the change in Kohn-Sham
potential averaged
over the constrained region,
that  is the $\Delta V$ induced by a finite $V_\textit{c}$,
using the same weight function as 
used to calculate the constrained occupancy $N$.
The derivative $dN/dV$  corresponds 
to the average non-interacting charge response
of the constrained region, whereas $ dN/dV_\textit{c}$ is the 
 interacting response. 
Unlike the 
derivatives shown in Fig.~\ref{fig2}, these non-interacting 
(i.e., bare or independent-particle) derivatives 
demonstrate no equivalence.
The perturbation 
theory data points are as in Fig.~\ref{fig2}.}
\label{fig3}
\end{center}
\end{figure}

\begin{figure}
\begin{center}
\includegraphics[width=1\columnwidth]{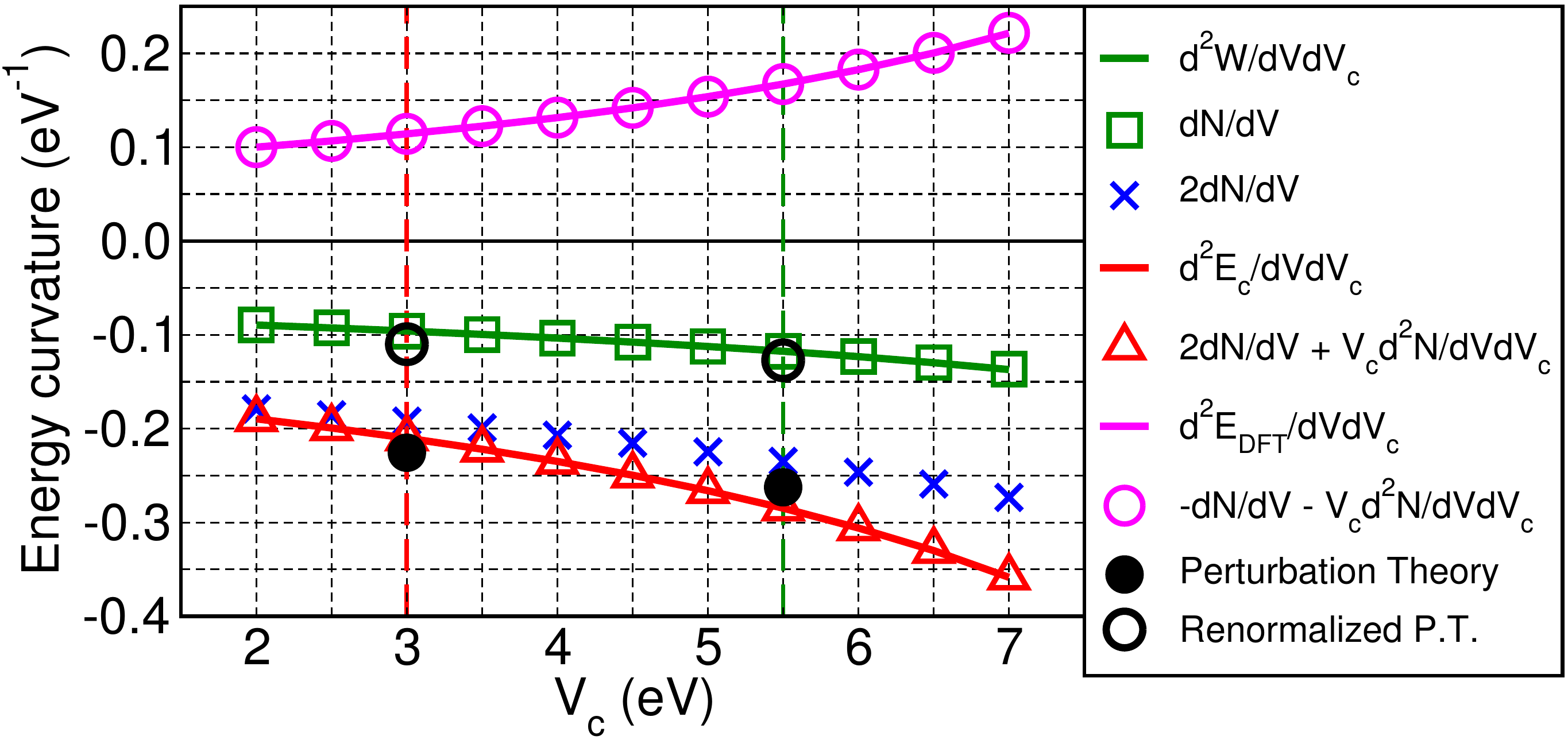}
\caption{(Color online) \rd{Following from Fig.~\ref{fig3}, 
the correspondence between energy curvatures and 
averaged non-interacting response functions
is recovered by using mixed interacting and 
non-interacting second derivatives, 
which restores their consistency.
The  sum-over-states perturbation theory (P.T.) data 
points following W\&VV are again shown (solid circles), together
with the ``Renormalized P.T.'' data points (open black circles)  described
in Appendix~\ref{App:summation}.}
}
\label{fig4}
\end{center}
\end{figure}

In Figs~\ref{fig3} and~\ref{fig4}, we \rd{illustrate} the relationship 
between energy curvatures and integrated density response 
functions with respect to the screened equivalent of the cDFT
Lagrange multiplier, that is the average \rd{change in} 
the Kohn-Sham potential
over the constrained region. 
This \rd{provides a   test of the  
 magnitude 
and potential importance
of dielectric screening effects in the  
energy versus Lagrange multiplier 
derivatives of self-consistent cDFT.}
The relevant weighted measure of the screened potential, for pure
density functionals and constraints,  is given by
\begin{align}
V^\sigma  = 
\left( \int   w_\textit{c}^\sigma \left( \mathbf{r} \right)
 v_\sigma^{\textrm{KS}} \left( \mathbf{r} \right) \; d\mathbf{r} \right)
\left( \int   w_\textit{c}^\sigma \left( \mathbf{r'} \right)
 \; d\mathbf{r'} \right)^{-1} .
 \label{Eq:averagepot}
\end{align}
Then, whereas \rd{$\chi = dN / dV_\textit{c}$} is the integrated
interacting density response function, 
\rd{we may define $\chi_0 = dN / dV$ as} its non-interacting 
(also known as independent-particle) counterpart.
\rd{If
$\chi^{\sigma \sigma'}_0 \left( \mathbf{r} , \mathbf{r'} \right)
= d \rho^\sigma \left( \mathbf{r}\right) / 
d v^\textrm{KS}_\sigma \left( \mathbf{r'}\right) $, then
\begin{align}
\chi^{\sigma \sigma'}_0
={}& 
\frac{dN^\sigma }{ dV^{\sigma'} } 
=
\frac{
d \int   w_\textit{c}^{\sigma} \left( \mathbf{r} \right)
 \rho^{\sigma} \left( \mathbf{r}  \right) 
\; d\mathbf{r} 
}{d 
\int   w_\textit{c}^{\sigma'} \left( \mathbf{r'} \right)
v_{\sigma'}^{\textrm{KS}} \left( \mathbf{r'} \right)
\; d\mathbf{r'} } 
\int w_\textit{c}^{\sigma'} \left( \mathbf{r''} \right)
\; d\mathbf{r''} \nonumber \\
 \approx{}&
\iint   w_\textit{c}^\sigma \left( \mathbf{r} \right)
\chi^{\sigma \sigma'}_0 \left( \mathbf{r} , \mathbf{r'} \right)
w_\textit{c}^{\sigma'} \left( \mathbf{r'} \right) 
\; d\mathbf{r} \; d\mathbf{r'},
\end{align}
where the final approximation, 
a consequence of neglected local-field effects, becomes 
an equality in the special case that 
$ w_\textit{c}^\sigma \left( \mathbf{r} \right)
d v^\textrm{KS}_\sigma \left( \mathbf{r}\right)
= w_\textit{c}^\sigma \left( \mathbf{r} \right) d V$ for all $\mathbf{r}$.
}

Fig.~\ref{fig3} shows that 
replacing $V_\textit{c}$ by $V$
does not preserve any equivalence between energy curvatures
and response functions. 
Mixed derivatives with respect to $V_\textit{c}$ and $V$,
\rd{the results of which are shown in Fig.~\ref{fig4},
are required for consistency of screening.
Firstly,} returning with the result $d W / d V_\textit{c} = C$
for \rd{orthonormal} states, \rd{we find that}
\begin{align}
\frac{d^2 W}{dV d V_\textit{c}} = \frac{d C}{d V}
= \frac{d N}{d V} \rd{\equiv \chi_0
= \frac{1}{2} \sum_{\sigma \sigma'} \chi_0^{\sigma \sigma'} },
\label{Eq:nonintercurv}
\end{align}
which is numerically confirmed via  Fig.~\ref{fig4}.
Next, we have
\begin{align}
\frac{d^2 E_\textit{c}}{dV d V_\textit{c}} ={}&  \frac{d^2 
\left( V_\textit{c} C\right)}{dV d V_\textit{c}} 
= \frac{d }{d V} \left[ C + V_\textit{c} \frac{d C}{d V_\textit{c}}\right]
 \nonumber  \\ 
={}& 
 \frac{d C}{d V}  + \frac{d V_\textit{c}}{d V}  \frac{d C}{d V_\textit{c}}
 + V_\textit{c} \frac{d^2 C}{dV d V_\textit{c} }
 \nonumber  \\ 
={}&
 2 \frac{d N}{d V}  + V_\textit{c} \frac{d^2 N}{d V d V_\textit{c} }
 \rd{= 2 \chi_0 + V_\textit{c} \frac{d \chi_0}{d V_\textit{c}}}, 
\end{align}
\rd{since the derivatives with respect to the external and
internal potentials commute.}
Finally, for 
\rd{the DFT component of the total-energy,} we deduce that
\begin{align}
\frac{d^2 E_\textrm{DFT} }{d V d V_\textit{c}}
={}& 
\frac{d^2 \left( W - E_\textit{c} \right) }{d V d V_\textit{c}}
=
- \frac{d N }{ d V} - V_\textit{c} \frac{d^2 N }{d V  dV_\textit{c}} \nonumber \\
={}&
\rd{  - \chi_0  - V_\textit{c} \frac{d \chi}{ d V }}
=
- \chi_0  - V_\textit{c} \frac{d \chi_0}{ d V_\textit{c} }.
\label{Eq:DFT2ndderiv}
\end{align} 
In practice, these mixed derivatives \rd{are} calculated 
simultaneously with the previously detailed curvatures
with respect to $V_\textit{c}$, 
by monitoring the variation of the weighted, \rd{fully relaxed} Kohn-Sham 
potential of Eq.~\ref{Eq:averagepot} in self-consistent  cDFT calculations.
\rd{As before, we observe a precise agreement between the numerically
evaluated energy curvatures and response functions.  
This confirms Eq.~\ref{Eq:nonintercurv}, namely, 
that the averaged non-interacting (i.e., independent-particle)
response function of DFT may be expressed
as an total-energy landscape property.
We return  to discuss the unscreened
sum-over-states perturbation theory results 
shown in Figs.~\ref{fig3} and~\ref{fig4} 
in Appendix~\ref{App:summation}.
}

\rd{The process of Lagrange multiplier 
optimization offers ready access to the physical response 
properties of the constrained region, 
which are not limited to those  of the  target state.
The simplest such quantity is  the 
subspace inverse  dielectric constant (i.e., screening factor), 
\begin{align}
 \label{Eq:mydielectric}
 \epsilon^{-1}_{ \sigma \sigma'} \equiv{}& 
 \iint   w_\textit{c}^\sigma \left( \mathbf{r} \right)
\epsilon^{-1}_{ \sigma \sigma'} \left( \mathbf{r} , \mathbf{r'} \right)
w_\textit{c}^{\sigma'} \left( \mathbf{r'} \right) 
\; d\mathbf{r} \; d\mathbf{r'} \nonumber \\
{}&\times \Big( \int   w_\textit{c}^{\sigma'} 
\left( \mathbf{r''} \right) \; d\mathbf{r''} \Big)^{-1}.
\end{align}
This may be calculated directly  using the
 cDFT integrated response functions, since, by definition, 
\begin{align}
d v_\sigma^\textrm{KS} \left( \mathbf{r} \right) 
={}& 
\int \epsilon^{-1}_{\sigma \sigma'} \left( \mathbf{r},  \mathbf{r'} \right) 
d v_{\sigma'}^\textrm{external}  \left( \mathbf{r'} \right) 
\; d\mathbf{r'} \nonumber \\
={}& 
 \int \epsilon^{-1}_{\sigma \sigma'} \left( \mathbf{r},  \mathbf{r'} \right) 
w_\textit{c}^{\sigma'} \left( \mathbf{r'} \right)  d V_\textit{c}^{\sigma'} 
\; d\mathbf{r'} \nonumber \\
\Rightarrow \quad d V^\sigma
={}& 
\epsilon^{-1}_{ \sigma \sigma'} d V_\textit{c}^{\sigma'} .
\end{align}
From this, it is clear that $ \epsilon^{-1}_{ \sigma \sigma'} $ 
 is a property of the constrained ground-state density 
 and is not explicitly dependent on unoccupied states.
Next, we may apply the chain rule via the  
   the constrained property $N^\sigma$, introducing the notation
 $\chi_0^{-1 \sigma \sigma''} = d V^\sigma / d N^{\sigma''}$,
which provides that
 \begin{align}
\epsilon^{-1}_{ \sigma \sigma'}  ={}& 
 \frac{d V^\sigma}{d V_\textit{c}^{\sigma'}}
  = \sum_{\sigma''}
  \chi_0^{-1 \sigma \sigma''} \chi^{\sigma'' \sigma'}.
\end{align}
%
%
The inverse of this quantity
is the subspace-averaged dielectric constant, 
neglecting local-field effects, 
given by
\begin{align}
\epsilon_{ \sigma \sigma'} ={}& 
 \frac{d V_\textit{c}^{\sigma'}}{d V^\sigma}
  = \sum_{\sigma''}
  \chi_0^{ \sigma \sigma''} \chi^{-1 \sigma'' \sigma'}.
\end{align}
%
cDFT thus provides an efficient means of
estimating the dielectric constants of  spatial regions, which 
are central in  implicit solvation~\cite{0295-5075-95-4-43001,
:/content/aip/journal/jcp/136/6/10.1063/1.3676407}, supercell
convergence acceleration of excitation 
energies~\cite{PhysRevB.93.165102}, and  
in high-frequency optical response, in terms of
by-products of its optimization.
}

\section{Stable cDFT solutions are energy maxima
with respect to their Lagrange multiplier}
\label{Sec:curvaturesign}

\rd{
The response function based approach
to cDFT analysis is  used in this Section
to globally characterize its stationary points.
W\&VV have shown that a non-negative total-energy curvature 
with respect to the cDFT Lagrange multiplier is guaranteed
in cases where non-degenerate perturbation theory is 
applicable~\cite{PhysRevA.72.024502},  
in the absence of screening effects.
This regime may hold during cDFT optimization within the 
density-update loop of self-consistent field 
codes, where the Kohn-Sham potential is 
fixed~\cite{wu:2498,doi:10.1021/ct0503163}.
More generally, or physically, the Kohn-Sham potential
is relaxed self-consistently for each $V_\textit{c}$,
and sum-over-states perturbation theory becomes 
inconclusive as discussed in the text surrounding Eq.~\ref{Eq:bare}.
We provide a  general, non-perturbative 
proof that stable cDFT solutions always occur at
energy maxima with respect to their Lagrange multiplier.
%
%
We then extend this proof to the multivariate cDFT regime
in Section~\ref{Sec:multiple}.
}


We have shown that the curvatures of the total-energy
with respect to the cDFT Lagrange
multiplier $V_\textit{c}$ are equal to integrated 
density response functions.
\rd{Next, we will show that, similarly,}
the curvatures with respect to the cDFT occupancy $N$, 
for a particular target $N_\textit{c}$, are equal to 
\rd{the inverses of these} response functions.
The occupancy $N$ is not a free parameter, and so,
in order to analyze these curvatures,
it is convenient to initially take derivatives 
with respect to \rd{the cDFT target occupancy 
$N_\textit{c}$, subject to the condition that
$N = N_\textit{c}$ for a suitable Lagrange multiplier
$V_\textit{c} \left( N_\textit{c} \right)$.
Echoing Eq.~\ref{Eq:first}, assuming
that the ground-state is located for all $V_\textit{c}$, 
\begin{align}
\frac{d W}{d N_\textit{c} } ={}& \sum_\sigma \sum_i^{N_\sigma}
\textrm{Tr} \left[ 
\left( \frac{\delta W}{\delta \phi_{i \sigma}^{\ast}} 
\frac{d \phi_{i \sigma}^{\ast}}{d V_\textit{c}} 
+ \textit{c.c.} \right) \frac{d V_\textit{c}}{d N_\textit{c}} \right]
+\frac{\partial W}{\partial N_\textit{c} } \nonumber \\
={}&
\sum_\sigma \sum_i^{N_\sigma}
\textrm{Tr}  \left[ \left( \left( \hat{H}_\sigma \phi_{i \sigma}  \right)
\frac{d \phi_{i \sigma}^{\ast}}{d V_\textit{c} } 
+ \textit{c.c.} \right)  \frac{d V_\textit{c}}{d N_\textit{c} } 
\right] - V_\textit{c}
\nonumber \\
={}& - V_\textit{c}  \left( N_\textit{c} \right)
\quad \Rightarrow \quad \frac{d^2 W \left( N_\textit{c} \right)
}{d N_\textit{c}^{ 2} } 
=
- \frac{ d V_\textit{c}  }{ d N_\textit{c} }.
\label{Eq:occupancycurvaturefirst}
\end{align}
Since the constraint $N = N_\textit{c}$ is satisfied
for each $N_\textit{c}$, the constraint energy 
always vanishes and $E_\textrm{DFT} \left( N \right) = W 
\left( V_\textit{c} \left( N_\textit{c} \right) \right)$ along the curve.
Thus, we may write, for the 
occupancy curvature of the DFT contribution, that
\begin{align}
\frac{d^2 E_\textrm{DFT} \left( N \right)
}{d N^2 }  
=
\frac{d^2 W \left( V_\textit{c} \left( N_\textit{c} \right) \right)
}{d N_\textit{c}^{ 2} } 
= 
- \frac{ d V_\textit{c}  }{ d N } = - \chi^{-1}.
\label{Eq:occupancycurvature}
\end{align}
As previously established by 
W\&VV for the unscreened case~\cite{PhysRevA.72.024502}, 
the combination of Eqs.~\ref{Eq:d2WdVc2} 
and~\ref{Eq:occupancycurvature}}
provides that the curvature of the total-energy
with respect to Lagrange multiplier is  directly related to the
curvature of the DFT energy with respect to the 
 occupancy, namely
\rd{ \begin{align} 
 \left( \frac{d^2 W}{d V_\textit{c}^2 } \right)^{-1}
&{}=
 \left( \frac{d N} {d V_\textit{c} } \right)^{-1}
= 
\chi^{-1} \nonumber \\ &{}= 
  \frac{d V_\textit{c} }{d N} 
 =
 -  \frac{d^2 E_\textrm{DFT} }{ d N^2 } .
  \label{Eq:finalsingle}
\end{align} 
}

\begin{figure}
\begin{center}
\includegraphics[width=1\columnwidth]{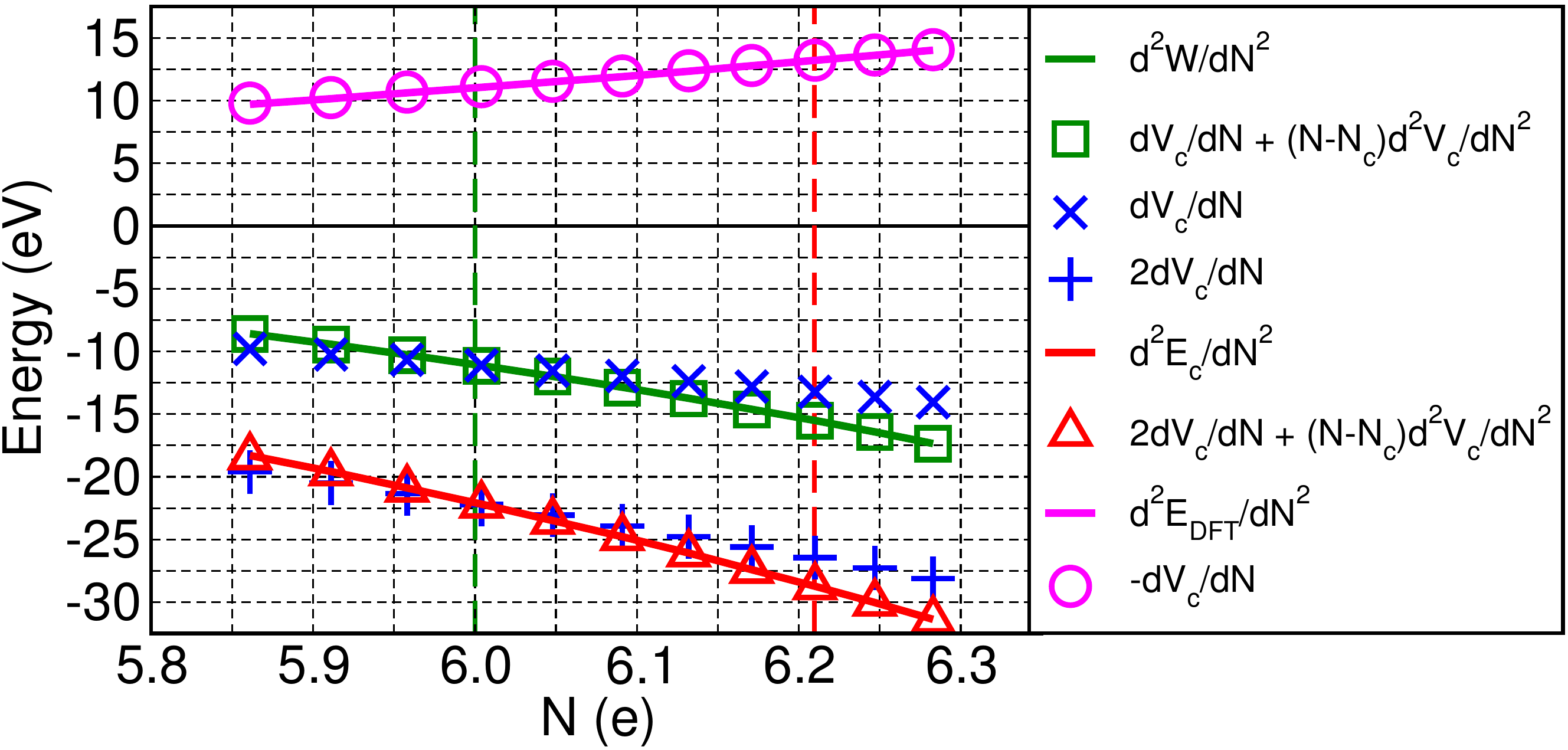}
\caption{(Color online) Numerically evaluated
second derivatives of the cDFT total-energy $W$, 
its constraint $E_\textit{c}$ and DFT
$E_\textrm{DFT}$ components, 
with respect to the constrained occupancy $N$, for the  system
shown in Fig.~\ref{fig1} and occupancy target $N_\textit{c} =
 6~\textrm{e}$. Dashed vertical lines show the occupancy
 at which $W$ (green) and $E_\textit{c}$ (red) are maximized. 
 The corresponding  occupancy 
 derivatives of the Lagrange multiplier
(averaged inverse screened response functions) are shown, 
as well as their non-corresponding linear-response approximations
(crosses).}
\label{fig5}
\end{center}
\end{figure}

\rd{We may extend Eq.~\ref{Eq:finalsingle} 
to the  general cDFT total-energy $W$, 
no longer subject the constraint that the target occupancy is attained, 
by freeing the target $N_\textit{c}$
after optimization of $V_\textit{c}$.
This has no bearing on $E_\textrm{DFT}$, by definition,
and so Eq.~\ref{Eq:occupancycurvature} holds irrespective of
whether the constraint is satisfied.
For $W \left( V_\textit{c}\right) $
with $N$ not necessarily equal to $N_\textit{c}$, it is sufficient to add}
the curvature of the now non-vanishing
constraint energy term $E_\textit{c}$,  given  by
\begin{align}
\frac{d^2 E_\textit{c} }{d N^2}
={}&  \frac{d^2 \left( V_\textit{c} C \right) }{d N^2}
= \frac{d }{d N} \left[ V_\textit{c}
\frac{d C}{d N} +   \label{Eq:myecdfteq}
C \frac{d V_\textit{c} }{d N}   \right]  \\ \nonumber
={}& 
2 \frac{d V_\textit{c} }{d N}  
  \frac{d C}{d N}  +   C \frac{d^2 V_\textit{c}}{d N^2 }  
=
2 \frac{d V_\textit{c} }{d N}  + \left( N - N_\textit{c} \right)
  \frac{d^2 V_\textit{c}}{d N^2 }  .
\end{align} 
The total-energy curvature is  provided by the sum
\begin{align}
\frac{d^2 W}{d N^2} ={}& 
\frac{d^2 E_\textrm{DFT}}{d N^2} +
\frac{d^2 
E_\textit{c} }{d N^2} \nonumber \\
={}& \frac{d V_\textit{c} }{d N}  + \left( N - N_\textit{c} \right)
  \frac{d^2 V_\textit{c}}{d N^2 }.
  \label{Eq:myweq}
\end{align}
\rd{
Then, by combining Eqs.~\ref{Eq:occupancycurvature} to~\ref{Eq:myweq}
and by applying the constraint
condition $N = N_\textit{c} $,
we arrive at our central result that, valid for all cDFT stationary points
as defined,
  \begin{align}
 \left( \frac{d^2 W}{d V_\textit{c}^2 } \right)^{-1}
&{}=
 \frac{1}{2} \frac{d^2 
E_\textit{c} }{d N^2 }
=
- \frac{d^2 E_\textrm{DFT}
}{d N^2 }
= \frac{d^2 W
}{d N^2 } .
\label{Eq:mycentral}
\end{align} 
This is numerically confirmed via Figs.~\ref{fig2}
and~\ref{fig5},  and the inaccuracy of
linear-response approximations in Fig.~\ref{fig5} (shown with crosses), 
except at $N = N_\textit{c}$,  is  clear. }

\rd{
Next, let us consider the possibility of 
simulating a stable constrained ground-state at a 
given value of $V_\textit{c}$, which
may be a cDFT solution but is not necessarily so.
Stability implies that
the energy is minimized with respect to the density, locally at least,
and put more precisely, that the energy is locally 
strictly convex. Mathematically, this condition is
represented by a positive definite
matrix $d^2 W / d \rho^{\sigma'} \left( \mathbf{r'} \right) 
d \rho^\sigma \left( \mathbf{r} \right)$,
evaluated at a fixed $V_\textit{c}$, 
where the coordinate pairs
$\left\lbrace \sigma, \mathbf{r} \right\rbrace$ form  
the basis vectors.
Separating the DFT and constraint energy terms, 
the stability of constrained ground-states is then determined by
\begin{align} 
\left. \frac{ d^2 W }{ d \rho^{\sigma'} \left( \mathbf{r'} \right) 
d \rho^{\sigma} \left( \mathbf{r} \right) } 
\right|_{V_\textit{c}, \hat{w}_\textit{c}} 
&{}= 
 \frac{ d^2 E_\textrm{DFT} 
}{ d \rho^{\sigma'} \left( \mathbf{r'} \right) d \rho^{\sigma} \left( \mathbf{r} \right) },
 \end{align}
  so that the latter is also positive definite. 
The underlying DFT energy landscape therefore alone determines
the stability of constrained ground-states and thus, while 
cDFT may relocate the minima of $W$ within the simply connected
domains of stability of the unconstrained DFT problem, 
it cannot deform those domains. 
This is a consequence of the linearity of $E_\textit{c}$ 
in the density and thus its vanishing 
contribution to the fixed-$V_\textit{c}$
curvature of $W$.
The constraint may, however cause  abrupt transitions of energy
minima across disconnected domains, as observed 
in the ``OS(-)'' to ``OS(+)''
transition of Fig.~\ref{fig0} (albeit in the more general case of 
a spin-dependent constraint).
}

\rd{In order to analyze the sign of the quantity  
expressed in Eq.~\ref{Eq:mycentral}, we may  focus on the DFT energy curvature with
respect to the subspace occupancy, 
$d^2 E_\textrm{DFT} / d N^2$.
Let us suppose that stable ground-states are
observed, in a particular constrained system, 
as $V_\textit{c}$ is continuously varied
within an interval.
The local
special case of Eq.~\ref{Eq:occupancycurvaturefirst} provides that 
$ d E_\textrm{DFT} / d \rho^\sigma \left( \mathbf{r} \right)
= - v^\textrm{external}_\sigma  \left(  \mathbf{r} \right)
= - w_\textit{c} \left( \mathbf{r}\right)
  V_\textit{c} $. 
Combining this with the general 
properties of a positive definite matrix, i.e., that
its inverse is positive definite and
 that its diagonal elements are positive, we find that
\begin{align}
0 &{}<  \left( \frac{ d^2 E_\textrm{DFT} }{ d \rho^{\sigma 2}  } \right)^{-1} 
\left( \mathbf{r} , \mathbf{r} \right) 
=  \left( -  \frac{ d v_{\sigma}^\textrm{external} }{ d \rho^{\sigma}  } \right)^{-1} 
\left( \mathbf{r} , \mathbf{r} \right) \nonumber \\ 
&{}= - \chi^{\sigma \sigma} \left( \mathbf{r} , \mathbf{r} \right)  =
- \frac{d \rho^\sigma  \left( \mathbf{r}\right) }{ d v_{\sigma}^\textrm{external}  \left( \mathbf{r}\right)}
 = - \frac{d \rho^\sigma \left( \mathbf{r}\right) }{ 
 w^\sigma_\textit{c} \left( \mathbf{r}\right)
 d V_\textit{c}}.
 \label{Eq:mylocal}
\end{align}
Following from this, we
may multiply by 
$\left ( w^\sigma_\textit{c} \left( \mathbf{r} \right) \right)^2$,
allowing $w^\sigma_\textit{c} \left( \mathbf{r} \right)$ 
to take zero (but not at all $\mathbf{r}$) and negative values, 
as is necessary, for example, when constraining the difference
 in charge between two regions. 
Then, integrating over all space, we arrive at the result
\begin{align}
0 &{}> 
 \sum_\sigma \int \left( w^\sigma_\textit{c} \left( \mathbf{r}  \right) \right)^2
\frac{ d \rho^\sigma \left( \mathbf{r} 
\right) }{  w^\sigma_\textit{c} \left( \mathbf{r}  \right) d V_\textit{c} } 
\; d\mathbf{r}
=  \frac{ d N  }{  d V_\textit{c} }  =  \chi .
 \label{Eq:myintegrated}
\end{align}
Thus, Eq.~\ref{Eq:mylocal}, which states that
the local part of the interacting response function is negative
about stable ground-states, 
is generalized in Eq.~\ref{Eq:myintegrated} to
the cDFT response function  for an arbitrary weight function,
irrespective of its profile or even its sign.
Recalling Eq.~\ref{Eq:d2WdVc2}, we arrive at the 
sought-after result that $d^2 W / d V_\textit{c}^2 < 0 $
over stable ground-states. We note that this result is valid up to 
arbitrary order in response, it is general to 
degenerate and non-degenerate systems, 
and it  accounts for screening. 
}

\rd{
Conversely, let us briefly suppose that $d^2 W / d V_\textit{c}^2 
 \ge 0 $  for some value of $V_\textit{c}$.
 Then, by virtue of Eq.~\ref{Eq:d2WdVc2}, we have
  $ d N / d V_\textit{c}  \ge 0 $ 
 and, combining Eq.~\ref{Eq:mylocal}, Eq.~\ref{Eq:myintegrated},
and the non-negativity of 
 $ \left( w^\sigma_\textit{c} \left( \mathbf{r}  \right) \right)^2$,
there exist    
$\mathbf{r}$ and $\sigma$ for which
$w^\sigma_\textit{c} \left( \mathbf{r}  \right) \ne 0$ and 
$0 \ge 
\left( d^2 E_\textrm{DFT} / d \rho^{\sigma 2} 
 \right)^{-1} 
\left( \mathbf{r} , \mathbf{r} \right) $.
Then, $\big( d^2 E_\textrm{DFT} / d \rho^{\sigma } 
d \rho^{\sigma' } \big)^{-1}
\left( \mathbf{r} , \mathbf{r'}  \right) $ 
may not be positive definite since it has a non-positive 
diagonal element. Therefore, its inverse
$\big( d^2 E_\textrm{DFT} / d \rho^{\sigma } 
d \rho^{\sigma' } \big)
\left( \mathbf{r} , \mathbf{r'}  \right) $ may also not
be positive definite, and 
 this guarantees that the state is unstable  
with respect to spin-density perturbations. Then,  
only meta-stable states such as the ``CS'' 
state depicted in Fig.~\ref{fig0} 
may be observed in practice.
There are two ways in which 
cDFT inflection points may arise, that is, cases where
$d^2 W / d V_\textit{c}^2  =
d N / d V_\textit{c} = 0 $. First, the constraint may couple 
pathologically to the density or not couple to it at all.
Trivial examples of this are constraints
 where, for all $\sigma$ and $\mathbf{r}$, 
 $w^\sigma_\textit{c} \left( \mathbf{r} \right)
= 1$ or $w^\sigma_\textit{c} \left( \mathbf{r} \right) 
\rho^\sigma \left( \mathbf{r} \right) = 0$, which result
 in a vanishing response $\chi$. Second, the 
 state may be degenerate on a line or higher surface, 
 and the constraint may be contrived so as not to
  break this degeneracy. 
}
 
\rd{
Combining these arguments, we conclude that the stability
of ground-states over 
an interval of $V_\textit{c}$ both implies and requires that  
 $d^2 W / d V_\textit{c}^2$ remains negative and finite 
at constrained ground-states within that interval.
In practice, this means that intervals of positive curvature
are not numerically observable, because unstable or meta-stable
states cannot be sampled in a continuous manner. 
It is however possible,  
as demonstrated in Fig.~\ref{fig0}, to observe numerical 
discontinuities in the otherwise concave total-energy $W$, 
for fixed $N_\textit{c}$, when the response, and hence
energy curvature, diverges at a phase transition. 
Such transitions occur at vanishing values of 
the quantity expressed in Eq.~\ref{Eq:mycentral}, i.e., at inflection
points of $E_\textrm{DFT} \left( N \right)$.
For this, it is necessary for $\chi^{\sigma \sigma } \left( \mathbf{r}, \mathbf{r}\right)$ 
to diverge for some $\sigma$ and $\mathbf{r}$ where
$w^\sigma_\textit{c} \left( \mathbf{r} \right) \ne 0$, by virtue
of Eqs.~\ref{Eq:finalsingle},~\ref{Eq:mylocal}
and~\ref{Eq:myintegrated}.
}

\section{Multiple constraints, 
the cDFT condition number, and the extension 
to self-consistent field cDFT}
\label{Sec:multiple}

\rd{In this section, we consider, in detail, 
the extent to which our principal 
results may be generalized to
multiple constraints, and thereafter to 
 self-consistent field cDFT. 
For  the former, it is helpful to consider 
a vector Lagrange multiplier $\mathbf{V}_\textit{c}$,
acting upon a vector of constraint functionals
$\mathbf{C} = \mathbf{N} - \mathbf{N}_\textit{c}$, yielding 
$E_\textit{c} = \mathbf{V}_\textit{c} \cdot \mathbf{C}$.
The multivariate generalization of Eq.~\ref{Eq:finalsingle}
is  the equation 
\begin{align} 
 \left( \frac{d^2 W}{d \mathbf{V}_\textit{c}^2 } \right)^{-1}
&{}=
 \left( \frac{d \mathbf{N} } {d \mathbf{V}_\textit{c} } \right)^{-1}
= 
\boldsymbol{\chi}^{-1} \nonumber \\ &{}= 
  \frac{d \mathbf{V}_\textit{c} }{d \mathbf{N} } 
 =
 -  \frac{d^2 E_\textrm{DFT} }{ d \mathbf{N}^2 }, 
  \label{Eq:finalresult}
\end{align}
where the negative exponent denotes matrix inversion,
and the matrix is symmetric due to  
the symmetry of $\chi \left( \mathbf{r},\mathbf{r'} \right)$
under co-ordinate exchange 
in Eq.~\ref{Eq:mydefintegratedchi}.
We may also consider the non-interacting analogue
of Eq.~\ref{Eq:finalresult}, viz.
\begin{align} 
 \left( \frac{d^2 W}{d \mathbf{V} d \mathbf{V}_\textit{c} } \right)^{-1}
&{}=
 \left( \frac{d \mathbf{N} } {d \mathbf{V} } \right)^{-1}
= 
\boldsymbol{\chi}^{-1}_0. 
\end{align}
Similar expressions may be derived for the second
derivatives of  $E_\textrm{DFT}$ and $E_\textit{c}$.
For example, Eq.~\ref{Eq:DFT2ndderiv} becomes,  
noting that the final object is a rank-three tensor,
\begin{align}
 \frac{d^2 E_\textrm{DFT} }{ d \mathbf{V} d \mathbf{V}_\textit{c} }
 = - \boldsymbol{\chi}_0  - \mathbf{V}_\textit{c} \cdot
 \frac{d \boldsymbol{\chi}_0 }{d \mathbf{V}_\textit{c} }.
\end{align}
The   dielectric function may also be 
straightforwardly generalised in order
to describe the coupling of dielectric screening between 
constrained subspaces. In the spin-degenerate 
or spin-averaged case, 
it takes the particularly simple 
matrix form \begin{align}
\boldsymbol{\epsilon} ={}& 
 \frac{d \textbf{V}_\textit{c} }{d \textbf{V} }
  = \frac{1}{2}
 \boldsymbol{\chi}_0 \cdot \boldsymbol{\chi}^{-1 }.
\end{align}
}

\rd{
The  behaviour of multivariate cDFT optimisation  
depends on the nature of the energy landscape
with respect to $\textbf{V}_\textit{c}$, and it may 
be complicated by the presence of multiple extrema 
or by  surfaces which are  elongatated in some 
directions relatively to others.
The curvature of the energy landscape, 
about a particular value of $\textbf{V}_\textit{c}$, 
is characterized by the eigenvalues of the Hessian 
matrix $\boldsymbol{\chi}$.
These must be identical
for optimal  extremization of $W \left[ \mathbf{V}_\textit{c} \right]$
by simple conjugate gradients. 
More generally, the positive-valued condition number $k$, 
defined by
\begin{align}
k^2 =  \sum_{I J} \left(  \left[
\frac{d^2 W}{d \mathbf{V}_\textit{c}^2  }  \right]_{I J}  \right)^2
 \sum_{K L}  \left( \left[
 \frac{d^2 W}{d \mathbf{V}_\textit{c}^2  } 
\right]_{K L}^{-1} \right)^2, 
\label{Eq:mycondition}
\end{align}
where capitalized letters index constraints, 
gives a  measure of the spread of the Hessian eigenvalues.
The optimization problem is said to be ill-conditioned 
if $k$ is much larger than one.
Then, conjugate gradients may be expected to 
perform poorly, and Newton or quasi-Newton
methods may be considered. 
We may rephrase $k$  as
\begin{align}
\label{Eq:variouscondition}
k^2 &{}=  \sum_{I J} \left( \left[
\frac{d\mathbf{N}}{d \mathbf{V}_\textit{c}  } \right]_{I J} \right)^2
 \sum_{K L}  \left( \left[
 \frac{d \mathbf{V}_\textit{c}  }{d \mathbf{N}} 
\right]_{K L} \right)^2  \\ \nonumber
&{}=  \sum_{I J} \left( \left[
\boldsymbol{\chi} \right]_{I J} \right)^2
 \sum_{K L}  \left( \left[
 \boldsymbol{\chi}^{-1 }
\right]_{K L} \right)^2  \\ \nonumber
&{}=  \sum_{I J} \left( \left[
 \left( \frac{d^2 E_\textrm{DFT}  }{d \mathbf{N}^2} 
 \right)^{-1} \right]_{I J} \right)^2
 \sum_{K L}  \left( \left[
 \frac{d^2 E_\textrm{DFT}  }{d \mathbf{N}^2} 
\right]_{K L} \right)^2, 
\end{align}
through which it is clear that 
$W \left[ \mathbf{V}_\textit{c} \right]$ has the same condition
number as $E_\textrm{DFT} \left[ \mathbf{N}_\textit{c} \right]$.
If optimising $\mathbf{V}_\textit{c}$ during the
density update step of 
self-consistent field DFT 
codes~\cite{wu:2498,doi:10.1021/ct0503163}, 
the corresponding condition number $k_0$ may be calculated
using the fixed-potential equivalent of Eq.~\ref{Eq:variouscondition}.
Both $k$ and $k_0$ are amenable to evaluation by finite differences.
If the response $\boldsymbol{\chi}$ or inverse response
$\boldsymbol{\chi}^{-1}$ matrices are singular, $k$ diverges,
 the problem  is said to be ill-posed, and 
  its definition must be reconsidered.
We refer the reader to Ref.~\onlinecite{Rondinelli2007345} for a more
general discussion of condition numbers 
in the context of electronic structure.
}

\rd{
The connection between the stability of constrained states
and the negativity of $d^2 W / d V_\textit{c}^2 $, 
as demonstrated in Section~\ref{Sec:curvaturesign},
is here generalized to the multivariate case. 
More precisely, the stability of multiply-constrained
ground-states both implies and requires that the
 matrix $ d^2 W / d \mathbf{V}_\textit{c}^2 $ has  strictly negative
diagonal  elements. If the matrix is strictly diagonally
dominant, furthermore, stability implies and requires
that $W \left(  \mathbf{V}_\textit{c} \right)$ is
strictly concave locally. 
To see this, let us assume that stable ground-states
are observed as $\mathbf{V}_\textit{c}$ is varied within some
simply connected volume.
Then, taking the diagonal matrix elements of 
$ d^2 W / d \mathbf{V}_\textit{c}^2 $ individually, while following
Eqs.~\ref{Eq:mylocal} and~\ref{Eq:myintegrated}
 for each constraint labelled $I$, we find that
\begin{align}
0 &{}> 
 \sum_\sigma \int \left( w^\sigma_{\textit{c} I} \left( \mathbf{r}  \right) \right)^2
\frac{ d \rho^\sigma \left( \mathbf{r} 
\right) }{  w^\sigma_{\textit{c}  I }\left( \mathbf{r}  \right) d V_{\textit{c} I}} 
\; d\mathbf{r}
=  \left[ \frac{d\mathbf{N}}{d \mathbf{V}_{ \textit{c} } } \right]_{I I}.
\label{Eq:myalmostfinal}
\end{align}
Combining this with Eq.~\ref{Eq:finalresult}, we deduce 
that the diagonal matrix elements of 
$ d^2 W / d \mathbf{V}_\textit{c}^2 $ are all negative.
This is a necessary but not sufficient condition for 
$ d^2 W / d \mathbf{V}_\textit{c}^2 $ to be negative definite locally.
However, it becomes a sufficient condition for negative
definiteness if the symmetric matrix is also 
diagonally dominant~\cite{harville2011matrix}.
Diagonal dominance is a physically reasonable, albeit 
system-dependent, condition related to the extent of non-locality
in $\chi \left( \mathbf{r}, \mathbf{r'}\right)$. If it holds, stability 
implies that 
$W \left(  \mathbf{V}_\textit{c} \right)$ is
strictly concave locally. 
Additionally, if 
$w^\sigma_{\textit{c}  J}\left( \mathbf{r}  \right)
w^\sigma_{\textit{c}  I }\left( \mathbf{r}  \right) > 0 $ for all
$I$, $J$, $\sigma$, and $\mathbf{r}$ (such
as if $w^\sigma_{\textit{c}  I }\left( \mathbf{r}  \right)
> 0$ for all $I$ and $\sigma$), then 
\begin{align}
0 &{}> 
 \sum_\sigma \int  w^\sigma_{\textit{c} J} \left( \mathbf{r}  \right)
 w^\sigma_{\textit{c} I}
\left( \mathbf{r}  \right)
\frac{ d \rho^\sigma \left( \mathbf{r} 
\right) }{  w^\sigma_{\textit{c}  I }\left( \mathbf{r}  \right) d V_{\textit{c} I}} 
\; d\mathbf{r}, 
\end{align}
and the Hessian is a strictly negative matrix, but that alone
does not imply its negative definiteness.
Conversely, if $d^2 W / d \mathbf{V}_\textit{c}^2 $ does not
exhibit all negative diagonal elements, 
as is possible if it is not negative definite, for example, 
then, $\left[ d \mathbf{N} / d \mathbf{V}_\textit{c}
\right]_{I I} \ge 0$ for one or more values of $I$.
Then, by the non-negativity of 
$\left( w^\sigma_{\textit{c}  I }\left( \mathbf{r}  \right)  \right)^2$
for all $I$, $\sigma$, and $\mathbf{r}$, 
the state is unstable, as previously shown 
in the single-constraint case.
Thus, we conclude that if a ground-state subject
to multiple linear, non-trivial constraints is stable
with respect to perturbations, 
and hence  locatable by numerical cDFT optimization,
this implies and requires that the associated Hessian matrix
$d^2 W / d \mathbf{V}_\textit{c}$ has a strictly negative
diagonal, and that it is negative definite in the event that it
is diagonally dominant. If it is not strictly diagonally dominant, however,
we cannot rule out the possibility of stable ground-states
occurring at saddle points of the self-consistent
$W \left(  \mathbf{V}_\textit{c} \right)$.
}



\rd{Finally, we consider the
energy landscape associated with cDFT Lagrange
multiplier optimization within the fixed-potential inner
loop of self-consistent field DFT codes. 
For this, let us first assume 
that stable minimum energy 
eigenstates are observed, 
for a given fixed DFT contribution 
$\hat{v}^\textrm{DFT}_\sigma$ to the
Kohn-Sham potential, 
 as $\mathbf{V}_\textit{c}$ is varied within some
simply connected volume. 
Again, we may analyze the diagonal matrix elements 
of $d^2 W / d \mathbf{V}_\textit{c}^2 $ individually, but 
in this case 
$d v_\sigma^\textrm{KS}  \left( \mathbf{r} \right)
= d v_\sigma^\textrm{external}
\left( \mathbf{r} \right) =  w_{\textit{c} I}^\sigma
\left( \mathbf{r} \right) d V_{\textit{c} I}^\sigma $, for each
$I$, and for all $\sigma$ and $\mathbf{r}$.
Then, Eq.~\ref{Eq:mylocal} may be modified 
for a fixed DFT potential, to give
\begin{align}
 0 &{}<  \left( 
\frac{ d^2 E_\textrm{DFT} }{ d \rho^{\sigma 2}  } 
\right)^{-1}_{\hat{v}^\textrm{DFT}_\sigma} 
\left( \mathbf{r} , \mathbf{r} \right) 
 = -  \left. \frac{ d \rho^\sigma \left( \mathbf{r}\right) }{ 
 w^\sigma_{ \textit{c} I} \left( \mathbf{r}\right)
 d V_{ \textit{c} I} } \right|_{\hat{v}^\textrm{DFT}_\sigma},
\end{align}
whereupon Eq.~\ref{Eq:myalmostfinal} may be similarly
adapted, yielding
\begin{align}
  0 &{}> 
   \sum_\sigma \int   \left.  \frac{ 
   w^\sigma_{\textit{c}  I }\left( \mathbf{r}  \right) 
 d \rho^\sigma \left( \mathbf{r} 
\right) 
}{ d V_{\textit{c} I}}  \right|_{\hat{v}^\textrm{DFT}_\sigma}
  \; d\mathbf{r} 
=  \left[ 
\frac{d\mathbf{N}}{d \mathbf{V}_{ \textit{c} } } 
\right]_{I I}^{ \hat{v}^\textrm{DFT}_\sigma}.
\end{align}
The identity $d^2 \mathbf{W} / d V_\textit{c}^2 = 
d \mathbf{N} / d \mathbf{V}_\textit{c}$ 
remains valid for a 
fixed $\hat{v}^\textrm{DFT}_\sigma$, since it relies only on
the orthonormality of Kohn-Sham eigenstates. Thus, 
the diagonal elements of 
$ d^2 \mathbf{W} / d V_\textit{c}^2 $
are individually negative, and 
the guarantee of strict local concavity  of  
 $W \left( \mathbf{V}_\textit{c} \right)$, 
 for any given  $\hat{v}_\sigma^\textrm{DFT}$, 
 is again conditional on a
 symmetric diagonally dominant~\cite{harville2011matrix}
 $ d^2 \mathbf{W} / d V_\textit{c}^2 $.
If the matrix does not exhibit a strictly negative
diagonal, conversely, then  $\left[ d\mathbf{N} / 
d \mathbf{V}_\textit{c}\right]_{I I} \ge 0$ for one or more 
values of $I$, and there exist $\mathbf{r}$ and $\sigma$
for which $w_\textit{c}^\sigma \left(\mathbf{r} \right) \ne 0$
and $0 \ge 
\left( d^2 E_\textrm{DFT} / d \rho^{\sigma 2} 
 \right)^{-1}
\left( \mathbf{r} , \mathbf{r} \right) $
at  fixed $\hat{v}_\sigma^\textrm{DFT}$.
Then, the non-self-consistent 
$\big( d^2 E_\textrm{DFT} / d \rho^{\sigma } 
d \rho^{\sigma' } \big)
\left( \mathbf{r} , \mathbf{r'}  \right) $ 
cannot be positive definite, and the state is unstable.
We conclude that for a minimum-energy eigenstate, 
given a fixed DFT potential and subject
to one or more linear, non-trivial constraints, to be stable
with respect to perturbations 
and hence locatable by SCF-type cDFT optimization,
it implies and requires
 the associated fixed-$\hat{v}_\sigma^\textrm{DFT}$
 Hessian matrix
$d^2 W / d \mathbf{V}_\textit{c}$ to have a strictly
 negative diagonal, 
and for it to be negative definite in the event that it
is diagonally dominant.
This generalizes the results of Refs.~\onlinecite{PhysRevA.72.024502,
doi:10.1021/ct0503163, doi10.1021/cr200148b} 
beyond the first-order non-degenerate perturbative regime, 
extending their validity  to systems which  exhibit 
non-linear response or orbital degeneracy.}

\section{Conclusion and Summary}
\label{Sec:Conclusion}

Constrained DFT is an flexible, potent approach that
broadens the scope and flexibility of DFT-based atomistic
simulation.
A growing number of software implementations of
cDFT are now appearing~\cite{doi10.1021/cr200148b,
Valiev20101477, QCHEM4, Oberhofer09, Oberhofer10jcp,
PhysRevB.88.165112, PhysRevB.91.195438, doi:10.1021/ct200570u,
PhysRevB.69.024415, PhysRevB.72.045121, WCMS:WCMS1159, PhysRevB.91.054420},
including linear-scaling implementations designed 
for application
to large systems~\cite{
doi:10.1021/ct100601n, :/content/aip/journal/jcp/142/23/10.1063/1.4922378}.
It is inherently
parallelizable, and thus potentially suitable for use in combination
with high-throughput materials screening
infrastructures~\cite{Jain20112295,curtarolo}.
Fundamental developments will be required to 
bring cDFT into the realm of such very routine use.
For transferability and comparability between codes,
for example, the \rd{standardized},  
automated selection of  population analysis and
targeting schemes, 
ideally but not necessarily
based on energy considerations,
would surely be beneficial. 
Methods based on 
promolecule densities~\cite{:/content/aip/journal/jcp/127/16/10.1063/1.2800022} 
or self-consistent 
Wannier functions~\cite{PhysRevB.82.081102}
are promising possibilities in this direction. 
\rd{
The key findings and conclusions of the present work are that: 
\begin{itemize}
\item For any self-consistent energy second-derivative with respect
to a constrained expectation value,
there is an equivalent integrated linear response or 
inverse-response function which is 
more convenient to calculate. This provides a 
basis for future quasi-Newton or 
preconditioning approaches for cDFT.
\item A negative-diagonal self-consistent  cDFT  Hessian is
implied by the local stability of the system 
with respect to perturbations about the point of evaluation.
A Hessian lacking this property cannot 
be observed by finite differences, and
so may be excluded in automation.
This generalizes W\&VV's result
to the self-consistent cDFT problem, the non-linear response
regime, and to  degenerate orbitals.
\item Concave regions in the 
DFT energy versus spin-density
landscape cannot be explored using cDFT.
\item  Integrated response  and dielectric functions 
may be evaluated as  by-products of cDFT optimization,
without  sums over empty Kohn-Sham eigenstates.
\item Existing cDFT optimization schemes based on the self-consistent
field approach, which update the cDFT Lagrange multipliers with
fixed DFT potentials, may be readily adapted for the self-consistent
gradients required in direct-minimization DFT.
\item cDFT does not change the domains of stability of the
underlying DFT energy landscape, it moves the solutions around,
or between, these domains. When there are two or more
such domains in cDFT, it is possible to observe multiple solutions, 
hysteresis, and energy discontinuities at such  transitions.
\end{itemize}
We expect that this work may  
facilitate the advanced automation of 
cDFT Lagrange multiplier optimization,
particularly in the high-throughput, 
molecular dynamics, and linear-scaling regimes.
Our general framework for treating energy landscapes
in terms of integrated response functions
now enables the extension of
cDFT to new areas of atomistic and continuum simulation.}

\section*{Acknowledgements}

This work was enabled by the
Royal Irish Academy -- Royal Society 
International Exchange Cost Share Programme (IE131505). 
GT acknowledges support from EPSRC UK 
(EP/I004483/1 and EP/K013610/1). 
\rd{We acknowledge and thank the ONETEP Developer's Group
for hosting workshops that facilitated this development, and
Qin Wu for a helpful communication and discussion.}

\appendix

\section{Approximate connection of
the cDFT non-interacting response function 
to unscreened sum-over-states
perturbation theory}
\label{App:summation}

\rd{In this Appendix, we investigate  how the  
non-interacting linear-response function defined by
$\chi_0 = d N / dV = d^2 W / d V d V_\textit{c}$ 
in Eq.~\ref{Eq:nonintercurv}
relates to the 
unscreened sum-over-states perturbation theory expression
given by Eq.~7 of Ref.~\onlinecite{PhysRevA.72.024502}.
The latter is obtained by replacing
$\left( \varepsilon^{-1} w_\textit{c} \right)^{\sigma}$ by 
$ w_\textit{c}^\sigma$ in  Eq.~\ref{Eq:bare}.
It yields the linear response of the targeted 
density $N$ to a change in 
the subspace-averaged Kohn-Sham potential $V$ with all other
aspects of the Kohn-Sham potential kept fixed.} Its 
evaluation for two values of $V_\textit{c}$, 
 close to those at which $E_\textit{c}$ and $W$ 
 assume  maxima, 
is shown in Fig.~\ref{fig4} (solid circles).
 The proximity of these data points to 
 $d^2 E_\textit{c} / d V d V_\textit{c}$ 
 is  \rd{perhaps not entirely coincidental, but} they sit  
rather far from their most closely 
related quantity, $d N / dV$.
The difference between the sum-over-states expression
and $d N / dV$ is subtle, since both are non-interacting 
response functions with the same integration.
It arises due to complex \rd{spatial} fluctuations in 
$v_\sigma^{\textrm{KS}} \left( \mathbf{r} \right)$ both within
and without the weighted region in the case of $d N / dV$,
\rd{in contrast to the effective $\delta v_\sigma^{\textrm{KS}} \left( \mathbf{r} \right) 
= w^\sigma_\textit{c}  \left( \mathbf{r} \right) \delta V $
implied in the sum-over-states expression. 
More precisely, $d N / d V$ is the complete
non-interacting linear-response function,
whereas the sum-over-states is
the non-interacting linear-response function truncated at
first-order in perturbation theory.
}

Referring  to Eq.~\ref{Eq:bare}, 
the $d N / dV$ calculated using \rd{self-consistent cDFT simplifies to
the perturbation theory} expression
 in the case that both
 $d v_\sigma^{\textrm{KS}} \left( \mathbf{r} \right) / d V^\sigma 
 = w^\sigma_\textit{c}  \left( \mathbf{r} \right) $ and
 the change in the average weighted 
 Kohn-Sham potential $V^\sigma$ is 
the wholly responsible for the 
change in subspace occupancy $N$, 
with all other degrees of freedom \rd{fixed}.
The first condition and Eq.~\ref{Eq:averagepot} together imply that \begin{align}
 1 \equiv{}&  \int \left( \frac{d V^\sigma}{d v_\sigma^{\textrm{KS}} \left( \mathbf{r} \right)}
\right)
 \left( \frac{d v_\sigma^{\textrm{KS}} \left( \mathbf{r} \right)}{d V^\sigma} \right)
  d\mathbf{r} \nonumber \\
 ={}& \left( \int   w_\textit{c}^\sigma \left( \mathbf{r'} \right)
 \; d\mathbf{r'} \right)^{-1}
  \int  \left( w^\sigma_\textit{c}  \left( \mathbf{r} \right) \right)^2
 \; d\mathbf{r},
 \end{align}
 which would be readily satisfied only if 
 $w_\textit{c}^\sigma \left( \mathbf{r} \right)$ were an
abrupt three dimensional unit step function.
 
 The second condition is  more unrealistic, since 
we may directly vary the Lagrange
 multiplier $V_\textit{c}$  when constraining DFT, 
 but not, at least directly, the average
 subspace Kohn-Sham potential $V^\sigma$. 
As $V_\textit{c}$ is varied, self-consistent changes in 
 $v_\sigma^{\textrm{KS}} \left( \mathbf{r} \right)$ 
 outside of the weighted region may substantially 
 mitigate the charge transfer due to changes within it.
 Such  effects are absent \rd{at first order in perturbation theory, 
 and so it} may  be expected to typically 
 overestimate the magnitude of
 the \rd{non-interacting} response $dN / dV$,
a situation \rd{which is} exemplified in Fig.~\ref{fig4}.
  
 \rd{ Since the differences in definition between the  
  non-interacting linear and sum-over-states  
  response functions are related, albeit not entirely,  to effects
  outside of the weighted region, 
  they cannot be reconciled.}
  An approximate reconciling 
  renormalization may 
  be made, however, 
   by  down-scaling  one of the two 
  functions \rd{$w^\sigma_\textit{c} \left( \mathbf{r} \right)$} in 
  the unscreened analogue of Eq.~\ref{Eq:bare} 
  (one applies the perturbation, 
   the other measures the charge) by a factor of 
  \rd{$\sum_\sigma \left( \int w^\sigma_\textit{c}  \left( \mathbf{r} \right) \; d\mathbf{r} \right)
  / \left( \int \; d\mathbf{r'} \right) $}.
  This scaling mimics the compensating background changes
  in the Kohn-Sham potential in the realistic cDFT calculation, 
   by  reducing or ``redistributing'' the  change in 
   potential \rd{$ w_\textit{c}^\sigma \left( \mathbf{r} \right) \delta V^\sigma$} 
   in proportion to the  system volume.
  The result of this renormalization, 
  suitably adapted for orbital-based population analysis
  \rd{(based on  orbital count, 
  with the advantage that the rescaling is not extensive with respect to 
   the volume of the vacuum region)},
  is shown in open black circles in Fig.~\ref{fig4}.
  We find that the unscreened sum-over-states 
  expression is thus brought into fair, but not exact, 
  agreement with the  
  \rd{non-perturbative}  non-interacting response $d N / d V $.
  A further downscaling by a factor of
  the integrated inverse microscopic dielectric function, given
  by Eq.~\ref{Eq:mydielectric}, 
  is then required to approximately
  \rd{match} the unscreened sum-over-states 
  perturbation theory  \rd{and  self-consistent} 
  cDFT total-energy curvatures.

\bibliography{latest-ok-gibo-cDFT_at_T_PRL.bib}

\begin{thebibliography}{81}%
\makeatletter
\providecommand \@ifxundefined [1]{%
 \@ifx{#1\undefined}
}%
\providecommand \@ifnum [1]{%
 \ifnum #1\expandafter \@firstoftwo
 \else \expandafter \@secondoftwo
 \fi
}%
\providecommand \@ifx [1]{%
 \ifx #1\expandafter \@firstoftwo
 \else \expandafter \@secondoftwo
 \fi
}%
\providecommand \natexlab [1]{#1}%
\providecommand \enquote  [1]{``#1''}%
\providecommand \bibnamefont  [1]{#1}%
\providecommand \bibfnamefont [1]{#1}%
\providecommand \citenamefont [1]{#1}%
\providecommand \href@noop [0]{\@secondoftwo}%
\providecommand \href [0]{\begingroup \@sanitize@url \@href}%
\providecommand \@href[1]{\@@startlink{#1}\@@href}%
\providecommand \@@href[1]{\endgroup#1\@@endlink}%
\providecommand \@sanitize@url [0]{\catcode `\\12\catcode `\$12\catcode
  `\&12\catcode `\#12\catcode `\^12\catcode `\_12\catcode `\%12\relax}%
\providecommand \@@startlink[1]{}%
\providecommand \@@endlink[0]{}%
\providecommand \url  [0]{\begingroup\@sanitize@url \@url }%
\providecommand \@url [1]{\endgroup\@href {#1}{\urlprefix }}%
\providecommand \urlprefix  [0]{URL }%
\providecommand \Eprint [0]{\href }%
\providecommand \doibase [0]{http://dx.doi.org/}%
\providecommand \selectlanguage [0]{\@gobble}%
\providecommand \bibinfo  [0]{\@secondoftwo}%
\providecommand \bibfield  [0]{\@secondoftwo}%
\providecommand \translation [1]{[#1]}%
\providecommand \BibitemOpen [0]{}%
\providecommand \bibitemStop [0]{}%
\providecommand \bibitemNoStop [0]{.\EOS\space}%
\providecommand \EOS [0]{\spacefactor3000\relax}%
\providecommand \BibitemShut  [1]{\csname bibitem#1\endcsname}%
\let\auto@bib@innerbib\@empty
\bibitem [{\citenamefont {Kaduk}\ \emph {et~al.}(2012)\citenamefont {Kaduk},
  \citenamefont {Kowalczyk},\ and\ \citenamefont
  {Van~Voorhis}}]{doi10.1021/cr200148b}%
  \BibitemOpen
  \bibfield  {author} {\bibinfo {author} {\bibfnamefont {B.}~\bibnamefont
  {Kaduk}}, \bibinfo {author} {\bibfnamefont {T.}~\bibnamefont {Kowalczyk}}, \
  and\ \bibinfo {author} {\bibfnamefont {T.}~\bibnamefont {Van~Voorhis}},\
  }\href {\doibase 10.1021/cr200148b} {\bibfield  {journal} {\bibinfo
  {journal} {Chemical Reviews}\ }\textbf {\bibinfo {volume} {112}},\ \bibinfo
  {pages} {321} (\bibinfo {year} {2012})}\BibitemShut {NoStop}%
\bibitem [{\citenamefont {Hohenberg}\ and\ \citenamefont
  {Kohn}(1964)}]{PhysRev.136.B864}%
  \BibitemOpen
  \bibfield  {author} {\bibinfo {author} {\bibfnamefont {P.}~\bibnamefont
  {Hohenberg}}\ and\ \bibinfo {author} {\bibfnamefont {W.}~\bibnamefont
  {Kohn}},\ }\href {\doibase 10.1103/PhysRev.136.B864} {\bibfield  {journal}
  {\bibinfo  {journal} {Phys. Rev.}\ }\textbf {\bibinfo {volume} {136}},\
  \bibinfo {pages} {B864} (\bibinfo {year} {1964})}\BibitemShut {NoStop}%
\bibitem [{\citenamefont {Kohn}\ and\ \citenamefont
  {Sham}(1965)}]{PhysRev.140.A1133}%
  \BibitemOpen
  \bibfield  {author} {\bibinfo {author} {\bibfnamefont {W.}~\bibnamefont
  {Kohn}}\ and\ \bibinfo {author} {\bibfnamefont {L.~J.}\ \bibnamefont
  {Sham}},\ }\href {\doibase 10.1103/PhysRev.140.A1133} {\bibfield  {journal}
  {\bibinfo  {journal} {Phys. Rev.}\ }\textbf {\bibinfo {volume} {140}},\
  \bibinfo {pages} {A1133} (\bibinfo {year} {1965})}\BibitemShut {NoStop}%
\bibitem [{\citenamefont {Wu}\ and\ \citenamefont
  {Van~Voorhis}(2006{\natexlab{a}})}]{doi:10.1021/ct0503163}%
  \BibitemOpen
  \bibfield  {author} {\bibinfo {author} {\bibfnamefont {Q.}~\bibnamefont
  {Wu}}\ and\ \bibinfo {author} {\bibfnamefont {T.}~\bibnamefont
  {Van~Voorhis}},\ }\href {\doibase 10.1021/ct0503163} {\bibfield  {journal}
  {\bibinfo  {journal} {Journal of Chemical Theory and Computation}\ }\textbf
  {\bibinfo {volume} {2}},\ \bibinfo {pages} {765} (\bibinfo {year}
  {2006}{\natexlab{a}})}\BibitemShut {NoStop}%
\bibitem [{\citenamefont {Wu}\ and\ \citenamefont
  {Van~Voorhis}(2005)}]{PhysRevA.72.024502}%
  \BibitemOpen
  \bibfield  {author} {\bibinfo {author} {\bibfnamefont {Q.}~\bibnamefont
  {Wu}}\ and\ \bibinfo {author} {\bibfnamefont {T.}~\bibnamefont
  {Van~Voorhis}},\ }\href {\doibase 10.1103/PhysRevA.72.024502} {\bibfield
  {journal} {\bibinfo  {journal} {Phys. Rev. A}\ }\textbf {\bibinfo {volume}
  {72}},\ \bibinfo {pages} {024502} (\bibinfo {year} {2005})}\BibitemShut
  {NoStop}%
\bibitem [{\citenamefont {Segal}\ \emph {et~al.}(2007)\citenamefont {Segal},
  \citenamefont {Singh}, \citenamefont {Rivoire}, \citenamefont {Difley},
  \citenamefont {Van~Voorhis},\ and\ \citenamefont
  {Baldo}}]{doi:10.1038/nmat1885}%
  \BibitemOpen
  \bibfield  {author} {\bibinfo {author} {\bibfnamefont {M.}~\bibnamefont
  {Segal}}, \bibinfo {author} {\bibfnamefont {M.}~\bibnamefont {Singh}},
  \bibinfo {author} {\bibfnamefont {K.}~\bibnamefont {Rivoire}}, \bibinfo
  {author} {\bibfnamefont {S.}~\bibnamefont {Difley}}, \bibinfo {author}
  {\bibfnamefont {T.}~\bibnamefont {Van~Voorhis}}, \ and\ \bibinfo {author}
  {\bibfnamefont {M.~A.}\ \bibnamefont {Baldo}},\ }\href@noop {} {\bibfield
  {journal} {\bibinfo  {journal} {Nature Materials}\ }\textbf {\bibinfo
  {volume} {6}},\ \bibinfo {pages} {374} (\bibinfo {year} {2007})}\BibitemShut
  {NoStop}%
\bibitem [{\citenamefont {Kowalczyk}\ \emph {et~al.}(2010)\citenamefont
  {Kowalczyk}, \citenamefont {Lin},\ and\ \citenamefont
  {Voorhis}}]{doi:10.1021/jp103153a}%
  \BibitemOpen
  \bibfield  {author} {\bibinfo {author} {\bibfnamefont {T.}~\bibnamefont
  {Kowalczyk}}, \bibinfo {author} {\bibfnamefont {Z.}~\bibnamefont {Lin}}, \
  and\ \bibinfo {author} {\bibfnamefont {T.~V.}\ \bibnamefont {Voorhis}},\
  }\href {\doibase 10.1021/jp103153a} {\bibfield  {journal} {\bibinfo
  {journal} {The Journal of Physical Chemistry A}\ }\textbf {\bibinfo {volume}
  {114}},\ \bibinfo {pages} {10427} (\bibinfo {year} {2010})}\BibitemShut
  {NoStop}%
\bibitem [{\citenamefont {Dederichs}\ \emph {et~al.}(1984)\citenamefont
  {Dederichs}, \citenamefont {Bl\"ugel}, \citenamefont {Zeller},\ and\
  \citenamefont {Akai}}]{PhysRevLett.53.2512}%
  \BibitemOpen
  \bibfield  {author} {\bibinfo {author} {\bibfnamefont {P.~H.}\ \bibnamefont
  {Dederichs}}, \bibinfo {author} {\bibfnamefont {S.}~\bibnamefont {Bl\"ugel}},
  \bibinfo {author} {\bibfnamefont {R.}~\bibnamefont {Zeller}}, \ and\ \bibinfo
  {author} {\bibfnamefont {H.}~\bibnamefont {Akai}},\ }\href {\doibase
  10.1103/PhysRevLett.53.2512} {\bibfield  {journal} {\bibinfo  {journal}
  {Phys. Rev. Lett.}\ }\textbf {\bibinfo {volume} {53}},\ \bibinfo {pages}
  {2512} (\bibinfo {year} {1984})}\BibitemShut {NoStop}%
\bibitem [{\citenamefont {Wu}\ and\ \citenamefont
  {Van~Voorhis}(2006{\natexlab{b}})}]{doi:10.1021/jp061848y}%
  \BibitemOpen
  \bibfield  {author} {\bibinfo {author} {\bibfnamefont {Q.}~\bibnamefont
  {Wu}}\ and\ \bibinfo {author} {\bibfnamefont {T.}~\bibnamefont
  {Van~Voorhis}},\ }\href {\doibase 10.1021/jp061848y} {\bibfield  {journal}
  {\bibinfo  {journal} {The Journal of Physical Chemistry A}\ }\textbf
  {\bibinfo {volume} {110}},\ \bibinfo {pages} {9212} (\bibinfo {year}
  {2006}{\natexlab{b}})}\BibitemShut {NoStop}%
\bibitem [{\citenamefont {Oberhofer}\ and\ \citenamefont
  {Blumberger}(2009{\natexlab{a}})}]{:/content/aip/journal/jcp/131/6/10.1063/1.3190169}%
  \BibitemOpen
  \bibfield  {author} {\bibinfo {author} {\bibfnamefont {H.}~\bibnamefont
  {Oberhofer}}\ and\ \bibinfo {author} {\bibfnamefont {J.}~\bibnamefont
  {Blumberger}},\ }\href@noop {} {\bibfield  {journal} {\bibinfo  {journal}
  {The Journal of Chemical Physics}\ }\textbf {\bibinfo {volume} {131}},\
  \bibinfo {eid} {064101} (\bibinfo {year} {2009}{\natexlab{a}})}\BibitemShut
  {NoStop}%
\bibitem [{\citenamefont {Sit}\ \emph {et~al.}(2006)\citenamefont {Sit},
  \citenamefont {Cococcioni},\ and\ \citenamefont
  {Marzari}}]{PhysRevLett.97.028303}%
  \BibitemOpen
  \bibfield  {author} {\bibinfo {author} {\bibfnamefont {P.~H.-L.}\
  \bibnamefont {Sit}}, \bibinfo {author} {\bibfnamefont {M.}~\bibnamefont
  {Cococcioni}}, \ and\ \bibinfo {author} {\bibfnamefont {N.}~\bibnamefont
  {Marzari}},\ }\href@noop {} {\bibfield  {journal} {\bibinfo  {journal} {Phys.
  Rev. Lett.}\ }\textbf {\bibinfo {volume} {97}},\ \bibinfo {pages} {028303}
  (\bibinfo {year} {2006})}\BibitemShut {NoStop}%
\bibitem [{\citenamefont {Tim~Kowalczyk}\ \emph {et~al.}(2011)\citenamefont
  {Tim~Kowalczyk}, \citenamefont {Wang},\ and\ \citenamefont
  {Van~Voorhis}}]{doi:10.1021/jp204962k}%
  \BibitemOpen
  \bibfield  {author} {\bibinfo {author} {\bibfnamefont {T.}~\bibnamefont
  {Tim~Kowalczyk}}, \bibinfo {author} {\bibfnamefont {L.-P.}\ \bibnamefont
  {Wang}}, \ and\ \bibinfo {author} {\bibfnamefont {T.}~\bibnamefont
  {Van~Voorhis}},\ }\href {\doibase 10.1021/jp204962k} {\bibfield  {journal}
  {\bibinfo  {journal} {The Journal of Physical Chemistry B}\ }\textbf
  {\bibinfo {volume} {115}},\ \bibinfo {pages} {12135} (\bibinfo {year}
  {2011})}\BibitemShut {NoStop}%
\bibitem [{\citenamefont {Behler}\ \emph {et~al.}(2008)\citenamefont {Behler},
  \citenamefont {Reuter},\ and\ \citenamefont
  {Scheffler}}]{PhysRevB.77.115421}%
  \BibitemOpen
  \bibfield  {author} {\bibinfo {author} {\bibfnamefont {J.}~\bibnamefont
  {Behler}}, \bibinfo {author} {\bibfnamefont {K.}~\bibnamefont {Reuter}}, \
  and\ \bibinfo {author} {\bibfnamefont {M.}~\bibnamefont {Scheffler}},\ }\href
  {\doibase 10.1103/PhysRevB.77.115421} {\bibfield  {journal} {\bibinfo
  {journal} {Phys. Rev. B}\ }\textbf {\bibinfo {volume} {77}},\ \bibinfo
  {pages} {115421} (\bibinfo {year} {2008})}\BibitemShut {NoStop}%
\bibitem [{\citenamefont {Rudra}\ \emph {et~al.}(2006)\citenamefont {Rudra},
  \citenamefont {Wu},\ and\ \citenamefont
  {Van~Voorhis}}]{:/content/aip/journal/jcp/124/2/10.1063/1.2145878}%
  \BibitemOpen
  \bibfield  {author} {\bibinfo {author} {\bibfnamefont {I.}~\bibnamefont
  {Rudra}}, \bibinfo {author} {\bibfnamefont {Q.}~\bibnamefont {Wu}}, \ and\
  \bibinfo {author} {\bibfnamefont {T.}~\bibnamefont {Van~Voorhis}},\
  }\href@noop {} {\bibfield  {journal} {\bibinfo  {journal} {The Journal of
  Chemical Physics}\ }\textbf {\bibinfo {volume} {124}},\ \bibinfo {eid}
  {024103} (\bibinfo {year} {2006})}\BibitemShut {NoStop}%
\bibitem [{\citenamefont {Rudra}\ \emph {et~al.}(2007)\citenamefont {Rudra},
  \citenamefont {Wu},\ and\ \citenamefont
  {Van~Voorhis}}]{doi:10.1021/ic700871f}%
  \BibitemOpen
  \bibfield  {author} {\bibinfo {author} {\bibfnamefont {I.}~\bibnamefont
  {Rudra}}, \bibinfo {author} {\bibfnamefont {Q.}~\bibnamefont {Wu}}, \ and\
  \bibinfo {author} {\bibfnamefont {T.}~\bibnamefont {Van~Voorhis}},\ }\href
  {\doibase 10.1021/ic700871f} {\bibfield  {journal} {\bibinfo  {journal}
  {Inorganic Chemistry}\ }\textbf {\bibinfo {volume} {46}},\ \bibinfo {pages}
  {10539} (\bibinfo {year} {2007})}\BibitemShut {NoStop}%
\bibitem [{\citenamefont {Schmidt}\ \emph {et~al.}(2008)\citenamefont
  {Schmidt}, \citenamefont {Shenvi},\ and\ \citenamefont
  {Tully}}]{:/content/aip/journal/jcp/129/11/10.1063/1.2978168}%
  \BibitemOpen
  \bibfield  {author} {\bibinfo {author} {\bibfnamefont {J.~R.}\ \bibnamefont
  {Schmidt}}, \bibinfo {author} {\bibfnamefont {N.}~\bibnamefont {Shenvi}}, \
  and\ \bibinfo {author} {\bibfnamefont {J.~C.}\ \bibnamefont {Tully}},\
  }\href@noop {} {\bibfield  {journal} {\bibinfo  {journal} {The Journal of
  Chemical Physics}\ }\textbf {\bibinfo {volume} {129}},\ \bibinfo {eid}
  {114110} (\bibinfo {year} {2008})}\BibitemShut {NoStop}%
\bibitem [{\citenamefont {Difley}\ \emph {et~al.}(2008)\citenamefont {Difley},
  \citenamefont {Beljonne},\ and\ \citenamefont {Troy
  Van~Voorhis}}]{doi:10.1021/ja076125m}%
  \BibitemOpen
  \bibfield  {author} {\bibinfo {author} {\bibfnamefont {S.}~\bibnamefont
  {Difley}}, \bibinfo {author} {\bibfnamefont {D.}~\bibnamefont {Beljonne}}, \
  and\ \bibinfo {author} {\bibfnamefont {T.}~\bibnamefont {Troy Van~Voorhis}},\
  }\href {\doibase 10.1021/ja076125m} {\bibfield  {journal} {\bibinfo
  {journal} {Journal of the American Chemical Society}\ }\textbf {\bibinfo
  {volume} {130}},\ \bibinfo {pages} {3420} (\bibinfo {year}
  {2008})}\BibitemShut {NoStop}%
\bibitem [{\citenamefont {Sena}\ \emph {et~al.}(2011)\citenamefont {Sena},
  \citenamefont {Miyazaki},\ and\ \citenamefont
  {Bowler}}]{doi:10.1021/ct100601n}%
  \BibitemOpen
  \bibfield  {author} {\bibinfo {author} {\bibfnamefont {A.~M.~P.}\
  \bibnamefont {Sena}}, \bibinfo {author} {\bibfnamefont {T.}~\bibnamefont
  {Miyazaki}}, \ and\ \bibinfo {author} {\bibfnamefont {D.~R.}\ \bibnamefont
  {Bowler}},\ }\href {\doibase 10.1021/ct100601n} {\bibfield  {journal}
  {\bibinfo  {journal} {Journal of Chemical Theory and Computation}\ }\textbf
  {\bibinfo {volume} {7}},\ \bibinfo {pages} {884} (\bibinfo {year}
  {2011})}\BibitemShut {NoStop}%
\bibitem [{\citenamefont {Turban}\ \emph {et~al.}(2016)\citenamefont {Turban},
  \citenamefont {Teobaldi}, \citenamefont {O'Regan},\ and\ \citenamefont
  {Hine}}]{PhysRevB.93.165102}%
  \BibitemOpen
  \bibfield  {author} {\bibinfo {author} {\bibfnamefont {D.~H.~P.}\
  \bibnamefont {Turban}}, \bibinfo {author} {\bibfnamefont {G.}~\bibnamefont
  {Teobaldi}}, \bibinfo {author} {\bibfnamefont {D.~D.}\ \bibnamefont
  {O'Regan}}, \ and\ \bibinfo {author} {\bibfnamefont {N.~D.~M.}\ \bibnamefont
  {Hine}},\ }\href@noop {} {\bibfield  {journal} {\bibinfo  {journal} {Phys.
  Rev. B}\ }\textbf {\bibinfo {volume} {93}},\ \bibinfo {pages} {165102}
  (\bibinfo {year} {2016})}\BibitemShut {NoStop}%
\bibitem [{\citenamefont {Runge}\ and\ \citenamefont
  {Gross}(1984)}]{PhysRevLett.52.997}%
  \BibitemOpen
  \bibfield  {author} {\bibinfo {author} {\bibfnamefont {E.}~\bibnamefont
  {Runge}}\ and\ \bibinfo {author} {\bibfnamefont {E.~K.~U.}\ \bibnamefont
  {Gross}},\ }\href@noop {} {\bibfield  {journal} {\bibinfo  {journal} {Phys.
  Rev. Lett.}\ }\textbf {\bibinfo {volume} {52}},\ \bibinfo {pages} {997}
  (\bibinfo {year} {1984})}\BibitemShut {NoStop}%
\bibitem [{\citenamefont {Difley}\ and\ \citenamefont
  {Van~Voorhis}(2011)}]{doi:10.1021/ct100508y}%
  \BibitemOpen
  \bibfield  {author} {\bibinfo {author} {\bibfnamefont {S.}~\bibnamefont
  {Difley}}\ and\ \bibinfo {author} {\bibfnamefont {T.}~\bibnamefont
  {Van~Voorhis}},\ }\href {\doibase 10.1021/ct100508y} {\bibfield  {journal}
  {\bibinfo  {journal} {Journal of Chemical Theory and Computation}\ }\textbf
  {\bibinfo {volume} {7}},\ \bibinfo {pages} {594} (\bibinfo {year}
  {2011})}\BibitemShut {NoStop}%
\bibitem [{\citenamefont {Nakamura}\ \emph {et~al.}(2008)\citenamefont
  {Nakamura}, \citenamefont {Yoshimoto}, \citenamefont {Arita}, \citenamefont
  {Tsuneyuki},\ and\ \citenamefont {Imada}}]{PhysRevB.77.195126}%
  \BibitemOpen
  \bibfield  {author} {\bibinfo {author} {\bibfnamefont {K.}~\bibnamefont
  {Nakamura}}, \bibinfo {author} {\bibfnamefont {Y.}~\bibnamefont {Yoshimoto}},
  \bibinfo {author} {\bibfnamefont {R.}~\bibnamefont {Arita}}, \bibinfo
  {author} {\bibfnamefont {S.}~\bibnamefont {Tsuneyuki}}, \ and\ \bibinfo
  {author} {\bibfnamefont {M.}~\bibnamefont {Imada}},\ }\href {\doibase
  10.1103/PhysRevB.77.195126} {\bibfield  {journal} {\bibinfo  {journal} {Phys.
  Rev. B}\ }\textbf {\bibinfo {volume} {77}},\ \bibinfo {pages} {195126}
  (\bibinfo {year} {2008})}\BibitemShut {NoStop}%
\bibitem [{\citenamefont {Roychoudhury}\ \emph {et~al.}(2016)\citenamefont
  {Roychoudhury}, \citenamefont {Motta},\ and\ \citenamefont
  {Sanvito}}]{PhysRevB.93.045130}%
  \BibitemOpen
  \bibfield  {author} {\bibinfo {author} {\bibfnamefont {S.}~\bibnamefont
  {Roychoudhury}}, \bibinfo {author} {\bibfnamefont {C.}~\bibnamefont {Motta}},
  \ and\ \bibinfo {author} {\bibfnamefont {S.}~\bibnamefont {Sanvito}},\
  }\href@noop {} {\bibfield  {journal} {\bibinfo  {journal} {Phys. Rev. B}\
  }\textbf {\bibinfo {volume} {93}},\ \bibinfo {pages} {045130} (\bibinfo
  {year} {2016})}\BibitemShut {NoStop}%
\bibitem [{\citenamefont {Souza}\ \emph {et~al.}(2013)\citenamefont {Souza},
  \citenamefont {Rungger}, \citenamefont {Pemmaraju}, \citenamefont
  {Schwingenschloegl},\ and\ \citenamefont {Sanvito}}]{PhysRevB.88.165112}%
  \BibitemOpen
  \bibfield  {author} {\bibinfo {author} {\bibfnamefont {A.~M.}\ \bibnamefont
  {Souza}}, \bibinfo {author} {\bibfnamefont {I.}~\bibnamefont {Rungger}},
  \bibinfo {author} {\bibfnamefont {C.~D.}\ \bibnamefont {Pemmaraju}}, \bibinfo
  {author} {\bibfnamefont {U.}~\bibnamefont {Schwingenschloegl}}, \ and\
  \bibinfo {author} {\bibfnamefont {S.}~\bibnamefont {Sanvito}},\ }\href@noop
  {} {\bibfield  {journal} {\bibinfo  {journal} {Phys. Rev. B}\ }\textbf
  {\bibinfo {volume} {88}},\ \bibinfo {pages} {165112} (\bibinfo {year}
  {2013})}\BibitemShut {NoStop}%
\bibitem [{\citenamefont {Sau}\ \emph {et~al.}(2008)\citenamefont {Sau},
  \citenamefont {Neaton}, \citenamefont {Choi}, \citenamefont {Louie},\ and\
  \citenamefont {Cohen}}]{PhysRevLett.101.026804}%
  \BibitemOpen
  \bibfield  {author} {\bibinfo {author} {\bibfnamefont {J.~D.}\ \bibnamefont
  {Sau}}, \bibinfo {author} {\bibfnamefont {J.~B.}\ \bibnamefont {Neaton}},
  \bibinfo {author} {\bibfnamefont {H.~J.}\ \bibnamefont {Choi}}, \bibinfo
  {author} {\bibfnamefont {S.~G.}\ \bibnamefont {Louie}}, \ and\ \bibinfo
  {author} {\bibfnamefont {M.~L.}\ \bibnamefont {Cohen}},\ }\href {\doibase
  10.1103/PhysRevLett.101.026804} {\bibfield  {journal} {\bibinfo  {journal}
  {Phys. Rev. Lett.}\ }\textbf {\bibinfo {volume} {101}},\ \bibinfo {pages}
  {026804} (\bibinfo {year} {2008})}\BibitemShut {NoStop}%
\bibitem [{\citenamefont {Brooke}\ \emph {et~al.}(2015)\citenamefont {Brooke},
  \citenamefont {Vezzoli}, \citenamefont {Higgins}, \citenamefont {Zotti},
  \citenamefont {Palacios},\ and\ \citenamefont
  {Nichols}}]{PhysRevB.91.195438}%
  \BibitemOpen
  \bibfield  {author} {\bibinfo {author} {\bibfnamefont {C.}~\bibnamefont
  {Brooke}}, \bibinfo {author} {\bibfnamefont {A.}~\bibnamefont {Vezzoli}},
  \bibinfo {author} {\bibfnamefont {S.~J.}\ \bibnamefont {Higgins}}, \bibinfo
  {author} {\bibfnamefont {L.~A.}\ \bibnamefont {Zotti}}, \bibinfo {author}
  {\bibfnamefont {J.~J.}\ \bibnamefont {Palacios}}, \ and\ \bibinfo {author}
  {\bibfnamefont {R.~J.}\ \bibnamefont {Nichols}},\ }\href {\doibase
  10.1103/PhysRevB.91.195438} {\bibfield  {journal} {\bibinfo  {journal} {Phys.
  Rev. B}\ }\textbf {\bibinfo {volume} {91}},\ \bibinfo {pages} {195438}
  (\bibinfo {year} {2015})}\BibitemShut {NoStop}%
\bibitem [{\citenamefont {Wu}\ and\ \citenamefont
  {Van~Voorhis}(2006{\natexlab{c}})}]{:/content/aip/journal/jcp/125/16/10.1063/1.2360263}%
  \BibitemOpen
  \bibfield  {author} {\bibinfo {author} {\bibfnamefont {Q.}~\bibnamefont
  {Wu}}\ and\ \bibinfo {author} {\bibfnamefont {T.}~\bibnamefont
  {Van~Voorhis}},\ }\href@noop {} {\bibfield  {journal} {\bibinfo  {journal}
  {The Journal of Chemical Physics}\ }\textbf {\bibinfo {volume} {125}},\
  \bibinfo {eid} {164105} (\bibinfo {year} {2006}{\natexlab{c}})}\BibitemShut
  {NoStop}%
\bibitem [{\citenamefont {Ding}\ \emph {et~al.}(2010)\citenamefont {Ding},
  \citenamefont {Wang}, \citenamefont {Wu}, \citenamefont {Van~Voorhis},
  \citenamefont {Chen},\ and\ \citenamefont
  {Konopelski}}]{doi:10.1021/jp912049p}%
  \BibitemOpen
  \bibfield  {author} {\bibinfo {author} {\bibfnamefont {F.}~\bibnamefont
  {Ding}}, \bibinfo {author} {\bibfnamefont {H.}~\bibnamefont {Wang}}, \bibinfo
  {author} {\bibfnamefont {Q.}~\bibnamefont {Wu}}, \bibinfo {author}
  {\bibfnamefont {T.}~\bibnamefont {Van~Voorhis}}, \bibinfo {author}
  {\bibfnamefont {S.}~\bibnamefont {Chen}}, \ and\ \bibinfo {author}
  {\bibfnamefont {J.~P.}\ \bibnamefont {Konopelski}},\ }\href {\doibase
  10.1021/jp912049p} {\bibfield  {journal} {\bibinfo  {journal} {The Journal of
  Physical Chemistry A}\ }\textbf {\bibinfo {volume} {114}},\ \bibinfo {pages}
  {6039} (\bibinfo {year} {2010})}\BibitemShut {NoStop}%
\bibitem [{\citenamefont {Oberhofer}\ and\ \citenamefont
  {Blumberger}(2009{\natexlab{b}})}]{Oberhofer09}%
  \BibitemOpen
  \bibfield  {author} {\bibinfo {author} {\bibfnamefont {H.}~\bibnamefont
  {Oberhofer}}\ and\ \bibinfo {author} {\bibfnamefont {J.}~\bibnamefont
  {Blumberger}},\ }\href@noop {} {\bibfield  {journal} {\bibinfo  {journal}
  {{J. Chem. Phys.}}\ }\textbf {\bibinfo {volume} {131}},\ \bibinfo {pages}
  {064101} (\bibinfo {year} {2009}{\natexlab{b}})}\BibitemShut {NoStop}%
\bibitem [{\citenamefont {Oberhofer}\ and\ \citenamefont
  {Blumberger}(2010{\natexlab{a}})}]{Oberhofer10jcp}%
  \BibitemOpen
  \bibfield  {author} {\bibinfo {author} {\bibfnamefont {H.}~\bibnamefont
  {Oberhofer}}\ and\ \bibinfo {author} {\bibfnamefont {J.}~\bibnamefont
  {Blumberger}},\ }\href {\doibase 10.1063/1.3507878} {\bibfield  {journal}
  {\bibinfo  {journal} {{J. Chem. Phys.}}\ }\textbf {\bibinfo {volume} {133}},\
  \bibinfo {pages} {244105} (\bibinfo {year} {2010}{\natexlab{a}})}\BibitemShut
  {NoStop}%
\bibitem [{\citenamefont {Kubas}\ \emph {et~al.}(2014)\citenamefont {Kubas},
  \citenamefont {Hoffmann}, \citenamefont {Heck}, \citenamefont {Oberhofer},
  \citenamefont {Elstner},\ and\ \citenamefont {Blumberger}}]{Kubas14jcp}%
  \BibitemOpen
  \bibfield  {author} {\bibinfo {author} {\bibfnamefont {A.}~\bibnamefont
  {Kubas}}, \bibinfo {author} {\bibfnamefont {F.}~\bibnamefont {Hoffmann}},
  \bibinfo {author} {\bibfnamefont {A.}~\bibnamefont {Heck}}, \bibinfo {author}
  {\bibfnamefont {H.}~\bibnamefont {Oberhofer}}, \bibinfo {author}
  {\bibfnamefont {M.}~\bibnamefont {Elstner}}, \ and\ \bibinfo {author}
  {\bibfnamefont {J.}~\bibnamefont {Blumberger}},\ }\href {\doibase
  10.1063/1.4867077} {\bibfield  {journal} {\bibinfo  {journal} {{J. Chem.
  Phys.}}\ }\textbf {\bibinfo {volume} {140}},\ \bibinfo {pages} {104105}
  (\bibinfo {year} {2014})}\BibitemShut {NoStop}%
\bibitem [{\citenamefont {Kubas}\ \emph {et~al.}(2015)\citenamefont {Kubas},
  \citenamefont {Gajdos}, \citenamefont {Heck}, \citenamefont {Oberhofer},
  \citenamefont {Elstner},\ and\ \citenamefont {Blumberger}}]{Kubas15pccp}%
  \BibitemOpen
  \bibfield  {author} {\bibinfo {author} {\bibfnamefont {A.}~\bibnamefont
  {Kubas}}, \bibinfo {author} {\bibfnamefont {F.}~\bibnamefont {Gajdos}},
  \bibinfo {author} {\bibfnamefont {A.}~\bibnamefont {Heck}}, \bibinfo {author}
  {\bibfnamefont {H.}~\bibnamefont {Oberhofer}}, \bibinfo {author}
  {\bibfnamefont {M.}~\bibnamefont {Elstner}}, \ and\ \bibinfo {author}
  {\bibfnamefont {J.}~\bibnamefont {Blumberger}},\ }\href {\doibase
  10.1039/C4CP04749D} {\bibfield  {journal} {\bibinfo  {journal} {{Phys. Chem.
  Chem. Phys.}}\ }\textbf {\bibinfo {volume} {17}},\ \bibinfo {pages} {14342}
  (\bibinfo {year} {2015})}\BibitemShut {NoStop}%
\bibitem [{\citenamefont {Oberhofer}\ and\ \citenamefont
  {Blumberger}(2010{\natexlab{b}})}]{Oberhofer10acie}%
  \BibitemOpen
  \bibfield  {author} {\bibinfo {author} {\bibfnamefont {H.}~\bibnamefont
  {Oberhofer}}\ and\ \bibinfo {author} {\bibfnamefont {J.}~\bibnamefont
  {Blumberger}},\ }\href {\doibase 10.1002/anie.200906455} {\bibfield
  {journal} {\bibinfo  {journal} {{Angew. Chem. Int. Ed.}}\ }\textbf {\bibinfo
  {volume} {49}},\ \bibinfo {pages} {3631} (\bibinfo {year}
  {2010}{\natexlab{b}})}\BibitemShut {NoStop}%
\bibitem [{\citenamefont {Oberhofer}\ and\ \citenamefont
  {Blumberger}(2012)}]{Oberhofer12}%
  \BibitemOpen
  \bibfield  {author} {\bibinfo {author} {\bibfnamefont {H.}~\bibnamefont
  {Oberhofer}}\ and\ \bibinfo {author} {\bibfnamefont {J.}~\bibnamefont
  {Blumberger}},\ }\href {\doibase 10.1039/C2CP41348E} {\bibfield  {journal}
  {\bibinfo  {journal} {{Phys. Chem. Chem. Phys.}}\ }\textbf {\bibinfo {volume}
  {14}},\ \bibinfo {pages} {13846} (\bibinfo {year} {2012})}\BibitemShut
  {NoStop}%
\bibitem [{\citenamefont {McKenna}\ and\ \citenamefont
  {Blumberger}(2012)}]{McKenna12prb}%
  \BibitemOpen
  \bibfield  {author} {\bibinfo {author} {\bibfnamefont {K.~P.}\ \bibnamefont
  {McKenna}}\ and\ \bibinfo {author} {\bibfnamefont {J.}~\bibnamefont
  {Blumberger}},\ }\href {\doibase 10.1103/PhysRevB.86.245110} {\bibfield
  {journal} {\bibinfo  {journal} {{Phys. Rev. B}}\ }\textbf {\bibinfo {volume}
  {86}},\ \bibinfo {pages} {245110} (\bibinfo {year} {2012})}\BibitemShut
  {NoStop}%
\bibitem [{\citenamefont {Blumberger}\ and\ \citenamefont
  {McKenna}(2013)}]{Blumberger13}%
  \BibitemOpen
  \bibfield  {author} {\bibinfo {author} {\bibfnamefont {J.}~\bibnamefont
  {Blumberger}}\ and\ \bibinfo {author} {\bibfnamefont {K.~P.}\ \bibnamefont
  {McKenna}},\ }\href {\doibase 10.1039/C2CP42537H} {\bibfield  {journal}
  {\bibinfo  {journal} {{Phys. Chem. Chem. Phys.}}\ }\textbf {\bibinfo {volume}
  {15}},\ \bibinfo {pages} {2184} (\bibinfo {year} {2013})}\BibitemShut
  {NoStop}%
\bibitem [{\citenamefont {Yeganeh}\ and\ \citenamefont
  {Van~Voorhis}(2010)}]{doi:10.1021/jp106989t}%
  \BibitemOpen
  \bibfield  {author} {\bibinfo {author} {\bibfnamefont {S.}~\bibnamefont
  {Yeganeh}}\ and\ \bibinfo {author} {\bibfnamefont {T.}~\bibnamefont
  {Van~Voorhis}},\ }\href {\doibase 10.1021/jp106989t} {\bibfield  {journal}
  {\bibinfo  {journal} {The Journal of Physical Chemistry C}\ }\textbf
  {\bibinfo {volume} {114}},\ \bibinfo {pages} {20756} (\bibinfo {year}
  {2010})}\BibitemShut {NoStop}%
\bibitem [{\citenamefont {Yost}\ \emph {et~al.}(2014)\citenamefont {Yost},
  \citenamefont {Lee}, \citenamefont {Wilson}, \citenamefont {Wu},
  \citenamefont {McMahon}, \citenamefont {Parkhurst}, \citenamefont {Thompson},
  \citenamefont {Congreve}, \citenamefont {Rao}, \citenamefont {Johnson},
  \citenamefont {Sfeir}, \citenamefont {Bawendi}, \citenamefont {Swager},
  \citenamefont {Friend}, \citenamefont {Baldo},\ and\ \citenamefont
  {Van~Voorhis}}]{doi:10.1038/nchem.1945}%
  \BibitemOpen
  \bibfield  {author} {\bibinfo {author} {\bibfnamefont {S.~R.}\ \bibnamefont
  {Yost}}, \bibinfo {author} {\bibfnamefont {J.}~\bibnamefont {Lee}}, \bibinfo
  {author} {\bibfnamefont {M.~W.~B.}\ \bibnamefont {Wilson}}, \bibinfo {author}
  {\bibfnamefont {T.}~\bibnamefont {Wu}}, \bibinfo {author} {\bibfnamefont
  {D.~P.}\ \bibnamefont {McMahon}}, \bibinfo {author} {\bibfnamefont {R.~R.}\
  \bibnamefont {Parkhurst}}, \bibinfo {author} {\bibfnamefont {N.~J.}\
  \bibnamefont {Thompson}}, \bibinfo {author} {\bibfnamefont {D.~N.}\
  \bibnamefont {Congreve}}, \bibinfo {author} {\bibfnamefont {A.}~\bibnamefont
  {Rao}}, \bibinfo {author} {\bibfnamefont {K.}~\bibnamefont {Johnson}},
  \bibinfo {author} {\bibfnamefont {M.~Y.}\ \bibnamefont {Sfeir}}, \bibinfo
  {author} {\bibfnamefont {M.~G.}\ \bibnamefont {Bawendi}}, \bibinfo {author}
  {\bibfnamefont {T.~M.}\ \bibnamefont {Swager}}, \bibinfo {author}
  {\bibfnamefont {R.~H.}\ \bibnamefont {Friend}}, \bibinfo {author}
  {\bibfnamefont {M.~A.}\ \bibnamefont {Baldo}}, \ and\ \bibinfo {author}
  {\bibfnamefont {T.}~\bibnamefont {Van~Voorhis}},\ }\href
  {http://dx.doi.org/10.1038/nchem.1945} {\bibfield  {journal} {\bibinfo
  {journal} {Nat. Chem.}\ }\textbf {\bibinfo {volume} {6}},\ \bibinfo {pages}
  {492} (\bibinfo {year} {2014})}\BibitemShut {NoStop}%
\bibitem [{\citenamefont {Evans}\ \emph {et~al.}(2008)\citenamefont {Evans},
  \citenamefont {Cheng},\ and\ \citenamefont
  {Van~Voorhis}}]{PhysRevB.78.165108}%
  \BibitemOpen
  \bibfield  {author} {\bibinfo {author} {\bibfnamefont {J.~S.}\ \bibnamefont
  {Evans}}, \bibinfo {author} {\bibfnamefont {C.-L.}\ \bibnamefont {Cheng}}, \
  and\ \bibinfo {author} {\bibfnamefont {T.}~\bibnamefont {Van~Voorhis}},\
  }\href {\doibase 10.1103/PhysRevB.78.165108} {\bibfield  {journal} {\bibinfo
  {journal} {Phys. Rev. B}\ }\textbf {\bibinfo {volume} {78}},\ \bibinfo
  {pages} {165108} (\bibinfo {year} {2008})}\BibitemShut {NoStop}%
\bibitem [{\citenamefont {Anisimov}\ \emph {et~al.}(1991)\citenamefont
  {Anisimov}, \citenamefont {Zaanen},\ and\ \citenamefont
  {Andersen}}]{PhysRevB.44.943}%
  \BibitemOpen
  \bibfield  {author} {\bibinfo {author} {\bibfnamefont {V.~I.}\ \bibnamefont
  {Anisimov}}, \bibinfo {author} {\bibfnamefont {J.}~\bibnamefont {Zaanen}}, \
  and\ \bibinfo {author} {\bibfnamefont {O.~K.}\ \bibnamefont {Andersen}},\
  }\href {\doibase 10.1103/PhysRevB.44.943} {\bibfield  {journal} {\bibinfo
  {journal} {Phys. Rev. B}\ }\textbf {\bibinfo {volume} {44}},\ \bibinfo
  {pages} {943} (\bibinfo {year} {1991})}\BibitemShut {NoStop}%
\bibitem [{\citenamefont {Meider}\ and\ \citenamefont
  {Springborg}(1998)}]{0953-8984-10-31-012}%
  \BibitemOpen
  \bibfield  {author} {\bibinfo {author} {\bibfnamefont {H.}~\bibnamefont
  {Meider}}\ and\ \bibinfo {author} {\bibfnamefont {M.}~\bibnamefont
  {Springborg}},\ }\href {http://stacks.iop.org/0953-8984/10/i=31/a=012}
  {\bibfield  {journal} {\bibinfo  {journal} {Journal of Physics: Condensed
  Matter}\ }\textbf {\bibinfo {volume} {10}},\ \bibinfo {pages} {6953}
  (\bibinfo {year} {1998})}\BibitemShut {NoStop}%
\bibitem [{\citenamefont {Danielsen}(1986)}]{0022-3719-19-31-004}%
  \BibitemOpen
  \bibfield  {author} {\bibinfo {author} {\bibfnamefont {P.~L.}\ \bibnamefont
  {Danielsen}},\ }\href {http://stacks.iop.org/0022-3719/19/i=31/a=004}
  {\bibfield  {journal} {\bibinfo  {journal} {Journal of Physics C: Solid State
  Physics}\ }\textbf {\bibinfo {volume} {19}},\ \bibinfo {pages} {L741}
  (\bibinfo {year} {1986})}\BibitemShut {NoStop}%
\bibitem [{\citenamefont {Meider}\ and\ \citenamefont
  {Springborg}(1999)}]{Meider1999339}%
  \BibitemOpen
  \bibfield  {author} {\bibinfo {author} {\bibfnamefont {H.}~\bibnamefont
  {Meider}}\ and\ \bibinfo {author} {\bibfnamefont {M.}~\bibnamefont
  {Springborg}},\ }\href {\doibase
  http://dx.doi.org/10.1016/S0009-2614(98)01386-4} {\bibfield  {journal}
  {\bibinfo  {journal} {Chemical Physics Letters}\ }\textbf {\bibinfo {volume}
  {300}},\ \bibinfo {pages} {339} (\bibinfo {year} {1999})}\BibitemShut
  {NoStop}%
\bibitem [{\citenamefont {Hybertsen}\ \emph {et~al.}(1989)\citenamefont
  {Hybertsen}, \citenamefont {Schl\"uter},\ and\ \citenamefont
  {Christensen}}]{PhysRevB.39.9028}%
  \BibitemOpen
  \bibfield  {author} {\bibinfo {author} {\bibfnamefont {M.~S.}\ \bibnamefont
  {Hybertsen}}, \bibinfo {author} {\bibfnamefont {M.}~\bibnamefont
  {Schl\"uter}}, \ and\ \bibinfo {author} {\bibfnamefont {N.~E.}\ \bibnamefont
  {Christensen}},\ }\href {\doibase 10.1103/PhysRevB.39.9028} {\bibfield
  {journal} {\bibinfo  {journal} {Phys. Rev. B}\ }\textbf {\bibinfo {volume}
  {39}},\ \bibinfo {pages} {9028} (\bibinfo {year} {1989})}\BibitemShut
  {NoStop}%
\bibitem [{\citenamefont {Petukhov}\ \emph {et~al.}(2003)\citenamefont
  {Petukhov}, \citenamefont {Mazin}, \citenamefont {Chioncel},\ and\
  \citenamefont {Lichtenstein}}]{PhysRevB.67.153106}%
  \BibitemOpen
  \bibfield  {author} {\bibinfo {author} {\bibfnamefont {A.~G.}\ \bibnamefont
  {Petukhov}}, \bibinfo {author} {\bibfnamefont {I.~I.}\ \bibnamefont {Mazin}},
  \bibinfo {author} {\bibfnamefont {L.}~\bibnamefont {Chioncel}}, \ and\
  \bibinfo {author} {\bibfnamefont {A.~I.}\ \bibnamefont {Lichtenstein}},\
  }\href {\doibase 10.1103/PhysRevB.67.153106} {\bibfield  {journal} {\bibinfo
  {journal} {Phys. Rev. B}\ }\textbf {\bibinfo {volume} {67}},\ \bibinfo
  {pages} {153106} (\bibinfo {year} {2003})}\BibitemShut {NoStop}%
\bibitem [{\citenamefont {Wu}\ \emph {et~al.}(2007)\citenamefont {Wu},
  \citenamefont {Cheng},\ and\ \citenamefont
  {Van~Voorhis}}]{:/content/aip/journal/jcp/127/16/10.1063/1.2800022}%
  \BibitemOpen
  \bibfield  {author} {\bibinfo {author} {\bibfnamefont {Q.}~\bibnamefont
  {Wu}}, \bibinfo {author} {\bibfnamefont {C.-L.}\ \bibnamefont {Cheng}}, \
  and\ \bibinfo {author} {\bibfnamefont {T.}~\bibnamefont {Van~Voorhis}},\
  }\href@noop {} {\bibfield  {journal} {\bibinfo  {journal} {The Journal of
  Chemical Physics}\ }\textbf {\bibinfo {volume} {127}},\ \bibinfo {eid}
  {164119} (\bibinfo {year} {2007})}\BibitemShut {NoStop}%
\bibitem [{\citenamefont {Wu}\ \emph {et~al.}(2009)\citenamefont {Wu},
  \citenamefont {Kaduk},\ and\ \citenamefont
  {Van~Voorhis}}]{:/content/aip/journal/jcp/130/3/10.1063/1.3059784}%
  \BibitemOpen
  \bibfield  {author} {\bibinfo {author} {\bibfnamefont {Q.}~\bibnamefont
  {Wu}}, \bibinfo {author} {\bibfnamefont {B.}~\bibnamefont {Kaduk}}, \ and\
  \bibinfo {author} {\bibfnamefont {T.}~\bibnamefont {Van~Voorhis}},\
  }\href@noop {} {\bibfield  {journal} {\bibinfo  {journal} {The Journal of
  Chemical Physics}\ }\textbf {\bibinfo {volume} {130}},\ \bibinfo {eid}
  {034109} (\bibinfo {year} {2009})}\BibitemShut {NoStop}%
\bibitem [{\citenamefont {Kaduk}\ and\ \citenamefont
  {Van~Voorhis}(2010)}]{:/content/aip/journal/jcp/133/6/10.1063/1.3470106}%
  \BibitemOpen
  \bibfield  {author} {\bibinfo {author} {\bibfnamefont {B.}~\bibnamefont
  {Kaduk}}\ and\ \bibinfo {author} {\bibfnamefont {T.}~\bibnamefont
  {Van~Voorhis}},\ }\href@noop {} {\bibfield  {journal} {\bibinfo  {journal}
  {The Journal of Chemical Physics}\ }\textbf {\bibinfo {volume} {133}},\
  \bibinfo {eid} {061102} (\bibinfo {year} {2010})}\BibitemShut {NoStop}%
\bibitem [{\citenamefont {Kaduk}\ \emph {et~al.}(2014)\citenamefont {Kaduk},
  \citenamefont {Tsuchimochi},\ and\ \citenamefont
  {Van~Voorhis}}]{:/content/aip/journal/jcp/140/18/10.1063/1.4862497}%
  \BibitemOpen
  \bibfield  {author} {\bibinfo {author} {\bibfnamefont {B.}~\bibnamefont
  {Kaduk}}, \bibinfo {author} {\bibfnamefont {T.}~\bibnamefont {Tsuchimochi}},
  \ and\ \bibinfo {author} {\bibfnamefont {T.}~\bibnamefont {Van~Voorhis}},\
  }\href@noop {} {\bibfield  {journal} {\bibinfo  {journal} {The Journal of
  Chemical Physics}\ }\textbf {\bibinfo {volume} {140}} (\bibinfo {year}
  {2014})}\BibitemShut {NoStop}%
\bibitem [{\citenamefont {Jain}\ \emph {et~al.}(2011)\citenamefont {Jain},
  \citenamefont {Hautier}, \citenamefont {Moore}, \citenamefont {Ong},
  \citenamefont {Fischer}, \citenamefont {Mueller}, \citenamefont {Persson},\
  and\ \citenamefont {Ceder}}]{Jain20112295}%
  \BibitemOpen
  \bibfield  {author} {\bibinfo {author} {\bibfnamefont {A.}~\bibnamefont
  {Jain}}, \bibinfo {author} {\bibfnamefont {G.}~\bibnamefont {Hautier}},
  \bibinfo {author} {\bibfnamefont {C.~J.}\ \bibnamefont {Moore}}, \bibinfo
  {author} {\bibfnamefont {S.~P.}\ \bibnamefont {Ong}}, \bibinfo {author}
  {\bibfnamefont {C.~C.}\ \bibnamefont {Fischer}}, \bibinfo {author}
  {\bibfnamefont {T.}~\bibnamefont {Mueller}}, \bibinfo {author} {\bibfnamefont
  {K.~A.}\ \bibnamefont {Persson}}, \ and\ \bibinfo {author} {\bibfnamefont
  {G.}~\bibnamefont {Ceder}},\ }\href {\doibase
  http://dx.doi.org/10.1016/j.commatsci.2011.02.023} {\bibfield  {journal}
  {\bibinfo  {journal} {Computational Materials Science}\ }\textbf {\bibinfo
  {volume} {50}},\ \bibinfo {pages} {2295 } (\bibinfo {year}
  {2011})}\BibitemShut {NoStop}%
\bibitem [{\citenamefont {Curtarolo}\ \emph {et~al.}(2013)\citenamefont
  {Curtarolo}, \citenamefont {Hart}, \citenamefont {Nardelli}, \citenamefont
  {Mingo}, \citenamefont {Sanvito},\ and\ \citenamefont {Levy}}]{curtarolo}%
  \BibitemOpen
  \bibfield  {author} {\bibinfo {author} {\bibfnamefont {S.}~\bibnamefont
  {Curtarolo}}, \bibinfo {author} {\bibfnamefont {G.~L.~W.}\ \bibnamefont
  {Hart}}, \bibinfo {author} {\bibfnamefont {M.~B.}\ \bibnamefont {Nardelli}},
  \bibinfo {author} {\bibfnamefont {N.}~\bibnamefont {Mingo}}, \bibinfo
  {author} {\bibfnamefont {S.}~\bibnamefont {Sanvito}}, \ and\ \bibinfo
  {author} {\bibfnamefont {O.}~\bibnamefont {Levy}},\ }\href
  {http://dx.doi.org/10.1038/nmat3568} {\bibfield  {journal} {\bibinfo
  {journal} {Nat Mater}\ }\textbf {\bibinfo {volume} {12}},\ \bibinfo {pages}
  {191} (\bibinfo {year} {2013})}\BibitemShut {NoStop}%
\bibitem [{\citenamefont {\u{R}ez\'{a}\u{c}}\ \emph {et~al.}(2012)\citenamefont
  {\u{R}ez\'{a}\u{c}}, \citenamefont {L\'{e}vy}, \citenamefont {Demachy},\ and\
  \citenamefont {de~la Lande}}]{doi:10.1021/ct200570u}%
  \BibitemOpen
  \bibfield  {author} {\bibinfo {author} {\bibfnamefont {J.}~\bibnamefont
  {\u{R}ez\'{a}\u{c}}}, \bibinfo {author} {\bibfnamefont {B.}~\bibnamefont
  {L\'{e}vy}}, \bibinfo {author} {\bibfnamefont {I.}~\bibnamefont {Demachy}}, \
  and\ \bibinfo {author} {\bibfnamefont {A.}~\bibnamefont {de~la Lande}},\
  }\href {\doibase 10.1021/ct200570u} {\bibfield  {journal} {\bibinfo
  {journal} {Journal of Chemical Theory and Computation}\ }\textbf {\bibinfo
  {volume} {8}},\ \bibinfo {pages} {418} (\bibinfo {year} {2012})}\BibitemShut
  {NoStop}%
\bibitem [{\citenamefont {Ratcliff}\ \emph {et~al.}(2015)\citenamefont
  {Ratcliff}, \citenamefont {Genovese}, \citenamefont {Mohr},\ and\
  \citenamefont
  {Deutsch}}]{:/content/aip/journal/jcp/142/23/10.1063/1.4922378}%
  \BibitemOpen
  \bibfield  {author} {\bibinfo {author} {\bibfnamefont {L.~E.}\ \bibnamefont
  {Ratcliff}}, \bibinfo {author} {\bibfnamefont {L.}~\bibnamefont {Genovese}},
  \bibinfo {author} {\bibfnamefont {S.}~\bibnamefont {Mohr}}, \ and\ \bibinfo
  {author} {\bibfnamefont {T.}~\bibnamefont {Deutsch}},\ }\href@noop {}
  {\bibfield  {journal} {\bibinfo  {journal} {The Journal of Chemical Physics}\
  }\textbf {\bibinfo {volume} {142}},\ \bibinfo {pages} {234105} (\bibinfo
  {year} {2015})}\BibitemShut {NoStop}%
\bibitem [{Note1()}]{Note1}%
  \BibitemOpen
  \bibinfo {note} {The ``curvature'' is used here as a convenient shorthand for
  the second derivative. We do not imply the geometric curvature, which equals
  the second derivative only at stationary points.}\BibitemShut {Stop}%
\bibitem [{\citenamefont {Wu}(2016)}]{wuprivate}%
  \BibitemOpen
  \bibfield  {author} {\bibinfo {author} {\bibfnamefont {Q.}~\bibnamefont
  {Wu}},\ }\href@noop {} {}\bibinfo {howpublished} {personal communication}
  (\bibinfo {year} {2016})\BibitemShut {NoStop}%
\bibitem [{\citenamefont {Wu}\ and\ \citenamefont {Yang}(2003)}]{wu:2498}%
  \BibitemOpen
  \bibfield  {author} {\bibinfo {author} {\bibfnamefont {Q.}~\bibnamefont
  {Wu}}\ and\ \bibinfo {author} {\bibfnamefont {W.}~\bibnamefont {Yang}},\
  }\href {\doibase 10.1063/1.1535422} {\bibfield  {journal} {\bibinfo
  {journal} {The Journal of Chemical Physics}\ }\textbf {\bibinfo {volume}
  {118}},\ \bibinfo {pages} {2498} (\bibinfo {year} {2003})}\BibitemShut
  {NoStop}%
\bibitem [{\citenamefont {Skylaris}\ \emph {et~al.}(2005)\citenamefont
  {Skylaris}, \citenamefont {Haynes}, \citenamefont {Mostofi},\ and\
  \citenamefont {Payne}}]{:/content/aip/journal/jcp/122/8/10.1063/1.1839852}%
  \BibitemOpen
  \bibfield  {author} {\bibinfo {author} {\bibfnamefont {C.-K.}\ \bibnamefont
  {Skylaris}}, \bibinfo {author} {\bibfnamefont {P.~D.}\ \bibnamefont
  {Haynes}}, \bibinfo {author} {\bibfnamefont {A.~A.}\ \bibnamefont {Mostofi}},
  \ and\ \bibinfo {author} {\bibfnamefont {M.~C.}\ \bibnamefont {Payne}},\
  }\href@noop {} {\bibfield  {journal} {\bibinfo  {journal} {The Journal of
  Chemical Physics}\ }\textbf {\bibinfo {volume} {122}},\ \bibinfo {eid}
  {084119} (\bibinfo {year} {2005})}\BibitemShut {NoStop}%
\bibitem [{\citenamefont {Perdew}\ \emph {et~al.}(1996)\citenamefont {Perdew},
  \citenamefont {Burke},\ and\ \citenamefont
  {Ernzerhof}}]{PhysRevLett.77.3865}%
  \BibitemOpen
  \bibfield  {author} {\bibinfo {author} {\bibfnamefont {J.~P.}\ \bibnamefont
  {Perdew}}, \bibinfo {author} {\bibfnamefont {K.}~\bibnamefont {Burke}}, \
  and\ \bibinfo {author} {\bibfnamefont {M.}~\bibnamefont {Ernzerhof}},\
  }\href@noop {} {\bibfield  {journal} {\bibinfo  {journal} {Phys. Rev. Lett.}\
  }\textbf {\bibinfo {volume} {77}},\ \bibinfo {pages} {3865} (\bibinfo {year}
  {1996})}\BibitemShut {NoStop}%
\bibitem [{\citenamefont {Coulson}\ and\ \citenamefont
  {Fischer}(1949)}]{cfpoint}%
  \BibitemOpen
  \bibfield  {author} {\bibinfo {author} {\bibfnamefont {C.}~\bibnamefont
  {Coulson}}\ and\ \bibinfo {author} {\bibfnamefont {I.}~\bibnamefont
  {Fischer}},\ }\href@noop {} {\bibfield  {journal} {\bibinfo  {journal}
  {Philos. Mag.}\ }\textbf {\bibinfo {volume} {40}},\ \bibinfo {pages} {386}
  (\bibinfo {year} {1949})}\BibitemShut {NoStop}%
\bibitem [{\citenamefont {Ruzsinszky}\ \emph {et~al.}(2005)\citenamefont
  {Ruzsinszky}, \citenamefont {Perdew},\ and\ \citenamefont
  {Csonka}}]{doi:10.1021/jp0534479}%
  \BibitemOpen
  \bibfield  {author} {\bibinfo {author} {\bibfnamefont {A.}~\bibnamefont
  {Ruzsinszky}}, \bibinfo {author} {\bibfnamefont {J.~P.}\ \bibnamefont
  {Perdew}}, \ and\ \bibinfo {author} {\bibfnamefont {G.~I.}\ \bibnamefont
  {Csonka}},\ }\href@noop {} {\bibfield  {journal} {\bibinfo  {journal} {The
  Journal of Physical Chemistry A}\ }\textbf {\bibinfo {volume} {109}},\
  \bibinfo {pages} {11006} (\bibinfo {year} {2005})}\BibitemShut {NoStop}%
\bibitem [{\citenamefont {Dziedzic}\ \emph {et~al.}(2011)\citenamefont
  {Dziedzic}, \citenamefont {Helal}, \citenamefont {Skylaris}, \citenamefont
  {Mostofi},\ and\ \citenamefont {Payne}}]{0295-5075-95-4-43001}%
  \BibitemOpen
  \bibfield  {author} {\bibinfo {author} {\bibfnamefont {J.}~\bibnamefont
  {Dziedzic}}, \bibinfo {author} {\bibfnamefont {H.~H.}\ \bibnamefont {Helal}},
  \bibinfo {author} {\bibfnamefont {C.-K.}\ \bibnamefont {Skylaris}}, \bibinfo
  {author} {\bibfnamefont {A.~A.}\ \bibnamefont {Mostofi}}, \ and\ \bibinfo
  {author} {\bibfnamefont {M.~C.}\ \bibnamefont {Payne}},\ }\href
  {http://stacks.iop.org/0295-5075/95/i=4/a=43001} {\bibfield  {journal}
  {\bibinfo  {journal} {EPL (Europhysics Letters)}\ }\textbf {\bibinfo {volume}
  {95}},\ \bibinfo {pages} {43001} (\bibinfo {year} {2011})}\BibitemShut
  {NoStop}%
\bibitem [{\citenamefont {Andreussi}\ \emph {et~al.}(2012)\citenamefont
  {Andreussi}, \citenamefont {Dabo},\ and\ \citenamefont
  {Marzari}}]{:/content/aip/journal/jcp/136/6/10.1063/1.3676407}%
  \BibitemOpen
  \bibfield  {author} {\bibinfo {author} {\bibfnamefont {O.}~\bibnamefont
  {Andreussi}}, \bibinfo {author} {\bibfnamefont {I.}~\bibnamefont {Dabo}}, \
  and\ \bibinfo {author} {\bibfnamefont {N.}~\bibnamefont {Marzari}},\ }\href
  {http://scitation.aip.org/content/aip/journal/jcp/136/6/10.1063/1.3676407}
  {\bibfield  {journal} {\bibinfo  {journal} {The Journal of Chemical Physics}\
  }\textbf {\bibinfo {volume} {136}},\ \bibinfo {eid} {064102} (\bibinfo {year}
  {2012})}\BibitemShut {NoStop}%
\bibitem [{Note2()}]{Note2}%
  \BibitemOpen
  \bibinfo {note} {A single, strictly local occupancy constraint is considered
  in Ref.~\protect \rev@citealpnum {PhysRevA.72.024502}. We retain these
  restrictions so as not to obscure the fundamental aspects under
  consideration. These conditions are typically lifted in practical cDFT
  calculations, bringing us into multivariate optimisation of constrained
  Kohn-Sham spin-density functional theory, which may also be non-local or
  orbital dependent.}\BibitemShut {Stop}%
\bibitem [{\citenamefont {Baroni}\ \emph {et~al.}(1987)\citenamefont {Baroni},
  \citenamefont {Giannozzi},\ and\ \citenamefont
  {Testa}}]{PhysRevLett.58.1861}%
  \BibitemOpen
  \bibfield  {author} {\bibinfo {author} {\bibfnamefont {S.}~\bibnamefont
  {Baroni}}, \bibinfo {author} {\bibfnamefont {P.}~\bibnamefont {Giannozzi}}, \
  and\ \bibinfo {author} {\bibfnamefont {A.}~\bibnamefont {Testa}},\
  }\href@noop {} {\bibfield  {journal} {\bibinfo  {journal} {Phys. Rev. Lett.}\
  }\textbf {\bibinfo {volume} {58}},\ \bibinfo {pages} {1861} (\bibinfo {year}
  {1987})}\BibitemShut {NoStop}%
\bibitem [{\citenamefont {Baroni}\ \emph {et~al.}(2001)\citenamefont {Baroni},
  \citenamefont {de~Gironcoli}, \citenamefont {Dal~Corso},\ and\ \citenamefont
  {Giannozzi}}]{RevModPhys.73.515}%
  \BibitemOpen
  \bibfield  {author} {\bibinfo {author} {\bibfnamefont {S.}~\bibnamefont
  {Baroni}}, \bibinfo {author} {\bibfnamefont {S.}~\bibnamefont
  {de~Gironcoli}}, \bibinfo {author} {\bibfnamefont {A.}~\bibnamefont
  {Dal~Corso}}, \ and\ \bibinfo {author} {\bibfnamefont {P.}~\bibnamefont
  {Giannozzi}},\ }\href@noop {} {\bibfield  {journal} {\bibinfo  {journal}
  {Rev. Mod. Phys.}\ }\textbf {\bibinfo {volume} {73}},\ \bibinfo {pages} {515}
  (\bibinfo {year} {2001})}\BibitemShut {NoStop}%
\bibitem [{Note3()}]{Note3}%
  \BibitemOpen
  \bibinfo {note} {We use a lunate epsilon $\epsilon $ for the microscopic
  dielectric function in order to distinguish from it from the Kohn-Sham
  eigenvalues $\varepsilon $}\BibitemShut {NoStop}%
\bibitem [{\citenamefont {Becke}(1988)}]{PhysRevA.38.3098}%
  \BibitemOpen
  \bibfield  {author} {\bibinfo {author} {\bibfnamefont {A.~D.}\ \bibnamefont
  {Becke}},\ }\href@noop {} {\bibfield  {journal} {\bibinfo  {journal} {Phys.
  Rev. A}\ }\textbf {\bibinfo {volume} {38}},\ \bibinfo {pages} {3098}
  (\bibinfo {year} {1988})}\BibitemShut {NoStop}%
\bibitem [{\citenamefont {Lee}\ \emph {et~al.}(1988)\citenamefont {Lee},
  \citenamefont {Yang},\ and\ \citenamefont {Parr}}]{PhysRevB.37.785}%
  \BibitemOpen
  \bibfield  {author} {\bibinfo {author} {\bibfnamefont {C.}~\bibnamefont
  {Lee}}, \bibinfo {author} {\bibfnamefont {W.}~\bibnamefont {Yang}}, \ and\
  \bibinfo {author} {\bibfnamefont {R.~G.}\ \bibnamefont {Parr}},\ }\href@noop
  {} {\bibfield  {journal} {\bibinfo  {journal} {Phys. Rev. B}\ }\textbf
  {\bibinfo {volume} {37}},\ \bibinfo {pages} {785} (\bibinfo {year}
  {1988})}\BibitemShut {NoStop}%
\bibitem [{\citenamefont {Skylaris}\ \emph {et~al.}(2002)\citenamefont
  {Skylaris}, \citenamefont {Mostofi}, \citenamefont {Haynes}, \citenamefont
  {Di\'eguez},\ and\ \citenamefont {Payne}}]{PhysRevB.66.035119}%
  \BibitemOpen
  \bibfield  {author} {\bibinfo {author} {\bibfnamefont {C.-K.}\ \bibnamefont
  {Skylaris}}, \bibinfo {author} {\bibfnamefont {A.~A.}\ \bibnamefont
  {Mostofi}}, \bibinfo {author} {\bibfnamefont {P.~D.}\ \bibnamefont {Haynes}},
  \bibinfo {author} {\bibfnamefont {O.}~\bibnamefont {Di\'eguez}}, \ and\
  \bibinfo {author} {\bibfnamefont {M.~C.}\ \bibnamefont {Payne}},\ }\href@noop
  {} {\bibfield  {journal} {\bibinfo  {journal} {Phys. Rev. B}\ }\textbf
  {\bibinfo {volume} {66}},\ \bibinfo {pages} {035119} (\bibinfo {year}
  {2002})}\BibitemShut {NoStop}%
\bibitem [{\citenamefont {O'Regan}\ \emph {et~al.}(2012)\citenamefont
  {O'Regan}, \citenamefont {Payne},\ and\ \citenamefont
  {Mostofi}}]{PhysRevB.85.193101}%
  \BibitemOpen
  \bibfield  {author} {\bibinfo {author} {\bibfnamefont {D.~D.}\ \bibnamefont
  {O'Regan}}, \bibinfo {author} {\bibfnamefont {M.~C.}\ \bibnamefont {Payne}},
  \ and\ \bibinfo {author} {\bibfnamefont {A.~A.}\ \bibnamefont {Mostofi}},\
  }\href@noop {} {\bibfield  {journal} {\bibinfo  {journal} {Phys. Rev. B}\
  }\textbf {\bibinfo {volume} {85}},\ \bibinfo {pages} {193101} (\bibinfo
  {year} {2012})}\BibitemShut {NoStop}%
\bibitem [{Note4()}]{Note4}%
  \BibitemOpen
  \bibinfo {note} {The resulting constraint acts on the Kohn-Sham
  density-matrix rather than on the density. Our analytical findings extends to
  that case with minor notational changes.}\BibitemShut {Stop}%
\bibitem [{\citenamefont {Ratcliff}\ \emph {et~al.}(2011)\citenamefont
  {Ratcliff}, \citenamefont {Hine},\ and\ \citenamefont
  {Haynes}}]{PhysRevB.84.165131}%
  \BibitemOpen
  \bibfield  {author} {\bibinfo {author} {\bibfnamefont {L.~E.}\ \bibnamefont
  {Ratcliff}}, \bibinfo {author} {\bibfnamefont {N.~D.~M.}\ \bibnamefont
  {Hine}}, \ and\ \bibinfo {author} {\bibfnamefont {P.~D.}\ \bibnamefont
  {Haynes}},\ }\href@noop {} {\bibfield  {journal} {\bibinfo  {journal} {Phys.
  Rev. B}\ }\textbf {\bibinfo {volume} {84}},\ \bibinfo {pages} {165131}
  (\bibinfo {year} {2011})}\BibitemShut {NoStop}%
\bibitem [{\citenamefont {Rondinelli}\ \emph {et~al.}(2007)\citenamefont
  {Rondinelli}, \citenamefont {Deng},\ and\ \citenamefont
  {Marks}}]{Rondinelli2007345}%
  \BibitemOpen
  \bibfield  {author} {\bibinfo {author} {\bibfnamefont {J.~M.}\ \bibnamefont
  {Rondinelli}}, \bibinfo {author} {\bibfnamefont {B.}~\bibnamefont {Deng}}, \
  and\ \bibinfo {author} {\bibfnamefont {L.~D.}\ \bibnamefont {Marks}},\ }\href
  {\doibase http://dx.doi.org/10.1016/j.commatsci.2007.01.004} {\bibfield
  {journal} {\bibinfo  {journal} {Computational Materials Science}\ }\textbf
  {\bibinfo {volume} {40}},\ \bibinfo {pages} {345 } (\bibinfo {year}
  {2007})}\BibitemShut {NoStop}%
\bibitem [{\citenamefont {Harville}(2011)}]{harville2011matrix}%
  \BibitemOpen
  \bibfield  {author} {\bibinfo {author} {\bibfnamefont {D.}~\bibnamefont
  {Harville}},\ }\href {https://books.google.fr/books?id=bYUHCAAAQBAJ} {\emph
  {\bibinfo {title} {Matrix Algebra: Exercises and Solutions}}}\ (\bibinfo
  {publisher} {Springer New York},\ \bibinfo {year} {2011})\BibitemShut
  {NoStop}%
\bibitem [{\citenamefont {Valiev}\ \emph {et~al.}(2010)\citenamefont {Valiev},
  \citenamefont {Bylaska}, \citenamefont {Govind}, \citenamefont {Kowalski},
  \citenamefont {Straatsma}, \citenamefont {Dam}, \citenamefont {Wang},
  \citenamefont {Nieplocha}, \citenamefont {Apra}, \citenamefont {Windus},\
  and\ \citenamefont {de~Jong}}]{Valiev20101477}%
  \BibitemOpen
  \bibfield  {author} {\bibinfo {author} {\bibfnamefont {M.}~\bibnamefont
  {Valiev}}, \bibinfo {author} {\bibfnamefont {E.}~\bibnamefont {Bylaska}},
  \bibinfo {author} {\bibfnamefont {N.}~\bibnamefont {Govind}}, \bibinfo
  {author} {\bibfnamefont {K.}~\bibnamefont {Kowalski}}, \bibinfo {author}
  {\bibfnamefont {T.}~\bibnamefont {Straatsma}}, \bibinfo {author}
  {\bibfnamefont {H.~V.}\ \bibnamefont {Dam}}, \bibinfo {author} {\bibfnamefont
  {D.}~\bibnamefont {Wang}}, \bibinfo {author} {\bibfnamefont {J.}~\bibnamefont
  {Nieplocha}}, \bibinfo {author} {\bibfnamefont {E.}~\bibnamefont {Apra}},
  \bibinfo {author} {\bibfnamefont {T.}~\bibnamefont {Windus}}, \ and\ \bibinfo
  {author} {\bibfnamefont {W.}~\bibnamefont {de~Jong}},\ }\href {\doibase
  http://dx.doi.org/10.1016/j.cpc.2010.04.018} {\bibfield  {journal} {\bibinfo
  {journal} {Computer Physics Communications}\ }\textbf {\bibinfo {volume}
  {181}},\ \bibinfo {pages} {1477 } (\bibinfo {year} {2010})}\BibitemShut
  {NoStop}%
\bibitem [{\citenamefont {Shao}\ \emph {et~al.}(2015)\citenamefont {Shao},
  \citenamefont {Gan}, \citenamefont {Epifanovsky}, \citenamefont {Gilbert},
  \citenamefont {Wormit}, \citenamefont {Kussmann}, \citenamefont {Lange},
  \citenamefont {Behn}, \citenamefont {Deng}, \citenamefont {Feng},
  \citenamefont {Ghosh}, \citenamefont {Goldey}, \citenamefont {Horn},
  \citenamefont {Jacobson}, \citenamefont {Kaliman}, \citenamefont
  {Khaliullin}, \citenamefont {K\'us}, \citenamefont {Landau}, \citenamefont
  {Liu}, \citenamefont {Proynov}, \citenamefont {Rhee}, \citenamefont
  {Richard}, \citenamefont {Rohrdanz}, \citenamefont {Steele}, \citenamefont
  {Sundstrom}, \citenamefont {{Woodcock III}}, \citenamefont {Zimmerman},
  \citenamefont {Zuev}, \citenamefont {Albrecht}, \citenamefont {Alguire},
  \citenamefont {Austin}, \citenamefont {Beran}, \citenamefont {Bernard},
  \citenamefont {Berquist}, \citenamefont {Brandhorst}, \citenamefont
  {Bravaya}, \citenamefont {Brown}, \citenamefont {Casanova}, \citenamefont
  {Chang}, \citenamefont {Chen}, \citenamefont {Chien}, \citenamefont
  {Closser}, \citenamefont {Crittenden}, \citenamefont {Diedenhofen},
  \citenamefont {{DiStasio Jr.}}, \citenamefont {Dop}, \citenamefont {Dutoi},
  \citenamefont {Edgar}, \citenamefont {Fatehi}, \citenamefont
  {{Fusti-Molnar}}, \citenamefont {Ghysels}, \citenamefont
  {{Golubeva-Zadorozhnaya}}, \citenamefont {Gomes}, \citenamefont
  {{Hanson-Heine}}, \citenamefont {Harbach}, \citenamefont {Hauser},
  \citenamefont {Hohenstein}, \citenamefont {Holden}, \citenamefont {Jagau},
  \citenamefont {Ji}, \citenamefont {Kaduk}, \citenamefont {Khistyaev},
  \citenamefont {Kim}, \citenamefont {Kim}, \citenamefont {King}, \citenamefont
  {Klunzinger}, \citenamefont {Kosenkov}, \citenamefont {Kowalczyk},
  \citenamefont {Krauter}, \citenamefont {Lao}, \citenamefont {Laurent},
  \citenamefont {Lawler}, \citenamefont {Levchenko}, \citenamefont {Lin},
  \citenamefont {Liu}, \citenamefont {Livshits}, \citenamefont {Lochan},
  \citenamefont {Luenser}, \citenamefont {Manohar}, \citenamefont {Manzer},
  \citenamefont {Mao}, \citenamefont {Mardirossian}, \citenamefont {Marenich},
  \citenamefont {Maurer}, \citenamefont {Mayhall}, \citenamefont {Oana},
  \citenamefont {{Olivares-Amaya}}, \citenamefont {O'Neill}, \citenamefont
  {Parkhill}, \citenamefont {Perrine}, \citenamefont {Peverati}, \citenamefont
  {Pieniazek}, \citenamefont {Prociuk}, \citenamefont {Rehn}, \citenamefont
  {Rosta}, \citenamefont {Russ}, \citenamefont {Sergueev}, \citenamefont
  {Sharada}, \citenamefont {Sharmaa}, \citenamefont {Small}, \citenamefont
  {Sodt}, \citenamefont {Stein}, \citenamefont {St\"uck}, \citenamefont {Su},
  \citenamefont {Thom}, \citenamefont {Tsuchimochi}, \citenamefont {Vogt},
  \citenamefont {Vydrov}, \citenamefont {Wang}, \citenamefont {Watson},
  \citenamefont {Wenzel}, \citenamefont {White}, \citenamefont {Williams},
  \citenamefont {Vanovschi}, \citenamefont {Yeganeh}, \citenamefont {Yost},
  \citenamefont {You}, \citenamefont {Zhang}, \citenamefont {Zhang},
  \citenamefont {Zhou}, \citenamefont {Brooks}, \citenamefont {Chan},
  \citenamefont {Chipman}, \citenamefont {Cramer}, \citenamefont {{Goddard
  III}}, \citenamefont {Gordon}, \citenamefont {Hehre}, \citenamefont {Klamt},
  \citenamefont {{Schaefer III}}, \citenamefont {Schmidt}, \citenamefont
  {Sherrill}, \citenamefont {Truhlar}, \citenamefont {Warshel}, \citenamefont
  {Xua}, \citenamefont {{Aspuru-Guzik}}, \citenamefont {Baer}, \citenamefont
  {Bell}, \citenamefont {Besley}, \citenamefont {Chai}, \citenamefont {Dreuw},
  \citenamefont {Dunietz}, \citenamefont {Furlani}, \citenamefont {Gwaltney},
  \citenamefont {Hsu}, \citenamefont {Jung}, \citenamefont {Kong},
  \citenamefont {Lambrecht}, \citenamefont {Liang}, \citenamefont {Ochsenfeld},
  \citenamefont {Rassolov}, \citenamefont {Slipchenko}, \citenamefont
  {Subotnik}, \citenamefont {{Van Voorhis}}, \citenamefont {Herbert},
  \citenamefont {Krylov}, \citenamefont {Gill},\ and\ \citenamefont
  {{Head-Gordon}}}]{QCHEM4}%
  \BibitemOpen
  \bibfield  {author} {\bibinfo {author} {\bibfnamefont {Y.}~\bibnamefont
  {Shao}}, \bibinfo {author} {\bibfnamefont {Z.}~\bibnamefont {Gan}}, \bibinfo
  {author} {\bibfnamefont {E.}~\bibnamefont {Epifanovsky}}, \bibinfo {author}
  {\bibfnamefont {A.~T.~B.}\ \bibnamefont {Gilbert}}, \bibinfo {author}
  {\bibfnamefont {M.}~\bibnamefont {Wormit}}, \bibinfo {author} {\bibfnamefont
  {J.}~\bibnamefont {Kussmann}}, \bibinfo {author} {\bibfnamefont {A.~W.}\
  \bibnamefont {Lange}}, \bibinfo {author} {\bibfnamefont {A.}~\bibnamefont
  {Behn}}, \bibinfo {author} {\bibfnamefont {J.}~\bibnamefont {Deng}}, \bibinfo
  {author} {\bibfnamefont {X.}~\bibnamefont {Feng}}, \bibinfo {author}
  {\bibfnamefont {D.}~\bibnamefont {Ghosh}}, \bibinfo {author} {\bibfnamefont
  {M.}~\bibnamefont {Goldey}}, \bibinfo {author} {\bibfnamefont {P.~R.}\
  \bibnamefont {Horn}}, \bibinfo {author} {\bibfnamefont {L.~D.}\ \bibnamefont
  {Jacobson}}, \bibinfo {author} {\bibfnamefont {I.}~\bibnamefont {Kaliman}},
  \bibinfo {author} {\bibfnamefont {R.~Z.}\ \bibnamefont {Khaliullin}},
  \bibinfo {author} {\bibfnamefont {T.}~\bibnamefont {K\'us}}, \bibinfo
  {author} {\bibfnamefont {A.}~\bibnamefont {Landau}}, \bibinfo {author}
  {\bibfnamefont {J.}~\bibnamefont {Liu}}, \bibinfo {author} {\bibfnamefont
  {E.~I.}\ \bibnamefont {Proynov}}, \bibinfo {author} {\bibfnamefont {Y.~M.}\
  \bibnamefont {Rhee}}, \bibinfo {author} {\bibfnamefont {R.~M.}\ \bibnamefont
  {Richard}}, \bibinfo {author} {\bibfnamefont {M.~A.}\ \bibnamefont
  {Rohrdanz}}, \bibinfo {author} {\bibfnamefont {R.~P.}\ \bibnamefont
  {Steele}}, \bibinfo {author} {\bibfnamefont {E.~J.}\ \bibnamefont
  {Sundstrom}}, \bibinfo {author} {\bibfnamefont {H.~L.}\ \bibnamefont
  {{Woodcock III}}}, \bibinfo {author} {\bibfnamefont {P.~M.}\ \bibnamefont
  {Zimmerman}}, \bibinfo {author} {\bibfnamefont {D.}~\bibnamefont {Zuev}},
  \bibinfo {author} {\bibfnamefont {B.}~\bibnamefont {Albrecht}}, \bibinfo
  {author} {\bibfnamefont {E.}~\bibnamefont {Alguire}}, \bibinfo {author}
  {\bibfnamefont {B.}~\bibnamefont {Austin}}, \bibinfo {author} {\bibfnamefont
  {G.~J.~O.}\ \bibnamefont {Beran}}, \bibinfo {author} {\bibfnamefont {Y.~A.}\
  \bibnamefont {Bernard}}, \bibinfo {author} {\bibfnamefont {E.}~\bibnamefont
  {Berquist}}, \bibinfo {author} {\bibfnamefont {K.}~\bibnamefont
  {Brandhorst}}, \bibinfo {author} {\bibfnamefont {K.~B.}\ \bibnamefont
  {Bravaya}}, \bibinfo {author} {\bibfnamefont {S.~T.}\ \bibnamefont {Brown}},
  \bibinfo {author} {\bibfnamefont {D.}~\bibnamefont {Casanova}}, \bibinfo
  {author} {\bibfnamefont {C.-M.}\ \bibnamefont {Chang}}, \bibinfo {author}
  {\bibfnamefont {Y.}~\bibnamefont {Chen}}, \bibinfo {author} {\bibfnamefont
  {S.~H.}\ \bibnamefont {Chien}}, \bibinfo {author} {\bibfnamefont {K.~D.}\
  \bibnamefont {Closser}}, \bibinfo {author} {\bibfnamefont {D.~L.}\
  \bibnamefont {Crittenden}}, \bibinfo {author} {\bibfnamefont
  {M.}~\bibnamefont {Diedenhofen}}, \bibinfo {author} {\bibfnamefont {R.~A.}\
  \bibnamefont {{DiStasio Jr.}}}, \bibinfo {author} {\bibfnamefont
  {H.}~\bibnamefont {Dop}}, \bibinfo {author} {\bibfnamefont {A.~D.}\
  \bibnamefont {Dutoi}}, \bibinfo {author} {\bibfnamefont {R.~G.}\ \bibnamefont
  {Edgar}}, \bibinfo {author} {\bibfnamefont {S.}~\bibnamefont {Fatehi}},
  \bibinfo {author} {\bibfnamefont {L.}~\bibnamefont {{Fusti-Molnar}}},
  \bibinfo {author} {\bibfnamefont {A.}~\bibnamefont {Ghysels}}, \bibinfo
  {author} {\bibfnamefont {A.}~\bibnamefont {{Golubeva-Zadorozhnaya}}},
  \bibinfo {author} {\bibfnamefont {J.}~\bibnamefont {Gomes}}, \bibinfo
  {author} {\bibfnamefont {M.~W.~D.}\ \bibnamefont {{Hanson-Heine}}}, \bibinfo
  {author} {\bibfnamefont {P.~H.~P.}\ \bibnamefont {Harbach}}, \bibinfo
  {author} {\bibfnamefont {A.~W.}\ \bibnamefont {Hauser}}, \bibinfo {author}
  {\bibfnamefont {E.~G.}\ \bibnamefont {Hohenstein}}, \bibinfo {author}
  {\bibfnamefont {Z.~C.}\ \bibnamefont {Holden}}, \bibinfo {author}
  {\bibfnamefont {T.-C.}\ \bibnamefont {Jagau}}, \bibinfo {author}
  {\bibfnamefont {H.}~\bibnamefont {Ji}}, \bibinfo {author} {\bibfnamefont
  {B.}~\bibnamefont {Kaduk}}, \bibinfo {author} {\bibfnamefont
  {K.}~\bibnamefont {Khistyaev}}, \bibinfo {author} {\bibfnamefont
  {J.}~\bibnamefont {Kim}}, \bibinfo {author} {\bibfnamefont {J.}~\bibnamefont
  {Kim}}, \bibinfo {author} {\bibfnamefont {R.~A.}\ \bibnamefont {King}},
  \bibinfo {author} {\bibfnamefont {P.}~\bibnamefont {Klunzinger}}, \bibinfo
  {author} {\bibfnamefont {D.}~\bibnamefont {Kosenkov}}, \bibinfo {author}
  {\bibfnamefont {T.}~\bibnamefont {Kowalczyk}}, \bibinfo {author}
  {\bibfnamefont {C.~M.}\ \bibnamefont {Krauter}}, \bibinfo {author}
  {\bibfnamefont {K.~U.}\ \bibnamefont {Lao}}, \bibinfo {author} {\bibfnamefont
  {A.}~\bibnamefont {Laurent}}, \bibinfo {author} {\bibfnamefont {K.~V.}\
  \bibnamefont {Lawler}}, \bibinfo {author} {\bibfnamefont {S.~V.}\
  \bibnamefont {Levchenko}}, \bibinfo {author} {\bibfnamefont {C.~Y.}\
  \bibnamefont {Lin}}, \bibinfo {author} {\bibfnamefont {F.}~\bibnamefont
  {Liu}}, \bibinfo {author} {\bibfnamefont {E.}~\bibnamefont {Livshits}},
  \bibinfo {author} {\bibfnamefont {R.~C.}\ \bibnamefont {Lochan}}, \bibinfo
  {author} {\bibfnamefont {A.}~\bibnamefont {Luenser}}, \bibinfo {author}
  {\bibfnamefont {P.}~\bibnamefont {Manohar}}, \bibinfo {author} {\bibfnamefont
  {S.~F.}\ \bibnamefont {Manzer}}, \bibinfo {author} {\bibfnamefont {S.-P.}\
  \bibnamefont {Mao}}, \bibinfo {author} {\bibfnamefont {N.}~\bibnamefont
  {Mardirossian}}, \bibinfo {author} {\bibfnamefont {A.~V.}\ \bibnamefont
  {Marenich}}, \bibinfo {author} {\bibfnamefont {S.~A.}\ \bibnamefont
  {Maurer}}, \bibinfo {author} {\bibfnamefont {N.~J.}\ \bibnamefont {Mayhall}},
  \bibinfo {author} {\bibfnamefont {C.~M.}\ \bibnamefont {Oana}}, \bibinfo
  {author} {\bibfnamefont {R.}~\bibnamefont {{Olivares-Amaya}}}, \bibinfo
  {author} {\bibfnamefont {D.~P.}\ \bibnamefont {O'Neill}}, \bibinfo {author}
  {\bibfnamefont {J.~A.}\ \bibnamefont {Parkhill}}, \bibinfo {author}
  {\bibfnamefont {T.~M.}\ \bibnamefont {Perrine}}, \bibinfo {author}
  {\bibfnamefont {R.}~\bibnamefont {Peverati}}, \bibinfo {author}
  {\bibfnamefont {P.~A.}\ \bibnamefont {Pieniazek}}, \bibinfo {author}
  {\bibfnamefont {A.}~\bibnamefont {Prociuk}}, \bibinfo {author} {\bibfnamefont
  {D.~R.}\ \bibnamefont {Rehn}}, \bibinfo {author} {\bibfnamefont
  {E.}~\bibnamefont {Rosta}}, \bibinfo {author} {\bibfnamefont {N.~J.}\
  \bibnamefont {Russ}}, \bibinfo {author} {\bibfnamefont {N.}~\bibnamefont
  {Sergueev}}, \bibinfo {author} {\bibfnamefont {S.~M.}\ \bibnamefont
  {Sharada}}, \bibinfo {author} {\bibfnamefont {S.}~\bibnamefont {Sharmaa}},
  \bibinfo {author} {\bibfnamefont {D.~W.}\ \bibnamefont {Small}}, \bibinfo
  {author} {\bibfnamefont {A.}~\bibnamefont {Sodt}}, \bibinfo {author}
  {\bibfnamefont {T.}~\bibnamefont {Stein}}, \bibinfo {author} {\bibfnamefont
  {D.}~\bibnamefont {St\"uck}}, \bibinfo {author} {\bibfnamefont {Y.-C.}\
  \bibnamefont {Su}}, \bibinfo {author} {\bibfnamefont {A.~J.~W.}\ \bibnamefont
  {Thom}}, \bibinfo {author} {\bibfnamefont {T.}~\bibnamefont {Tsuchimochi}},
  \bibinfo {author} {\bibfnamefont {L.}~\bibnamefont {Vogt}}, \bibinfo {author}
  {\bibfnamefont {O.}~\bibnamefont {Vydrov}}, \bibinfo {author} {\bibfnamefont
  {T.}~\bibnamefont {Wang}}, \bibinfo {author} {\bibfnamefont {M.~A.}\
  \bibnamefont {Watson}}, \bibinfo {author} {\bibfnamefont {J.}~\bibnamefont
  {Wenzel}}, \bibinfo {author} {\bibfnamefont {A.}~\bibnamefont {White}},
  \bibinfo {author} {\bibfnamefont {C.~F.}\ \bibnamefont {Williams}}, \bibinfo
  {author} {\bibfnamefont {V.}~\bibnamefont {Vanovschi}}, \bibinfo {author}
  {\bibfnamefont {S.}~\bibnamefont {Yeganeh}}, \bibinfo {author} {\bibfnamefont
  {S.~R.}\ \bibnamefont {Yost}}, \bibinfo {author} {\bibfnamefont {Z.-Q.}\
  \bibnamefont {You}}, \bibinfo {author} {\bibfnamefont {I.~Y.}\ \bibnamefont
  {Zhang}}, \bibinfo {author} {\bibfnamefont {X.}~\bibnamefont {Zhang}},
  \bibinfo {author} {\bibfnamefont {Y.}~\bibnamefont {Zhou}}, \bibinfo {author}
  {\bibfnamefont {B.~R.}\ \bibnamefont {Brooks}}, \bibinfo {author}
  {\bibfnamefont {G.~K.~L.}\ \bibnamefont {Chan}}, \bibinfo {author}
  {\bibfnamefont {D.~M.}\ \bibnamefont {Chipman}}, \bibinfo {author}
  {\bibfnamefont {C.~J.}\ \bibnamefont {Cramer}}, \bibinfo {author}
  {\bibfnamefont {W.~A.}\ \bibnamefont {{Goddard III}}}, \bibinfo {author}
  {\bibfnamefont {M.~S.}\ \bibnamefont {Gordon}}, \bibinfo {author}
  {\bibfnamefont {W.~J.}\ \bibnamefont {Hehre}}, \bibinfo {author}
  {\bibfnamefont {A.}~\bibnamefont {Klamt}}, \bibinfo {author} {\bibfnamefont
  {H.~F.}\ \bibnamefont {{Schaefer III}}}, \bibinfo {author} {\bibfnamefont
  {M.~W.}\ \bibnamefont {Schmidt}}, \bibinfo {author} {\bibfnamefont {C.~D.}\
  \bibnamefont {Sherrill}}, \bibinfo {author} {\bibfnamefont {D.~G.}\
  \bibnamefont {Truhlar}}, \bibinfo {author} {\bibfnamefont {A.}~\bibnamefont
  {Warshel}}, \bibinfo {author} {\bibfnamefont {X.}~\bibnamefont {Xua}},
  \bibinfo {author} {\bibfnamefont {A.}~\bibnamefont {{Aspuru-Guzik}}},
  \bibinfo {author} {\bibfnamefont {R.}~\bibnamefont {Baer}}, \bibinfo {author}
  {\bibfnamefont {A.~T.}\ \bibnamefont {Bell}}, \bibinfo {author}
  {\bibfnamefont {N.~A.}\ \bibnamefont {Besley}}, \bibinfo {author}
  {\bibfnamefont {J.-D.}\ \bibnamefont {Chai}}, \bibinfo {author}
  {\bibfnamefont {A.}~\bibnamefont {Dreuw}}, \bibinfo {author} {\bibfnamefont
  {B.~D.}\ \bibnamefont {Dunietz}}, \bibinfo {author} {\bibfnamefont {T.~R.}\
  \bibnamefont {Furlani}}, \bibinfo {author} {\bibfnamefont {S.~R.}\
  \bibnamefont {Gwaltney}}, \bibinfo {author} {\bibfnamefont {C.-P.}\
  \bibnamefont {Hsu}}, \bibinfo {author} {\bibfnamefont {Y.}~\bibnamefont
  {Jung}}, \bibinfo {author} {\bibfnamefont {J.}~\bibnamefont {Kong}}, \bibinfo
  {author} {\bibfnamefont {D.~S.}\ \bibnamefont {Lambrecht}}, \bibinfo {author}
  {\bibfnamefont {W.}~\bibnamefont {Liang}}, \bibinfo {author} {\bibfnamefont
  {C.}~\bibnamefont {Ochsenfeld}}, \bibinfo {author} {\bibfnamefont {V.~A.}\
  \bibnamefont {Rassolov}}, \bibinfo {author} {\bibfnamefont {L.~V.}\
  \bibnamefont {Slipchenko}}, \bibinfo {author} {\bibfnamefont {J.~E.}\
  \bibnamefont {Subotnik}}, \bibinfo {author} {\bibfnamefont {T.}~\bibnamefont
  {{Van Voorhis}}}, \bibinfo {author} {\bibfnamefont {J.~M.}\ \bibnamefont
  {Herbert}}, \bibinfo {author} {\bibfnamefont {A.~I.}\ \bibnamefont {Krylov}},
  \bibinfo {author} {\bibfnamefont {P.~M.~W.}\ \bibnamefont {Gill}}, \ and\
  \bibinfo {author} {\bibfnamefont {M.}~\bibnamefont {{Head-Gordon}}},\
  }\href@noop {} {\bibfield  {journal} {\bibinfo  {journal} {Mol.\ Phys.}\
  }\textbf {\bibinfo {volume} {113}},\ \bibinfo {pages} {184} (\bibinfo {year}
  {2015})}\BibitemShut {NoStop}%
\bibitem [{\citenamefont {Kurz}\ \emph {et~al.}(2004)\citenamefont {Kurz},
  \citenamefont {F\"orster}, \citenamefont {Nordstr\"om}, \citenamefont
  {Bihlmayer},\ and\ \citenamefont {Bl\"ugel}}]{PhysRevB.69.024415}%
  \BibitemOpen
  \bibfield  {author} {\bibinfo {author} {\bibfnamefont {P.}~\bibnamefont
  {Kurz}}, \bibinfo {author} {\bibfnamefont {F.}~\bibnamefont {F\"orster}},
  \bibinfo {author} {\bibfnamefont {L.}~\bibnamefont {Nordstr\"om}}, \bibinfo
  {author} {\bibfnamefont {G.}~\bibnamefont {Bihlmayer}}, \ and\ \bibinfo
  {author} {\bibfnamefont {S.}~\bibnamefont {Bl\"ugel}},\ }\href {\doibase
  10.1103/PhysRevB.69.024415} {\bibfield  {journal} {\bibinfo  {journal} {Phys.
  Rev. B}\ }\textbf {\bibinfo {volume} {69}},\ \bibinfo {pages} {024415}
  (\bibinfo {year} {2004})}\BibitemShut {NoStop}%
\bibitem [{\citenamefont {Ozaki}\ and\ \citenamefont
  {Kino}(2005)}]{PhysRevB.72.045121}%
  \BibitemOpen
  \bibfield  {author} {\bibinfo {author} {\bibfnamefont {T.}~\bibnamefont
  {Ozaki}}\ and\ \bibinfo {author} {\bibfnamefont {H.}~\bibnamefont {Kino}},\
  }\href {\doibase 10.1103/PhysRevB.72.045121} {\bibfield  {journal} {\bibinfo
  {journal} {Phys. Rev. B}\ }\textbf {\bibinfo {volume} {72}},\ \bibinfo
  {pages} {045121} (\bibinfo {year} {2005})}\BibitemShut {NoStop}%
\bibitem [{\citenamefont {Hutter}\ \emph {et~al.}(2014)\citenamefont {Hutter},
  \citenamefont {Iannuzzi}, \citenamefont {Schiffmann},\ and\ \citenamefont
  {VandeVondele}}]{WCMS:WCMS1159}%
  \BibitemOpen
  \bibfield  {author} {\bibinfo {author} {\bibfnamefont {J.}~\bibnamefont
  {Hutter}}, \bibinfo {author} {\bibfnamefont {M.}~\bibnamefont {Iannuzzi}},
  \bibinfo {author} {\bibfnamefont {F.}~\bibnamefont {Schiffmann}}, \ and\
  \bibinfo {author} {\bibfnamefont {J.}~\bibnamefont {VandeVondele}},\ }\href
  {\doibase 10.1002/wcms.1159} {\bibfield  {journal} {\bibinfo  {journal}
  {Wiley Interdisciplinary Reviews: Computational Molecular Science}\ }\textbf
  {\bibinfo {volume} {4}},\ \bibinfo {pages} {15} (\bibinfo {year}
  {2014})}\BibitemShut {NoStop}%
\bibitem [{\citenamefont {Ma}\ and\ \citenamefont
  {Dudarev}(2015)}]{PhysRevB.91.054420}%
  \BibitemOpen
  \bibfield  {author} {\bibinfo {author} {\bibfnamefont {P.-W.}\ \bibnamefont
  {Ma}}\ and\ \bibinfo {author} {\bibfnamefont {S.~L.}\ \bibnamefont
  {Dudarev}},\ }\href {\doibase 10.1103/PhysRevB.91.054420} {\bibfield
  {journal} {\bibinfo  {journal} {Phys. Rev. B}\ }\textbf {\bibinfo {volume}
  {91}},\ \bibinfo {pages} {054420} (\bibinfo {year} {2015})}\BibitemShut
  {NoStop}%
\bibitem [{\citenamefont {O'Regan}\ \emph {et~al.}(2010)\citenamefont
  {O'Regan}, \citenamefont {Hine}, \citenamefont {Payne},\ and\ \citenamefont
  {Mostofi}}]{PhysRevB.82.081102}%
  \BibitemOpen
  \bibfield  {author} {\bibinfo {author} {\bibfnamefont {D.~D.}\ \bibnamefont
  {O'Regan}}, \bibinfo {author} {\bibfnamefont {N.~D.~M.}\ \bibnamefont
  {Hine}}, \bibinfo {author} {\bibfnamefont {M.~C.}\ \bibnamefont {Payne}}, \
  and\ \bibinfo {author} {\bibfnamefont {A.~A.}\ \bibnamefont {Mostofi}},\
  }\href@noop {} {\bibfield  {journal} {\bibinfo  {journal} {Phys. Rev. B}\
  }\textbf {\bibinfo {volume} {82}},\ \bibinfo {pages} {081102} (\bibinfo
  {year} {2010})}\BibitemShut {NoStop}%
\end{thebibliography}%


\end{document}